\documentclass[twocolumn,preprint]{aastex63}
\usepackage{amsmath}
\usepackage{mathtools}
\usepackage{multirow}
\usepackage[capitalise]{cleveref}
\usepackage{booktabs}
\usepackage{physics}
\usepackage{xspace}
\usepackage{dsfont}

\usepackage[squaren]{SIunits}

\def\bq{\begin{equation}} 
\def\eq{\end{equation}}
\newcommand{\bqa}{\begin{eqnarray}} 
\newcommand{\eqa}{\end{eqnarray}}

\defcitealias{frontiere2017}{FRO17}
\defcitealias{habib2016hacc}{H16}
\defcitealias{Monaghan:1983dn}{MG83}

\newcommand{\Wij}{\ensuremath{W_{ij}}}

\newcommand{\Wrij}{\ensuremath{\mathcal{W}_{ij}^R}}
\newcommand{\Wrji}{\ensuremath{\mathcal{W}_{ji}^R}}
\newcommand{\etacrit}{\ensuremath{\eta_{\text{crit}}}}
\newcommand{\etafold}{\ensuremath{\eta_{\text{fold}}}}

\addunit{\massh}{\mathit{h}^{-1}M_\odot}
\addunit{\Mpch}{\mathit{h}^{-1}Mpc}
\addunit{\Gpch}{\mathit{h}^{-1}Gpc}
\addunit{\kpch}{\mathit{h}^{-1}kpc}
\addunit{\kms}{km\ s^{-1}}
\addunit{\invMpch}{\mathit{h}\ Mpc^{-1}}
\addunit{\GHz}{GHz}
\addunit{\sqamin}{arcmin^2}
\addunit{\sqdeg}{deg^2}

\definecolor{darkgreen}{rgb}{0.05,0.4,0.05}

\creflabelformat{equation}{#2\textup{#1}#3}
\usepackage{scalefnt}
\newcommand\smaller[2][0.85]{{\scalefont{#1}#2}}

\newcommand{\CRKHACC}{\smaller{CRK-HACC}\xspace}
\newcommand{\HACC}{\smaller{HACC}\xspace}
\newcommand{\CRKSPH}{\smaller{CRKSPH}\xspace}
\newcommand{\SWFFT}{\smaller{SWFFT}\xspace}

\renewcommand{\thefootnote}{\fnsymbol{footnote}}

\makeatletter
\g@addto@macro\normalsize{%
  \setlength\abovedisplayskip{8pt}
  \setlength\belowdisplayskip{8pt}
  \setlength\abovedisplayshortskip{8pt}
  \setlength\belowdisplayshortskip{8pt}
}
\makeatother

\begin{document}

\title{Simulating Hydrodynamics in Cosmology with CRK-HACC}

\author{Nicholas~Frontiere}
\affiliation{CPS Division,  Argonne National Laboratory, Lemont, IL 60439, USA}
\affiliation{HEP Division, Argonne National Laboratory, Lemont, IL 60439, USA}
\email{nfrontiere@anl.gov}
\author{J.D.~Emberson}
\affiliation{CPS Division, Argonne National Laboratory, Lemont, IL 60439, USA}
\author{Michael~Buehlmann}
\affiliation{HEP Division, Argonne National Laboratory, Lemont, IL 60439, USA}
\author{Joseph Adamo}
\affiliation{HEP Division, Argonne National Laboratory, Lemont, IL 60439, USA}
\affiliation{Department of Astronomy and Steward Observatory, University of Arizona, Tucson, AZ 85721, USA}
\author{Salman~Habib}
\affiliation{CPS Division, Argonne National Laboratory, Lemont, IL 60439, USA}
\affiliation{HEP Division, Argonne National Laboratory, Lemont, IL 60439, USA}
\author{Katrin Heitmann}
\affiliation{HEP Division, Argonne National Laboratory, Lemont, IL 60439, USA}
\author{Claude--Andr\'e Faucher--Gigu\`ere}
\affiliation{Department of Physics and Astronomy and Center for Interdisciplinary Exploration and Research in Astrophysics (CIERA), Northwestern University, Evanston, IL 60201, USA}

\begin{abstract}
We introduce \CRKHACC, an extension of the Hardware/Hybrid Accelerated Cosmology Code (\HACC), to resolve gas hydrodynamics in large-scale structure formation simulations of the universe. The new framework couples the \HACC\ gravitational N-body solver with a modern smoothed particle hydrodynamics (SPH) approach called \CRKSPH. 
\underline{C}onservative \underline{R}eproducing \underline{K}ernel \underline{SPH} utilizes smoothing functions that exactly interpolate linear fields while manifestly preserving conservation laws (momentum, mass, and energy). 
The \CRKSPH\ method has been incorporated to accurately model baryonic effects in cosmology simulations -- an important addition targeting the generation of precise synthetic sky predictions for upcoming observational surveys. 
\CRKHACC\ inherits the codesign strategies of the \HACC solver and is built to run on modern GPU-accelerated supercomputers.
In this work, we summarize the primary solver components and present a number of standard validation tests to demonstrate code accuracy, including idealized hydrodynamic and cosmological setups, as well as self-similarity measurements. 
\end{abstract}

\keywords{methods: numerical -- cosmology: theory -- hydrodynamics}

\section{Introduction}
\label{sec:introduction} 
Numerical simulations play a vital role in precision cosmology by providing accurate theoretical predictions of structure formation in the universe. 
Mock realizations explore the nature of dark matter and dark energy, probe the origins of primordial fluctuations, and facilitate new ways of investigating fundamental physics, such as the imposition of cosmological limits on the neutrino mass sum. 
Importantly, simulations can produce realistic synthetic measurements to help solve the (statistical) inverse problem of determining cosmological parameters -- providing constraints on the $\Lambda$CDM model -- and to categorize astrophysical and experimental systematics, crucial for the planning, calibration, and validation of observational surveys.

Traditionally, gravity-only N-body simulations have been the primary means to generate mock sky catalogs, attributable to their efficient computational cost and lack of model uncertainties (\citealt{angulo2021} provide an excellent overview of state-of-the-art simulation methods and challenges). The Hardware/Hybrid Accelerated Cosmology Code (\HACC) was developed to performantly run gravity-only simulations on all modern supercomputing platforms, scaling to millions of cores and exploiting heterogeneous hardware such as GPUs (\citealt{habib2016hacc} hereafter \citetalias{habib2016hacc}). However, accurate treatment of baryonic gas and its interplay with dark matter physics has become increasingly important for cosmological predictions, owing to the precise measurements of current and upcoming observational surveys that are sensitive to these effects (e.g., \citealt{vanDaalen2011, harnois2015, copeland2018}).

The field of hydrodynamic astrophysics simulations is rich, and significant progress has been made over the last few decades relating to the modeling of galaxy formation (see \citealt{naab2017} and \citealt{vogelsberger2020} for recent reviews). In this paper, we study an extension of the \HACC framework to include baryonic effects by incorporating a conservative reproducing kernel smoothed particle hydrodynamics (\CRKSPH) solver described in \citet*{frontiere2017} (henceforth abbreviated as \citetalias{frontiere2017}). The combined framework, called \CRKHACC, was developed to maintain all performance design characteristics of \HACC, while improving its predictive capabilities. 

Smoothed particle hydrodynamics (SPH) methods have a long history since the pioneering work of \cite{Lucy1977} and \cite{Gingold1977}, and have become a staple Lagrangian-based approach in astrophysics codes (\citealt{Rosswog2009,springel2010review, price2012review}). Precise enforcement of conservation laws, Galilean invariance, trivial parallelization, and refinement that is inherently adaptive (where particles naturally evolve to high density regions), are some of the many advantages of SPH codes. 

As described in detail in \citetalias{frontiere2017}, \CRKSPH\ was developed to leverage the benefits of SPH, while actively addressing several of its main difficulties, including the inability to exactly reproduce zeroth-order fields, and overly aggressive artificial viscosity models. Recently, there have been further extensions to incorporate radiation transport discretizations as well (\citealt{bassett2021}). This study is devoted to the application of \CRKSPH\ in a cosmology framework, capitalizing on the accuracy of the solver, and serving as a basis for future work modeling numerous baryonic probes -- such as Sunyaev-Zel’dovich maps, gravitational weak lensing measurements, gas-evolved synthetic sky catalogs, and the Lyman-$\alpha$ forest, to name a few. The \emph{Borg Cube} simulation, described in \cite{emberson2019}, was the first application of the \CRKHACC\ solver utilized in this manner, specifically in a non-radiative setting.  

The paper is organized as follows. In Section~\ref{sec:Background}, we provide background information on the cosmological fluid equations, as well as a summary of the \CRKSPH\ approach. Section~\ref{sec:addons} details all the primary modifications and additions to the \HACC framework required in \CRKHACC. To confirm solver correctness, we present a series of validation experiments for idealized hydrodynamic and self-gravitating tests in Section~\ref{sec:evaluaton}. Furthermore, we measure self-similar scaling relations for scale-free initial condition cosmology simulations in Section~\ref{sec:selfsim}. Finally, we conclude with a summary and discussion of future work in Section~\ref{sec:conclude}. 
\section{Background Equations}
\label{sec:Background}
We begin our discussion with an overview of both the expanding universe cosmology equations (Section \ref{sec:CosmoEqns}), and the methodology behind the hydrodynamic \CRKSPH\ solver (Section \ref{sec:HydroEqns}). Unless otherwise specified, we denote the comoving spatial coordinate as $\vb{x}$, where ${\vb{r} = a(t) \vb{x}}$ is the proper position, and $a(t)$ is the cosmic scale factor. 
\subsection{Cosmology Equations}
\label{sec:CosmoEqns}
The set of equations for an ideal fluid in an expanding Friedmann-Robertson-Walker (FRW) universe can be written in comoving units  (\citealt{peebles1980}) as follows:
\begin{align}
    \dv{\rho}{t} &= -\frac{1}{a^2}\rho \div{\vb{p}}, \displaybreak[0]\label{eq:CosmoContEqn}\displaybreak[0]\\
    \dv{\vb{p}}{t} &= -a^2 \frac{\grad{P}}{\rho} - \frac{\grad{\phi}}{a}, \displaybreak[0]\label{eq:CosmoMomEqn}\displaybreak[0]\\
    \dv{u}{t} &= -\frac{1}{a^2}\frac{P}{\rho }\div{\vb{p}} - \frac{\dot{a}}{a}(3(\gamma - 1) + 2)u.\displaybreak[0] \label{eq:CosmoEngEqn} 
\end{align}
The pressure $P$, density $\rho$, and thermal energy per unit mass $u$ couple via the equation of state 
\begin{equation}
\label{eq:EOS}
P = (\gamma - 1)\rho u,
\end{equation}
with adiabatic index $\gamma$ (often set to the monatomic value $\gamma = 5/3$). In our notation, $\vb{p} \equiv a^2 \dot{\vb{x}} = a \vb{v}$, where $\vb{v}$ is the proper peculiar velocity. The scalar (comoving) potential $\phi$ satisfies the Poisson equation 
\begin{equation}
\label{eq:Poisson}
    \laplacian{\phi} = 4 \pi G [\rho(\vb{x},t) - \rho_m (t)],
\end{equation}
where the background matter density $\rho_m = \Omega_m \rho_c$, in a universe with critical density $\rho_c$ and matter fraction $\Omega_m$, including dark matter and baryons. 
The scalar potential is related to the proper gravitational potential ($\Phi$) as
\begin{equation}
   \phi = a\Phi + \frac{1}{2} a^2 \ddot{a}x^2. \displaybreak[0]
\end{equation}

We remark that the fluid equations are written in Lagrangian form, using the convective (material) derivative 
\begin{equation}
\label{eq:matderv}
 \mathrm{d}/\mathrm{d}t = \partial/\partial t + \dot{\vb{x}} \cdot \grad. \displaybreak[0]
\end{equation}
Moreover, the spatial derivatives are measured with respect to $\vb{x}$, e.g., $\nabla \equiv \nabla_x$. 

As is common for numerical solvers, the evolution equations in \CRKHACC\ are evaluated in dimensionless (``tilde'') units via the following transformations
\begin{align}
    \tilde{\vb{x}} \equiv \vb{x}/x_0, \;\tilde{t} \equiv tH_0,\;
    \tilde{\rho} \equiv \rho/\rho_m, \nonumber \displaybreak[0]\\
    \tilde{\text{G}} \equiv \frac{3}{2}\Omega_m \frac{1}{4\pi},\;
    \tilde{\phi} \equiv \phi/(x_0^2 H_0^2), \label{eq:codeunits}
\end{align}
where $x_0$ is the particle-mesh grid spacing $L/n_g$, the ratio of the comoving domain length (in \Mpch) and the number of grid cells in a single dimension. As described in Section \ref{sec:timestep}, the \CRKHACC\ time integrator incorporates steps with equal spacing in units of the scale factor. We introduce the temporal variable
\begin{equation}\label{eq:time}
y \equiv a^\alpha,
\end{equation}
where $\alpha$ is a constant and the differential operators are obtained using the chain rule,
\begin{equation}
\label{eq:chain}
\dv{}{\tilde{t}} = \dv{}{y} \dv{y}{\tilde{t}} = \dv{}{y} \alpha y \frac{H}{H_0}.\displaybreak[0]
\end{equation}
Our canonical choice of time units is defined by $\alpha = 1$.

\subsection{The CRKSPH Scheme}
\label{sec:HydroEqns}
SPH is a Lagrangian approach that discretizes gas into  fluid parcels traveling with the flow velocity. Arbitrary fluid quantities $\psi(\vb{x})$ are generally evaluated using an integral interpolant 
\begin{equation}
\label{eq:SPHint}
    \psi(\vb{x}) = \int{\psi(\vb{x}') W(\vb{x}-\vb{x}',h)\mathrm{d}\vb{x}'},
\end{equation}
where $W(\vb{x},h)$ is a kernel of compact support (finite non-zero extent) with characteristic radius parameterized by the smoothing length $h$. 
Common selections for $W$ are spherically symmetric functions, including B-splines \citep{Schoenberg1969} and Wendland kernels \citep{Wendland1995}. The \CRKHACC\ solver follows the recommendation of \cite{WalterAly2012}, and utilizes a 4th order ($C^4$) Wendland function:
\begin{equation}\label{eq:C4}
    W_{C^4}(\eta,h) = \frac{495}{32\pi h^3}(1-\eta)^6(1+6\eta+\frac{35}{3}\eta^2),
\end{equation}
with $\eta \equiv |\vb{x}|/h$, and unit kernel extent ($W_{C^4}=0$ for $\eta \ge 1$). 
In general, SPH kernels integrate to unity and are required to asymptotically approach a $\delta-$function when $h \rightarrow 0 $, where the interpolant equation becomes exactly reproducing in the continuum limit. 

\cref{eq:SPHint} can be written in a discretized form as a summation over particle interactions between neighbors within the kernel support, viz.
\begin{equation}
\label{eq:SPHdisc}
    \psi_i = \sum_{j}{\psi_j V_j W(|\vb{x}_i-\vb{x}_j|,h_i)} = \sum_{j}{\psi_j V_j \Wij},
\end{equation}
with $V_j$ denoting the volume associated with each particle. In \CRKHACC, we express the particle volume as the inverse of the SPH number density $n_i$, namely
\begin{equation}\label{eq:vol}
V_i^{-1} = n_i = \sum_j W(|\vb{x}_i-\vb{x}_j|,h_i).
\end{equation}
Other representations, such as Voronoi volumes (e.g., \citealt{Hess2010}), can be utilized as a substitute definition.      

\CRKSPH\ improves the accuracy of SPH interpolation by utilizing reproducing kernels (RKs), which exactly interpolate polynomial fields of a specified order \citep[e.g.,][]{Liu1995,Liu1998}. Specifically, we employ first-order RK functions, which are constructed from SPH kernels as follows:
\begin{align}
    \Wrij &= A_i (1 + \vb{B}_i\cdot\vb{x}_{ij})\Wij, \label{eq:RK} \displaybreak[0]\\
    \nabla \Wrij &= \nabla A_i (1 + \vb{B}_i\cdot\vb{x}_{ij})\Wij \notag \displaybreak[0]\\ &+ A_i ( \nabla \vb{B}_i\cdot\vb{x}_{ij} + \vb{B}_i) \Wij \notag \displaybreak[0]\\ &+ A_i (1 + \vb{B}_i\cdot\vb{x}_{ij}) \nabla \Wij \label{eq:gRK},
\end{align}
where $\vb{x}_{ij} \equiv \vb{x}_i - \vb{x}_j$. The scalar and vector coefficients $(A,\vb{B})$ are determined by enforcing the zeroth and first moments of $\Wrij$ integrate to 1 and 0, respectively; i.e., $\int \Wrij = 1$ and $\int \vb{x}_{ij} \Wrij = 0$ (see Appendix A in \citetalias{frontiere2017} for a complete derivation). The resulting kernels exactly reproduce linear fields over the particle interpolants. 

RK functions are in general no longer symmetric for arbitrary particle distributions. The spherical symmetry of traditional SPH kernels allows for a simple discretization of the fluid equations that exactly conserve energy and momentum (summarized in \citealt{Monaghan2005}), a desirable feature of SPH. 
Care must be taken when utilizing RK interpolation to ensure that the system preserves the same invariants.  Originally developed for the Moving Least-squares SPH (\smaller{MLSPH}) approach described in \cite{Dilts1999}, the following inviscid hydrodynamic evolution equations (written for an expanding universe) maintain exact momentum and energy conservation for non-symmetric kernels: 
\begin{align}
    m_i\dv{\vb{p}_i}{t}&=\sum_j \frac{a^2}{2}V_iV_j(P_i+P_j) (\nabla \Wrji - \nabla \Wrij), \label{eq:mom}\\
    m_i\dv{u_i}{t}&=\sum_j{\frac{1}{2a^2}V_iV_jP_j\dot{\vb{x}}_{ji}\cdot(\nabla \Wrji - \nabla \Wrij)}, \label{eq:eng}
\end{align}
where $m_i$ is the mass of each particle and $\dot{\vb{x}}_{ji} \equiv \dot{\vb{x}}_{j} - \dot{\vb{x}}_{i}$. These differential relations for momentum and energy describe the discretization of the first (pressure) term in fluid \cref{eq:CosmoMomEqn,eq:CosmoEngEqn}. We observe that for any interaction pair (i,j) in the summation, ${m_i (\dv{\vb{p}}{t})_{i \rightarrow j} = - m_j (\dv{\vb{p}}{t})_{j \rightarrow i}}$, obeying Newton's third law. 

Although the differential energy listed in \cref{eq:eng} is fully conservative and can be evolved in conjunction with the momentum update from \cref{eq:mom}, in practice \CRKSPH\ utilizes the ``compatible energy'' formalism of \cite{Owen2014} to evolve the internal energy, with modifications outlined in \citetalias{frontiere2017}. The procedure enforces energy conservation for a given momentum change provided by \cref{eq:mom}, as opposed to independently evolving the internal energy separately. The compatible energy update demonstrates favorable behavior in a number of adiabatic tests where entropy invariance is important, and is the fiducial energy integration procedure used in \CRKHACC. 

The compatible differencing of comoving energy $\Delta u_i$ for a particle over a time step $\Delta t$ is calculated as 
\begin{align}
    \Delta u_i &= \sum_j \Delta u_{ij}\Delta t + \frac{\dot{a}}{a}(3(\gamma - 1) + 2)u_i, \label{eq:compEng} \displaybreak[0]\\
    \Delta u_{ij} &= \frac{f_{ij}}{2a^4}\left[ \vb{p}_j + \vb{p}'_j - \vb{p}_i - \vb{p}'_i \right]\dv{\vb{p}_{ij}}{t},\label{eq:compEng2} \displaybreak[0]\\
    \Delta E_{ij} &= m_i\Delta u_{ij} + m_j \Delta u_{ji}, \label{eq:Eij} \displaybreak[0] \\ 
    \vb{p}'_i &= \vb{p}_i + \dv{\vb{p}_{i}}{t}\Delta t,\label{eq:ptdt}\\ 
    f_{ij} &= \left\{\begin{array}{@{}lr@{}}
        \multirow{2}{*}{$\dfrac{\min\left(s_i,s_j\right)}{s_i+s_j},$} & (\Delta E_{ij} \ge 0, s_i \ge s_j)\text{ or}\,\\
                               & (\Delta E_{ij} < 0, s_i < s_j); \;\:\:\:\,\\
        \multirow{2}{*}{$\dfrac{\max\left(s_i,s_j\right)}{s_i+s_j},$} & (\Delta E_{ij} \ge 0, s_i < s_j)\text{ or}\\
                               & (\Delta E_{ij} < 0, s_i \ge s_j), \;\:\:\:\,\\
        \end{array}\right. \label{eq:fij} \displaybreak[0]
\end{align}
where ${s \equiv P/\rho^\gamma}$ is the specific entropy, and ${\dv{\vb{p}_{ij}}{t}}$ is the pair-wise acceleration described by \cref{eq:mom} (using ${\dv{\vb{p}_{i}}{t} \equiv \sum_j \dv{\vb{p}_{ij}}{t}}$). The first term in \cref{eq:compEng} represents the energy change due to the hydrodynamic impulse of each neighbor interaction, while the second term accounts for the energy loss from expansion -- discretizations of the first and second terms in \cref{eq:CosmoEngEqn}. The derivation of $\Delta u_{ij}$ (\cref{eq:compEng2}) follows identically to \citetalias{frontiere2017} (Eq. 61 in that work), requiring a simple transformation to comoving variables ($\vb{v}\rightarrow \vb{\dot{x}}=\vb{p}/a^2, \; \vb{a}_{ij} \rightarrow \frac{1}{a^2} \dv{\vb{p}_{ij}}{t}$). 
The partition function $f_{ij}$ determines the amount of energy transferred between two nodes given their pair-wise momentum change over a given time step. Any definition of $f_{ij}$ will uphold energy conservation as long as $f_{ij} + f_{ji} = 1$. The specific entropy form utilized in \cref{eq:fij} results in a physically motivated energy balancing, where cooler nodes are heated or hotter nodes are cooled, depending on the sign of the change in work for the interaction ($\Delta E_{ij}$).

The SPH evolution equations require the addition of artificial viscosity to properly capture shock hydrodynamics. The traditionally used pair-wise viscosity prescription of \cite{Monaghan:1983dn} (henceforth \citetalias{Monaghan:1983dn}) encapsulates the bulk (linear) and Von Neumann-Richtmyer (quadratic) viscosity within a viscous pressure $Q$, defined in comoving units as:
\begin{align}
    Q_i &= \rho_i (-C_l c_{s,i} \mu_i + C_q \mu^2_i), \label{eq:visc} \\
    \mu_i &= \text{min}(0,\frac{\hat{h}_i}{a}\frac{\vb{u}_{ij}\cdot \vb{x}_{ij}}{|\vb{x}_{ij}|^2 + \hat{h}_i^2\epsilon^2}), \label{eq:mu}
\end{align}
where $\vb{u}_{ij} \equiv \vb{u}_i - \vb{u}_j$ is the difference in the proper velocities ($\vb{u} \equiv \vb{v}+\dot{a}\vb{x}$) of interacting particles, $c_s \equiv \sqrt{\gamma P/\rho}$ is the sound speed, $\epsilon = 0.1$ is a small number to avoid division by zero, and ($C_l, C_q$) are the viscous linear and quadratic coefficients set to (2.0,1.0) in \CRKSPH. Using a normalized smoothing length $\hat{h}_i \equiv h_i/n_h$ confines the spatial dissipation scale to be roughly the local inter-particle separation, where $n_h$ is a resolution parameter specifying the effective number of particles per smoothing length ($n_h = 4$ in our results, see Section \ref{sec:kernel}). The implementation of artificial viscosity is trivial within \cref{eq:mom,eq:eng} by replacing the pressure $P$ with the combined total pressure $P+Q$. 

While the viscosity treatment of \citetalias{Monaghan:1983dn} is simple and conservative, it has been shown to be excessively diffusive in certain situations, leading to the development of a number of modifications over the years \citep[e.g.,][]{Balsara1995,Morris1997,Cullen2010,Read2010}. 
The novel viscosity model described in \citetalias{frontiere2017} was motivated by the work of \cite{Christensen1990} to develop a ``limited'' viscosity variation of \citetalias{Monaghan:1983dn}. 
The peculiar velocity gradient is accurately measured with reproducing kernels by 
\begin{equation}\label{eq:gradv}
\nabla \vb{v}_i = -\sum_j V_j (\vb{v}_i - \vb{v}_j)\otimes\nabla\Wrij.
\end{equation}
The gradient can be utilized to project the proper velocity to the peculiar midpoint of a particle interaction pair, viz.
\begin{gather}
     \hat{\vb{v}}_i \equiv \vb{v}_i - \frac{1}{2}\phi_{ij}\nabla \vb{v}_i\cdot \vb{x}_{ij}, \nonumber\\
     \hat{\vb{u}}_i \equiv \hat{\vb{v}}_i + \dot{a}\vb{x}_i,\label{eq:hatvi}
\end{gather}
where an additional ``limiter function'' $\phi$ is introduced to vary the extent of the projection ($\phi \in [0,1]$). Once the limiter is specified, the implementation of the modified \citetalias{Monaghan:1983dn} viscosity is trivial by simply replacing $\vb{u}_{ij}$ in \cref{eq:mu} with $\hat{\vb{u}}_{ij} \equiv \hat{\vb{u}}_{i}- \hat{\vb{u}}_{j}$. 
Fundamentally, for linear velocity flows, the pairwise difference vanishes, shutting off the viscosity in a regime where numerical dissipation is unnecessary and undesirable.
We emphasize that the projection is restricted to the peculiar component of the particle velocity, and not applied to the Hubble expansion term $(\dot{a}\vb{x})$; otherwise, the total projection would evaluate to ${\hat{\vb{u}}_{ij} = \hat{\vb{v}}_{ij} + (1-\phi_{ij})\dot{a}\vb{x}_{ij}}$, which would erroneously remove the expansion component in limited ($\phi=1$) flows. 

The limiter described in \citetalias{frontiere2017} was constructed to exclusively project the velocity in smooth flows, while returning to the standard non-projected \citetalias{Monaghan:1983dn} viscosity for non-linear interactions. Explicitly, a modified symmetric van Leer limiter \citep{vanLeer1974,Toro1989} applied to the ratio of the velocity gradients of an interaction pair was utilized:
\begin{align}
  \label{eq:phi}
  \phi_{ij} &=  \phi_{VL} \times \phi_{ad}, \displaybreak[0]\\
  \phi_{VL} &= \max\left[0,\min\left[1,\frac{4r_{ij}}{(1+r_{ij})^2}\right]\right],\label{eq:phiVL}\displaybreak[0]\\
  r_{ij} &= \frac{\nabla \vb{v}_i \cdot \vb{x}_{ij} \cdot \vb{x}_{ij}}{\nabla \vb{v}_j \cdot \vb{x}_{ij} \cdot \vb{x}_{ij}}\label{eq:rij},\displaybreak[0]\\
  \phi_{ad} &= 
  \begin{cases} \exp{-\left(\frac{\eta_{ij}-\etacrit}{\etafold}\right)^2}, & \eta_{ij} < \etacrit; \displaybreak[0]\\
  1, &\eta_{ij} \ge \etacrit,
  \end{cases}\label{eq:phiad} \displaybreak[0]\\
  \eta_{ij} &= \min \left(|\vb{x}_{ij}|/h_i,|\vb{x}_{ij}|/h_j\right),
\end{align}
where $\phi_{VL}$ is the standard van Leer limiter, and the adjustment function $\phi_{ad}$ exponentially suppresses the limiter if particles are driven closer together than physically plausible; specifically, in units of $\eta = \vb{x}/h$, we expect particles to be spatially separated on order of $1/n_h$.
Accordingly, the adjustment parameters $(\etacrit,\etafold)$ are fiducially set to (1/$n_h$, 0.2/$n_h$), ensuring that the limiter be exclusively suppressed on scales smaller than the inter-particle separation. 
 
\section{Framework Components}
\label{sec:addons}
A complete description of the software features in \HACC\ is listed in \citetalias{habib2016hacc}. Design elements, such as performant I/O, manual memory management, and portability strategies, have all been carried over to \CRKHACC. In this section, we outline the primary simulation components that are unique to, or modified in, the extended framework. 
These descriptions include the time integrator (\S\ref{sec:timestep}), long and short-range force operators (\S\ref{sec:spec} and \S\ref{sec:short}), particle overloading (\S\ref{sec:overload}), initial condition generation (\S\ref{sec:ICs}), choice of SPH kernel (\S\ref{sec:kernel}), and lastly, summaries of additional physics models (\S\ref{sec:subgrid}), analysis codes (\S\ref{sec:analysistools}), and performance strategies (\S\ref{sec:performance}). 

\subsection{Time Integrator} 
\label{sec:timestep}
The separable Hamiltonian for an ideal fluid described in Section \ref{sec:CosmoEqns} is 
\begin{equation}\label{eq:Hamil}
    H = m\frac{\vb{p}^2}{2a^2} + U(\vb{x},\rho) + m\frac{\phi}{a} + S,
\end{equation}
where $U(\vb{x},\rho)$ is the internal energy, and $S$ represents arbitrary source terms from additional astrophysics models (see Section \ref{sec:subgrid}). 
The Hamiltonian can be split into gravitational $H_G = m\frac{\phi}{a} $ and hydrodynamic $H_H = H_H^K + H_H^T + H_H^S$ terms, where $H_H^K=m\frac{\vb{p}^2}{2a^2}$, $H_H^T = U(\vb{x},\rho)$ and $H_H^S = S$ are the kinetic, thermal and source components, respectively.
The gravitational Hamiltonian can be further decomposed into a sum of fast and slow changing potentials $\phi = \phi_F + \phi_S \rightarrow H_G = H_G^F + H_G^S$; as summarized in Sections \ref{sec:spec} \& \ref{sec:short}, $\phi_S$ represents a long-range potential, which is calculated by solving the Poisson equation (\cref{eq:Poisson}) using a high-order spectral particle-mesh (PM) approach (\citealt{hockney1988}; \citetalias{habib2016hacc}), and $\phi_F$ is a local short-range potential, computed by a modified TreePM method (\citealt{xu1995, bode2000, bagla2002}). 

For a general Hamiltonian system $H(z,t)$, with ${z=(q,p)}$ canonical coordinates, the equations of motion are expressed by ${\dot{z} = \{z,H\}_P}$, where $\{\cdot,\cdot\}_P$ is a Poisson bracket. The dynamic solution $z(t)$ is described by the time evolution operator ${\hat{U}(t) \equiv \exp(-t\{H,\cdot\}_P)}$, where ${z(t) = \hat{U}(t)z(0)}$. 

Let $\hat{U}_i(y)$ represent the time propagator for each individual Hamiltonian component of \cref{eq:Hamil} (e.g., ${\hat{U}_G(y) = \exp(-y\{H_G,\cdot\}_P)}$), where the temporal variable $y$ is specified in \cref{eq:time}.
The evolution operators can be expressed as discretized particle state updates over a given time step $\Delta y$ (similar to \citealt{quinn1997}, with the inclusion of source terms and \HACC units):

\begin{align}
    \hat{U}_G^S(\Delta y) &: \vb{p}_i \mapsto \vb{p}_i - \frac{\nabla \phi_S}{a} \dv{t}{y} \Delta y, \label{eq:UGS} \displaybreak[0]\\
    \hat{U}_G^F(\Delta y) &: \vb{p}_i \mapsto \vb{p}_i - \frac{\nabla \phi_F}{a} \dv{t}{y} \Delta y, \label{eq:UGF} \displaybreak[0]\\
    \hat{U}_H^T(\Delta y)&: \begin{cases}
        \vb{p}_i \mapsto \vb{p}_i + \dv{\vb{p}_i}{t}\dv{t}{y} \Delta y,  \displaybreak[0]\\
        u_i \mapsto u_i + \Delta u_i \dv{t}{y} \Delta y, 
    \end{cases} \label{eq:UHT} \displaybreak[0]\\
    \hat{U}_H^K(\Delta y) &: \vb{x}_i \mapsto \vb{x}_i + \frac{\vb{p}_i}{a^2} \dv{t}{y} \Delta y, \label{eq:UHK} \\
    \hat{U}_H^S(\Delta y)&: \begin{cases}
        \vb{p}_i \mapsto \vb{p}_i + S_p \dv{t}{y} \Delta y,  \displaybreak[0]\\
        u_i \mapsto u_i + S_u \dv{t}{y} \Delta y, \displaybreak[0]\\
        \vb{x}_i \mapsto \vb{x}_i + S_x \dv{t}{y} \Delta y, \displaybreak[0]\\
    \end{cases} \label{eq:UHS}
\end{align}
where we generally label the source terms $S_p$, $S_u$, and $S_x$ to represent any model that affects the relevant state variable. The changes for momentum ($\dv{\vb{p}_i}{t}$) and energy ($\Delta u_i$) are taken from \cref{eq:mom,eq:compEng}, respectively. Note, the derivative $\dv{y}{t}$ (\cref{eq:chain}) is used to convert the time stepping variables to $y-$units. 

We can construct a discretized time integrator $\hat{U}(\Delta y)$ by composing the evolution propagators as follows
\begin{align}
\hat{U}_{\text{SR}}(\Delta y) &= \hat{U}_G^F(\frac{\Delta y}{2}) \hat{U}_H(\Delta y) \hat{U}_G^F(\frac{\Delta y}{2}), \label{eq:USR}\\
\hat{U}(\Delta y) &= \hat{U}_G^S(\frac{\Delta y}{2}) \hat{U}_{\text{SR}}(\Delta y) \hat{U}_G^S(\frac{\Delta y}{2}),\label{eq:Utot}
\end{align}
where $\hat{U}_{\text{SR}}$ encapsulates all of the ``short-range'' operators. 
In the case of a simple inviscid fluid (no artificial viscosity or sources), it can be shown that for a fixed time step $\Delta y$, the above integrator is second-order convergent, and is both symplectic (a volume-preserving phase space map) and time reversible (see, e.g., \citealt{yoshida1990,forest1990,saha1992}). 

Similarly, we can expand the hydrodynamic ($\hat{U}_H$) operator as
\begin{align}
    \hat{U}_V (\Delta y) &= \hat{U}_H^T(\frac{\Delta y}{2})\hat{U}_H^K(\Delta y)\hat{U}_H^T(\frac{\Delta y}{2}), \label{eq:verlet} \\
    \hat{U}_H(\Delta y) &= \hat{U}_V (\Delta y)\hat{U}_H^S(\Delta y).\label{eq:UH}
\end{align} 
In the absence of sources ($H_H^S = 0,\hat{U}_H^S = \mathds{1}$), the fluid integrator becomes the commonly used second-order velocity Verlet operator ($\hat{U}_H = \hat{U}_V$), where our implementation is similar to \cite{Saitoh2016}:
\begin{equation}
    \hat{U}_V (\Delta y): \begin{cases}
        \vb{p}^{\frac{1}{2}} = \vb{p}^0 + \frac{1}{2}\dv{\vb{p}}{t}\left(\vb{x}^0,\vb{p}^0,u^0\right) \dv{t}{y} \Delta y, \displaybreak[0] \\
        u^{\frac{1}{2}} \:\!= u^0 + \frac{1}{2}\Delta u\left(\vb{x}^0,\vb{p}^0,u^0\right)  \dv{t}{y} \Delta y, \displaybreak[0]\\
        \vb{x}^1 \:\!\,\!\!\;= \vb{x}^0 + \vb{p}^{\frac{1}{2}}\frac{1}{a^2} \dv{t}{y} \Delta y, \displaybreak[0]\\
        \vb{p}_\text{p}^{1} \:\!= \vb{p}^{\frac{1}{2}} + \frac{1}{2}\dv{\vb{p}}{t}\left(\vb{x}^0,\vb{p}^0,u^0\right) \dv{t}{y} \Delta y, \displaybreak[0]\\
        u_\text{p}^{1} \:\!\:\!= u^{\frac{1}{2}} + \frac{1}{2}\Delta u\left(\vb{x}^0,\vb{p}^0,u^0\right)  \dv{t}{y} \Delta y, \displaybreak[0]\\
        \vb{p}^{1} \:\!= \vb{p}^{\frac{1}{2}} + \frac{1}{2}\dv{\vb{p}}{t}\left(\vb{x}^1,\vb{p}_\text{p}^1,u_\text{p}^1\right) \dv{t}{y} \Delta y, \displaybreak[0]\\
        u^{1} \:\!\:\!= u^{\frac{1}{2}} + \frac{1}{2}\Delta u\left(\vb{x}^1,\vb{p}_\text{p}^1,u_\text{p}^1\right)  \dv{t}{y} \Delta y, \displaybreak[0]\\
    \end{cases}\label{eq:verletOP}\displaybreak[0]
\end{equation}
using $\vb{p}_\text{p}^1$ and $u_\text{p}^1$ as ``predictive'' momentum and energy at the end of a given step; when coupled to the gravity solver, the input velocities for the derivative calculation include predictions that account for the PM and short-range gravitational accelerations as well.

All sources are completely encapsulated by the $\hat{U}_H^S$ operator in \cref{eq:UH}, which is composed with the Verlet integrator via first-order Strang splitting (\citealt{strang1968}). In future work, we will describe the inclusion of source terms to model the effects of galaxy formation physics.

Evaluating \cref{eq:Utot} using a global synchronized time stepper would require discretized intervals small enough to accurately resolve each operator for all particles. In reality, global stepping is computationally demanding and wasteful, as the dynamic time scales to properly integrate each particle can vary considerably. Given that the long-range gravity operator $\hat{U}_G^S(y)$ is slowly varying by definition, we can ``subcycle'' (\citealt{tuckerman1992, duncan1998}) the short-range operators as follows

\begin{equation}\label{eq:Usub}
    \hat{U}(\Delta y) =\hat{U}_G^S(\frac{\Delta y}{2}) {\underbrace{\Big[ \hat{U}_{\text{SR}}(\Delta y')\Big]}_{\mathclap{\hat{U}_{\text{SR}}(\Delta y') \underset{\text{m--times}}{\cdots} \hat{U}_{\text{SR}}(\Delta y') }}}^m \hat{U}_G^S(\frac{\Delta y}{2}),
\end{equation}
where $\Delta y' = \Delta y/m$, for $m-$integer subcycles. Consistent with the \HACC integrator, the PM step length $\Delta y$ is selected to be evenly spaced in $a$, keeping the number of gravity subcycles fixed (typically $m=4$) to maintain symplecticity.\footnote[2]{CRKHACC optionally supports adaptive gravity timestepping as well, similar to \cite{Springel2005}. Without uniform gravity subcycling, conservation of momentum is lost in addition to symplecticity. Recovering the former can be achieved using second-order hierarchical Hamiltonian splitting (\citealt{pelupessy2012, zhu2021, springel2021}).}

Following the integration approach of \cite{saitoh2010}, the hydrodynamic operator is further subcycled into smaller steps $\Delta y'' = \Delta y'/n$ via the map composition
\begin{equation}\label{eq:Usubsub}
    \hat{U}_{\text{SR}}(\Delta y') = \hat{U}_G^F(\frac{\Delta y'}{2}) {\underbrace{\Big[ \hat{U}_H(\Delta y'') \Big]}_{\mathclap{\hat{U}_H(\Delta y'') \underset{\text{n--times}}{\cdots} \hat{U}_H(\Delta y'') }}}^n \hat{U}_G^F(\frac{\Delta y'}{2}).
\end{equation}
Unlike the gravity operators, which are synchronously applied to all particles, the hydrodynamic integration is performed hierarchically in power-of-two ($n=2^l$) time bins, where the level $l$ is calculated for each particle based on a combined Courant-Friedrichs-Lewy (CFL) and acceleration time-stepping criteria (e.g., \citealt{Hernquist1989,Springel2005}). 
Specifically, gas particles measure minimum time intervals $\Delta y_\text{min,i}$ following
\begin{align}
    v_\text{max,i} &= \max ( v_\text{sig,i}, \; c_s, \; c_{sq}, \; \frac{\hat{h}_i}{a} |\nabla \cdot \vb{u}_i|),\displaybreak[0]\\
    \Delta y_\text{hyd,i} &= \frac{C_\text{CFL} \hat{h}_i}{ v_\text{max,i}} \dv{y}{t}, \displaybreak[0]\\
    \Delta y_\text{acc,i} &= \sqrt{\frac{\min(\hat{h}_i,r_{\rm soft})}{|\vb{a}_\text{tot} |}}\dv{y}{t}, \displaybreak[0]\\
    \Delta y_\text{min,i} &= \min (\Delta y_\text{hyd,i}, \Delta y_\text{acc,i}),
\end{align}
to assign a time step level satisfying $\Delta y_\text{min,i} > \Delta y'/2^l$. 
The Courant condition $\Delta y_\text{hyd,i}$, employs a CFL coefficient $C_\text{CFL}=0.25$, normalized smoothing length $\hat{h}_i$ (as in \cref{eq:mu}), maximum viscous sound speed ${c_{sq} = \sqrt{\zeta Q_i/\rho_i}}$ (\cref{eq:visc} and $\zeta = 4$), velocity divergence magnitude $|\nabla \cdot \vb{u}_i| = | \Tr \nabla \vb{v}_i + 3 \dot{a}|$ (using the trace of \cref{eq:gradv}),
and maximum neighbor signal velocity $v_\text{sig,i} = \max_j v_\text{sig,ij}$ with   
\begin{equation}
v_\text{sig,ij}= \min \left (\frac{(\dot{\vb{x}}_{ij} + \frac{\dot{a}}{a}\vb{x}_{ij})\cdot \vb{x}_{ij}}{|\vb{x}_{ij}|}, 0.0 \right).
\end{equation}
The acceleration criteria $\Delta y_\text{acc,i}$ is taken from \cite{durier2011}, using the combined hydrodynamic and gravity particle (peculiar) acceleration ${\vb{a}_\text{tot} = \frac{1}{a^2} (\dv{\vb{p}_i}{t} - \frac{\nabla \phi}{a})}$ (from \cref{eq:CosmoMomEqn,eq:mom}), and simulation parameter $r_{\rm soft}$ denoting the gravitational softening length.

As described in Section \ref{sec:short}, the short-range force solvers use a tree method, where the number of particles per leaf is roughly a few hundred. For time stepping, individual particles measure the minimum $\Delta y_\text{min,leaf}$ required to resolve all elements in a leaf, which in turn determines a leaf-level $l_\text{leaf}$. Consequently, the leaves themselves are hierarchically time step-binned as opposed to individual particles. Following \cite{durier2011}, leaves can move to smaller time bins as required, but can only jump to shallower levels every other time step. Furthermore, we utilize a time stepping limiter as described in \cite{saitoh2009}; consistent with that work, we enforce the placement of all neighboring interacting leaves in bins that are within $f_l = 2$ time step levels of each other ($l_\text{leaf,i} \le f_l l_\text{leaf,j})$, found to be a necessary condition for accurate propagation of shock waves using hierarchical time stepping. Without a limiter, the level of a neighboring leaf may be too shallow to properly resolve a high-velocity signal, a problem that would not be encountered with global time stepping. The \CRKHACC limiter is technically more conservative than \cite{saitoh2009}, as it is enforced for leaves as opposed to individual particles (see Section \ref{sec:sedovblast} for a demonstration of the limiter accuracy on the Sedov-Taylor blast wave). 

To summarize, at every PM step (length $\Delta y$), all particles apply the slowly varying long-range gravity operator $\hat{U}_G^S$. Each particle then evaluates the short-range propagator $\hat{U}_G^F$ over a smaller gravity-subcycle $\Delta y'$. Particle leaves are further integrated using hydro-subcycles $\hat{U}_H(\Delta y'')$ that can vary in length (level) depending on the dynamics of each particle, and are limited to ensure neighboring leaves are within two levels. In practice, all particles are integrated at the smallest hydro interval, where each particle position is updated using the ``drift'' operator $\hat{U}_H^K$; at every hierarchical subcycle, all (``active'') gas leaves that require force updates will evaluate the necessary derivatives and update their state-variables (following \cref{eq:UH}), while the remaining (``passive'') leaves are simply streamed.\footnote[2]{A note on streaming passive particles: instead of using the $\hat{U}_H^K$ map -- $\vb{x}_i \mapsto \vb{x}_i + \frac{\vb{p}_i}{a^2} \dv{t}{y} \Delta y,$ which simulates linear trajectories with constant velocity -- we use a correction to integrate parabolic trajectories via $\vb{x}_i \mapsto \vb{x}_i + \frac{\vb{p}_i}{a^2} \dv{t}{y} \Delta y + \frac{1}{2} \dv{\vb{p}_i}{t}\dv{t}{y} \left( \Delta y \right)^2$ with constant acceleration. The correction results in the same total drift at the end of an active step, but allows for a more accurate integration at the smaller time intervals where the particle is passive.} At the conclusion of the hydro subcycles, the next gravity subcycle operator is applied, and the process repeats until the entire PM step interval has been integrated.

\subsection{Gravitational Spectral Solver}
\label{sec:spec}
The long-range (LR) gravity force operator in \CRKHACC\ is lifted from the \HACC\ framework described in \citetalias{habib2016hacc}. Each species is mass deposited on a grid using Cloud-in-Cell (CIC) interpolation. The Poisson equation (\cref{eq:Poisson}) is solved with a spectral method that employs a Gauss-sinc $k$-space filter, a fourth-order Super-Lanczos spectral differentiator (\citealt{hamming1998}), and a sixth-order Green's function.
This approach provides a well-behaved decomposition of the gravitation potential into a long-range contribution from the PM solver, and a short-range component at a fixed handover scale $r_s$ (typically set to a few grid units). 

The LR force calculation is \smaller{MPI} communication dominant, requiring an efficient and scalable Fast Fourier Transform (FFT). The \HACC\ FFT package called \SWFFT\ is publicly available,\footnote[7]{\url{https://git.cels.anl.gov/hacc/SWFFT}} and uses a 3D ``pencil'' data decomposition for scalability. \SWFFT\ has run effectively on millions of \smaller{MPI} ranks and is a key component of the \HACC\ performance design strategy. As described in Section \ref{sec:timestep}, the LR gravity operator is slowly varying, allowing for subcycling of computationally demanding short-range solvers that are node-local and do not require \smaller{MPI} communication. Thus, \HACC\ and, by extension \CRKHACC, achieve ideal scaling using the portable spectral solver, while modularizing the short-range kernels to small code segments that are individually optimized (and potentially rewritten in domain specific languages such as CUDA) for the targeted hardware of a given machine.

\subsection{Short-range Force Solvers}
\label{sec:short}
The \CRKHACC\ short-range gravitational force operator is similar to the ``TreePM'' solver in \HACC. Particles are sorted into 3D chaining mesh (CM) bins (\citealt{hockney1988}) of side-length equal to the force handover scale $r_s$, whereby elements in each cell need only interact with themselves and neighboring bins. In parallel, each cell is further decomposed using a recursive coordinate bisection (RCB) tree (\citealt{gafton2011}), where leaves are recursively subdivided into two groups using a division line passing through the center of mass of a given enclosing leaf; the division line is drawn perpendicular to the longest side of each partition. The base leaves contain roughly $N_{ppl}$ particles, where $N_{ppl} \in [128,256]$ is typically optimal on GPU hardware. For efficiency, the tree construction is only performed once per gravitation PM step, where leaf metadata (bounding box information) is updated at the smallest time steps when particles are streamed. This means that the volume partitioning of leaves will overlap over the course of a PM step, and chaining mesh domains will become ``fuzzy.''

An important modification in \CRKHACC\ is the indexing of the leaves. As opposed to storing the full tree hierarchy, only the base leaves are kept in a flattened contiguous data structure. Retaining the entire tree allows for efficient interaction calculations that can exploit multipole approximations of parent leaves, where leaf pairs that do not satisfy an opening angle criterion can interact with their monopole/quadrupole moments without requiring deeper traversals (see, e.g., \citealt{gafton2011, springel2021}). While this approach is optimal on a CPU, traversing a tree data structure can be difficult on GPUs, which prefer contiguous simple data accesses. We, therefore, exploit the locality of the CM data structure, which already restricts all leaf communication to cell neighbors, and assemble interaction lists by processing the metadata of all base leaves in a given cell neighborhood. Moreover, since the leaves are relatively fat (as $N_{ppl}$ is large), multipole interactions of base leaves can still significantly improve efficiency by not requiring as many direct $N^2$ leaf interactions.

The \CRKHACC\ hydrodynamic force solver utilizes the same tree data structure described above for gravity. 
The leaf metadata includes a baryon interaction domain, calculated by extending the bounding box to enclose the maximal smoothing length of the particles within the leaf.  
At every hydro subcycle, each active leaf assembles all interaction leaf pairs by visiting neighboring CM cells within the designated interaction distance. Again, as the data structure is contiguous, this operation is optimal on GPUs. Moreover, all hydro operators utilize the same interaction list, which need only be assembled once per hydro subcycle, and reused for each kernel.

\subsection{Particle Overloading}
\label{sec:overload} 

A typical 3D domain decomposition in cosmology simulations assigns cuboid \smaller{MPI} rank (node) volumes with edge extents ranging from $\unit{32-500}{\Mpch}$. Particles within each local partition are ``alive'', and evolved following the integration approach detailed in Section \ref{sec:timestep}. To reduce neighboring communication between nodes required to resolve boundaries, \CRKHACC\ employs a particle overloading approach discussed in \citetalias{habib2016hacc}. Each node extends its local boundary by an ``overload'' length (commonly $\unit{2-10}{\Mpch}$), where tracers are replicated from the overlapping domains of neighboring ranks, and marked as ``ghost'' particles. The overload regions are analogous to ghost zones in grid codes, where in our case the alive particles are accurately updated using the ghost copies. 

During each PM step, particles are CIC interpolated to a grid, where the gravitational spectral solver (Section \ref{sec:spec}) calculates the long-range potential. For each of the smaller subcycles (both gravitational and hydrodynamic), elements in the alive node regions accurately calculate the short-range forces using a combination of the alive and ghost particle sets, without the need for communication from other ranks. The ghost particles are similarly updated, with the exception that particles near the extended domain edge do not update their local forces; specifically, particles within the gravitational handover length $r_s$ of the domain boundary are only streamed, to avoid significant errors due to the anisotropic distribution at the edges. Nevertheless, in baryonic simulations, we still update the geometric properties (volume and density) of edge particles, particularly if they interact with alive neighbors. 

The overload lengths are selected to be large enough, such that the integration errors incurred at the boundary do not ``leak'' into the alive regions. Periodically, ghost particles are ``refreshed'', where the existing replications are removed and replaced by updated alive particles from neighboring ranks. In \CRKHACC\ this is done every PM step to be conservative and memory efficient, however this can be extended to longer time intervals with appropriately enlarged overload lengths. 

We emphasize how the overloading procedure designed originally for the gravity forces in \HACC is appropriate for \CRKHACC, which has the inclusion of hydrodynamic interactions. Unlike the short-range gravity force, which is restricted to a finite interaction radius $r_s$, hydrodynamic forces use an approximately fixed neighbor extent and can therefore include longer spatial interactions. However, in the default configuration of cosmology simulations, $r_s$ is roughly the same as the smoothing length for a particle distribution of overdensity $\delta \sim 1$, and consequently, in clustered regimes the hydro forces are much smaller in extent than the gravitational length scales. As a result, the error leakages from the boundaries are typically much smaller for the hydro forces in dense distributions. In void regions ($\delta \ll 1$), smoothing lengths can grow very large, where naively requiring encompassing overload lengths would be highly memory inefficient. 
Conveniently, we have found that using the standard gravity-informed overload extents incur negligible errors in baryon simulations, as the void regions are already heavily undersampled, where relaxing resolution fidelity is in fact inconsequential. 

We conclude with two remarks about overloading. Firstly, the procedure is extensible to the inclusion of astrophysical subgrid sources (summarized in Section \ref{sec:subgrid}), where attention to feedback model extent and stochastic processes is required; we leave the discussion of the particular nuances involved in these models to a future communication. Secondly, in GPU-accelerated systems, device memory is commonly imbalanced compared to host RAM. In response, we further decompose local node extents into smaller (similarly overloaded) subvolumes that can be independently transferred and evolved on a GPU over a full PM step. This procedure has the additional benefit that the smaller domains can be optionally load-balanced across the machine, which can be particularly beneficial when simulations evolve into highly imbalanced low-redshift clustered regimes; this approach was first discussed in \citetalias{habib2016hacc} (Section 3.3 in that work), and has been incorporated into \CRKHACC.   

\subsection{Cosmological Initial Conditions}
\label{sec:ICs}

As described in \cite{heitmann2010}, the \HACC\ initial conditions (IC) generator distorts a grid arrangement of particles using the Zel'dovich approximation \citep{Zeldovich1970}; a random white noise generator\footnote[2]{We use the Threefry counter-based pseudorandom number generator described in \citet{salmon2011}, which has the benefit of being highly efficient and parallelizable while also producing machine-independent random numbers.} is convolved with a total matter transfer function (provided by e.g, {{\small CAMB}\xspace} \citealt{Lewis_2000} or 
{{\small CLASS}\xspace} \citealt{Blas2011}) and is subsequently mapped to a displacement field. The initial velocities in such an approach are proportional to the displacement, obviating the need for velocity transfer functions. 

The \CRKHACC\ multi-species extension of the IC generator allows for the flexibility to initialize both grid and glass particle distributions. In the latter case, we construct the glass by evolving a random particle distribution in an Einstein--de Sitter universe with repulsive gravity until a quasi-equilibrium state is reached \citep{white1994formation}. As the resulting particle configuration is pseudo-random, we perform an inverse CIC interpolation from the displacement field to the glass positions, where we minimize the interpolation error by convolving the white noise field with a spectral sinc filter (\citealt{hockney1988}). 

For multi-species grid initial conditions, we lay a secondary lattice for gas particles, generally offset by a distance $\Delta$ from the dark matter arrangement. To avoid the necessity of interpolation to the baryon locations, we apply the Fourier phase shift $e^{-ik\Delta}$ to the white noise field. This procedure circumvents the necessity to calculate displacements for both species on a larger grid with, e.g., $2 \times N^3$ elements in the case $\Delta=1/2$ in grid units (e.g., \citealt{Valkenburg2017}).

Multi-species ICs, in particular, come with further challenges. For example, it is a common approach in cosmological hydrodynamic simulations to initialize both the
dark matter and baryons using the total matter transfer function (as described above). However, this is technically incorrect as the two species have noticeably distinct transfer functions at typical initialization redshifts owing to evolutionary differences in the early universe \citep[e.g.,][]{Angulo2013}. Consequently, we have added the capability to generate dark matter and baryons distributions with their respective density and velocity transfer functions. Different studies (e.g., \citealt{Angulo2013,Bird2020,Hahn2021}) have reported that this approach can introduce  large-scale biases in the growth of each species when the force resolution is finer than the mean inter-particle separation. We have found the same behavior in \CRKHACC\ and plan to investigate accordingly in upcoming work. 

There are a number of additional complexities in both gravity-only and multi-species initial conditions that should be stated. Firstly, the initial displacements and velocities must be sufficiently precise, a quality dependent on the order of perturbation theory and the initial redshift $z_{in}$. 
Moreover, consideration of the force solver accuracy when the displacements are small is necessary, as well as avoidance of nonphysical grid-imprinting, both of which are again conditional on $z_{in}$. 
We refer the reader to, e.g.,  \cite{lukic2007,heitmann2010,LHuillier2014} as references, noting that we intend to explore these effects using \CRKHACC\ more thoroughly as well.

\subsection{Kernel Selection and Smoothing Length}
\label{sec:kernel}
At a given resolution, the choice of smoothing kernel $W(\vb{x},h)$ has been shown to be arbitrary in \CRKSPH\ (see Appendix D of \citetalias{frontiere2017}). Since the RK corrections specified in \cref{eq:RK,eq:gRK} will modify a general SPH kernel function to be first-order consistent, it is unsurprising that the particular selection of interpolant is largely inconsequential. As described in Section \ref{sec:HydroEqns}, \CRKHACC\ utilizes the $C^4$ Wendland Kernel (\cref{eq:C4}). This selection was made to avoid the piece-wise functional definition of B-splines (particularly ill-suited for GPUs), in addition to being the ``optimal'' SPH kernel for simulations with $N \sim 200$ particle neighbors (as studied in \citealt{WalterAly2012}). The standard resolution choice in \CRKHACC\ is for the smoothing length to radially enclose 4 particles ($n_h=4$, for $C^4$),  corresponding to approximately $N_{res}=268$ neighbors.  

The particle smoothing length $h_i$ is related to the total number of neighbors by
\begin{equation}\label{eq:hi}
N_i = \frac{4}{3} \pi h_i^3 n_i = \frac{4}{3} \pi h_i^3 \sum_j W(|\vb{x}_i-\vb{x}_j|,h_i),
\end{equation}
with $n_i$ approximating the particle number density using an SPH summation (see \cref{eq:vol}). This equation is commonly solved iteratively (e.g., \citealt{Cullen2010, Hopkins2015}) to determine $h_i$ given a target resolution $N_i = N_{res}$. While convergence of the root-finding can be fast, on a GPU it is programmatically simpler to evolve $h_i$ over time than to implicitly determine it. 

To ensure a well-behaved solution for the evolved smoothing length, we follow the approach of \cite{thacker2000}. On a given time step, we measure the current neighbor count $N_i$ from \cref{eq:hi}. The smoothing length is then adjusted following 
\begin{equation}\label{eq:thacker}
    \tilde{h}_i = h_i (1-\lambda+\lambda \Delta_N),
\end{equation}
where $\lambda$ is a weighting coefficient, and $\Delta_N^3 \equiv (N_{res}/N_i)$ is the ratio of the target and measured neighbor counts. If $\Delta_N=1$, $\tilde{h}_i = h_i$ by construction, leaving the smoothing length unaltered. The weighting function is given by
\begin{equation}
    \lambda = \begin{cases}
        0.2(1+\Delta_N^2), & \Delta_N< 1;\\
        0.2(1+1/\Delta_N^3), & \Delta_N\ge 1.
    \end{cases}
\end{equation}
We have confirmed that the evolved smoothing length procedure produces similarly accurate solutions compared to iterative solvers for the relevant hydrodynamic and cosmological problems studied using \CRKHACC.  

\subsection{Additional Physics}
\label{sec:subgrid}

Sub-resolution (``subgrid'') models are necessary to emulate astrophysical processes -- such as those involved in galaxy formation -- that occur at scales smaller than the resolution limit of a simulation. A detailed investigation of the validation and calibration of subgrid models is beyond the scope of this paper, where we restrict our focus to non-radiative simulations. However, we will briefly summarize the additional physics options currently implemented within \CRKHACC\ for completeness.

The framework includes a radiative cooling and heating model that assumes gas to be optically thin and present in a spatially uniform time-varying ultraviolet background taken from either \cite{haardt2012} or \cite{faucher2020}. 
Similar to \cite{wiersma2009chemical}, the cooling rates are calculated using the simulation code \smaller{CLOUDY},\footnote[2]{\url{https://gitlab.nublado.org/cloudy/cloudy/-/wikis/home}} which further accounts for the inclusion of metals, combined with inverse Compton cooling from the cosmic microwave background (CMB). In practice, metal-line cooling is calculated from the total gas metallicity, as opposed to tracking element-by-element contributions. 
This simplification is justified because we will not resolve the necessary scales to properly model individual chemical yields, in addition to element corrections marginally affecting the overall photoionization model, particularly on cosmological scales (see, e.g., \citealt{wiersma2009chemical}). 
For a given UV background model, we use \smaller{CLOUDY} to produce a complete cooling rate table in four dimensions -- metallicity, density, temperature, and helium abundance for a range of redshifts. 
The cooling function for each particle is interpolated from this table, where the evolved energy is calculated using the exact integration technique of \cite{townsend2009exact}. 

Star formation and supernova feedback are treated using the hybrid multi-phase model described in \cite{springel2003cosmological}; above a specified density threshold, interstellar medium (ISM) gas is evolved via an effective equation of state, where stars form stochastically. Additionally, galactic outflow (wind) models are incorporated to quench star formation. The generated winds are likewise stochastic and scale with the star formation rate, following similarly to the work presented in \cite{Vogelsberger2014} and \cite{pillepich2018}. Lastly, star particles chemically enrich the ISM gas over time. This is achieved by integrating the average stellar enrichment rates from the FIRE simulations (\citealt{hopkins2018fire}). 

Finally, we have incorporated a selection of active galactic nuclei (AGN) models based on the standard prescriptions described in \cite{springel2005AGN,sijacki2007,booth2009,weinberger2016simulating}. AGNs are represented by black hole ``sink'' particles that stochastically accrete nearby gas. The AGN mass grows proportionally to the Bondi-Hoyle-Lyttleton rate (\citealt{bondi1944,hoyle1939}), capped to never exceed the Eddington limit. Every halo above an assigned mass limit is seeded with a black hole, and proximate AGN particles are allowed to merge. AGN feedback is implemented using a combination of thermal and kinetic sources at a rate proportional to the mass accretion. In ``quasar'' or high accretion mode, black holes inject thermal energy to the surrounding gas. For ``radio'' low accretion black holes, kinetic energy is deposited stochastically on neighboring gas particles. 

\subsection{Analysis Capabilities}
\label{sec:analysistools}
\textsc{CosmoTools} is a software toolkit in \HACC\ used for both in situ and post-processing analyses. Standard queries involve friends-of-friends (FOF, \citealt{davis1985}) and spherical overdensity (SO, \citealt{lacey1994}) halo finding, including mass function and concentration measurements (\citealt{child2018}), as well as the generation of detailed property and profile catalogs. \HACC\ also includes tools for extracting particle light-cones, determining correlation functions and power spectra, merger tree construction, and halo ``core'' tracking (see, \citealt{heitmann2021} and \citealt{frontiere2022farpoint} for detailed descriptions of \HACC\ outputs from recent state-of-the-art \emph{Last Journey} and \emph{Farpoint} simulations). 

All analysis codes have been incorporated into \CRKHACC, where outputs are generated for each particle species when applicable -- such as stellar, gas and dark matter power spectra, light cones, and halo profiles. Baryon-specific measurements, including computing  galaxy catalogs (similar to \citealt{schaye2015eagle}), state variable (density, temperature, entropy) phase diagrams, global statistics (e.g., star formation rate density), and Lyman-$\alpha$ skewer tracings (following \citealt{lukic2014lyman}) have also been implemented. The tools have been developed for optimum performance on GPUs, where we utilize the publicly available ArborX software\footnote[2]{\url{https://github.com/arborx/ArborX}} (\citealt{arborx2020}) for a selection of computationally demanding tasks, particularly halo/galaxy finding; for example, the GPU-accelerated FOF halo finder is over an order of magnitude faster than an equivalent CPU-optimized threaded algorithm. 
A detailed description of the \CRKHACC\ baryon analysis pipeline and GPU-performance optimization will be reported in a forthcoming paper.

\subsection{Performance Strategy}
\label{sec:performance}

We conclude this section with a brief discussion regarding the general \smaller{MPI+`X'} performance approach employed within the \CRKHACC\ framework. 
As described in Section \ref{sec:spec}, the \smaller{MPI}-based spectral solver encompasses the primary communication dominated operations of the code, which in turn, dictates the weak-scaling performance. Utilizing the custom FFT package \SWFFT, \HACC\ achieves near ideal weak-scaling up to millions of ranks regardless of heterogeneous hardware concerns for a given machine (see, e.g., \citealt{habib2012universe} for parallel efficiency measurements). 

On the other hand, the time-to-solution and peak sustained FLOPS performance are dominated by the compute-bound short-range force kernels (Section \ref{sec:short}) that are restricted to node-local data operations. The data locality is further exploited using overloaded cuboid sub-volumes (Section \ref{sec:overload}), which can be optionally distributed across the machine using standard greedy load-balancing schemes. 
In practice, a limited set of modularized compute functions are optimized to target the native machine hardware in order to maximize performance of the force interactions. Each kernel is coded in the necessary domain-specific semantics (CUDA, OpenCL, SYCL, HIP, etc., which we generally label as `X') to facilitate comprehensive performance tuning. Common considerations include optimizing instruction and data caching, managing data transfers from logically separate domains, minimizing register pressure, and maximizing memory bandwidth in a multi-level cache configuration. The downside of this approach is the necessity to rewrite the kernels for each new hardware encountered, sacrificing ``strict'' portability for performance -- albeit mitigated by the limited scope of the compute-heavy kernels that are considered. 

Utilizing the described hybrid approach has facilitated the ability to run extreme-scale simulations on a diverse set of machines that utilize Nvidia GPU \citep{frontiere2022farpoint, heitmann2015q}, Intel KNL \citep{emberson2019}, IBM BG/Q \citep{heitmann2021, heitmann2019}, and IBM Cell hardware \citep{habib2009hybrid,pope2010accelerated}. While a detailed performance analysis of the \CRKHACC\ additions is beyond the scope of this paper, similar studies such as \cite{habib2012universe} and \cite{habib2013hacc} will be published, focusing particularly on the upcoming exascale machines that will target AMD and Intel GPU platforms.


\section{Standard Validation}
\label{sec:evaluaton}
With the framework components of \CRKHACC\ specified, we now present a range of conventional tests to evaluate the implementation of the new hydrodynamic solver.
\CRKHACC\ utilizes compile-time options that can simulate fluids in a non-cosmological background using natural units. We individually validate the \CRKSPH\ integrator with traditional fluid experiments (Section \ref{sec:evaluation_hydro}), in addition to presenting results from gravitational collapse and standardized cosmological simulations (Section \ref{sec:evaluation_gravhydro}). 

\subsection{Pure Hydrodynamic Tests}\label{sec:evaluation_hydro}
Idealized fluid simulations are important to establish efficacy of the SPH solver. Here, we consider three standard hydrodynamic test problems: the \hyperref[sec:sodproblem]{Sod shock tube}, \hyperref[sec:sedovblast]{Sedov-Taylor blastwave}, and the \hyperref[sec:kelvinhelmholtz]{Kelvin-Helmholtz instability}. These experiments probe fluid mixing, shock heating, and solver accuracy. While by no means exhaustive, they serve to validate the incorporation of the fluid solver into the \HACC\ framework. For a broader investigation of fluid experiments using \CRKSPH\ (and comparisons with other methods), we refer the reader to \citetalias{frontiere2017}. In general, all measurements presented in this section are in close agreement with \smaller{Spheral},\footnote[7]{\url{https://github.com/LLNL/spheral}} the code used for the investigations in that work. These results are important as the \CRKHACC\ framework includes implementation differences compared to \smaller{Spheral}, for example the use of hierarchical time stepping, which require validation. 

\begin{figure}[tbp]
    \includegraphics[width=0.9\columnwidth]{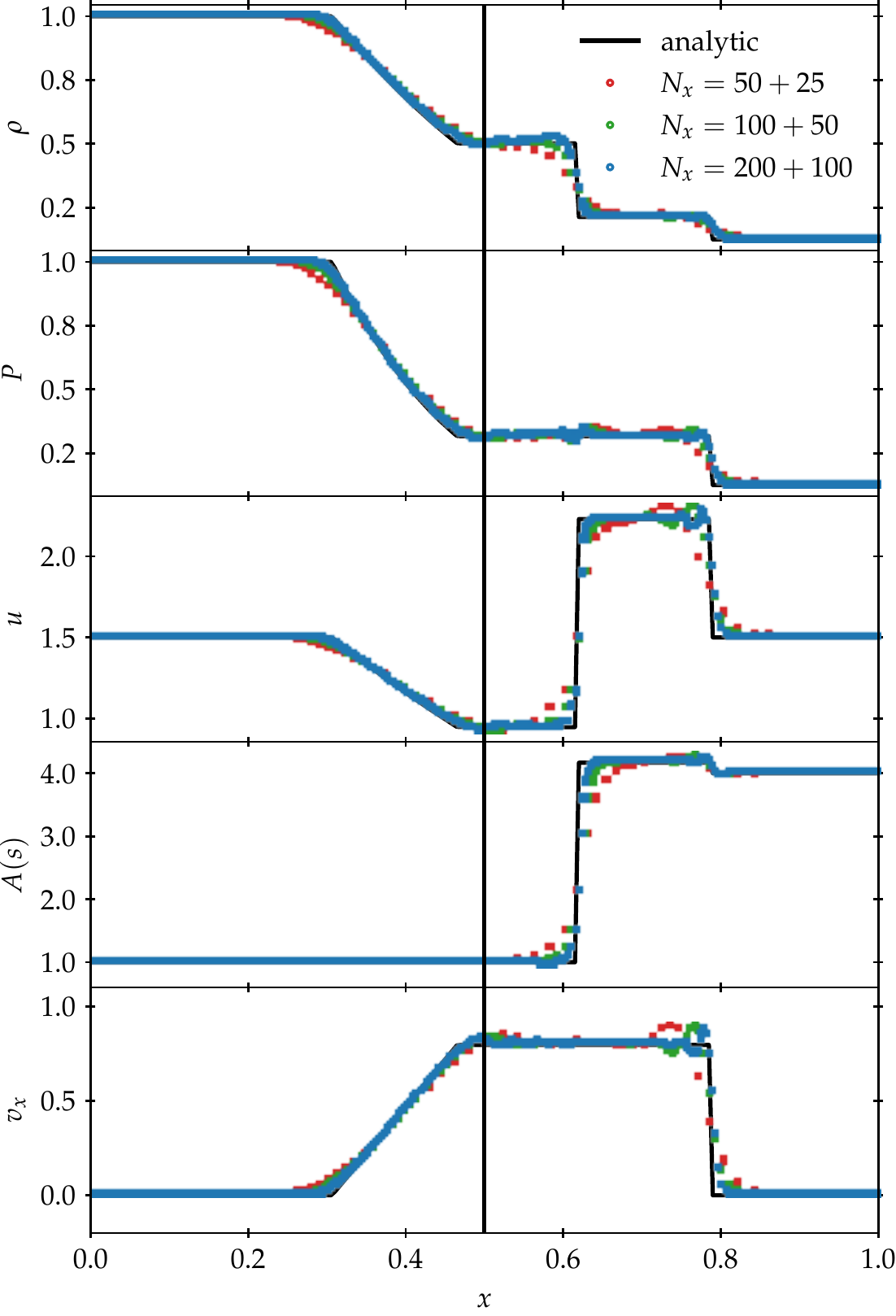}
    \caption{Results of the 3D Sod shock tube test run at $t=0.15$ with 1D-resolutions of $N_x=50+25$, $100+50$, and $200+100$ particles across a $[0, 1)$ domain, evenly replicated along the remaining dimensions. The simulations converge to the analytic prediction (black line). Although the panels include all particles, no scatter is visible at each x-coordinate, demonstrating that the 1D solution is well maintained.}
    \label{fig:sodproblem}
\end{figure}

\begin{figure}[tbp]
    \includegraphics[width=\columnwidth]{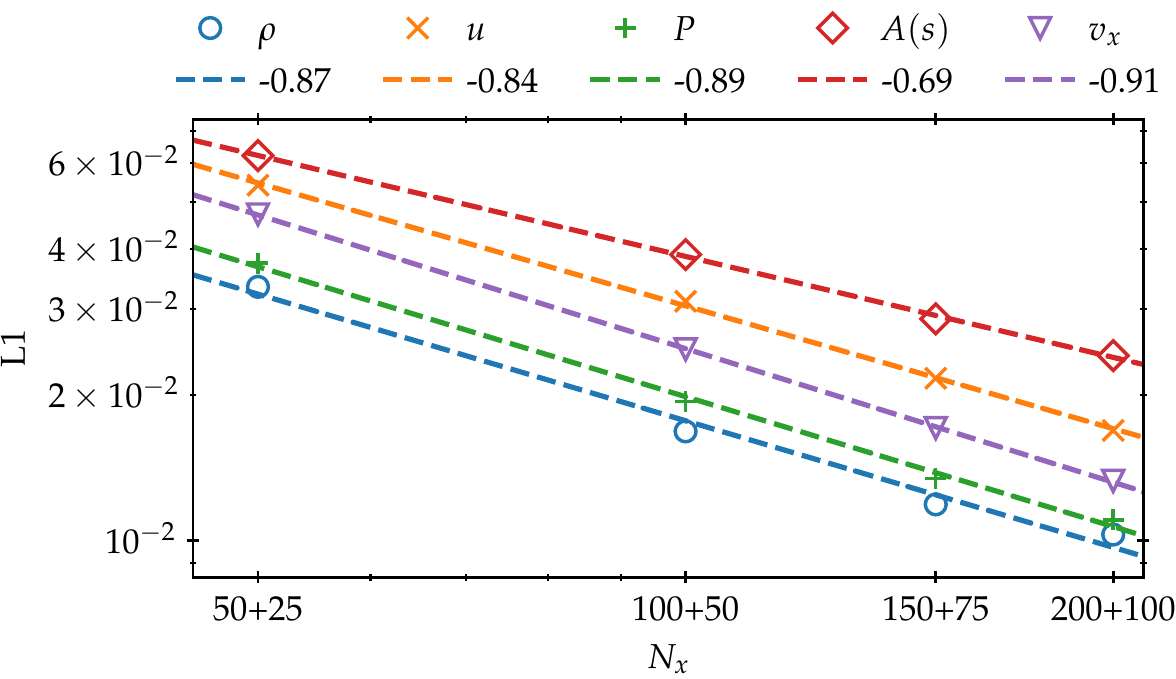}
    \caption{
    L1-norm convergence of the density, energy, pressure, entropy, and velocity to the analytic solution of the Sod shock tube problem with increasing resolution. The dashed lines show the best-fit power law $N^\alpha$, with exponents given in the legend. Appropriately, the state-variables converge roughly linearly ($N^{-1}$), as expected for a shock test.
    }
    \label{fig:sod_convergence}
\end{figure}

\subsubsection{Sod Shock Tube}\label{sec:sodproblem}

The Sod test \citep{Sod1978} is a classic 1D Riemann problem used to probe the accuracy and convergence of hydrodynamic codes in the presence of shocks. It studies the evolution of a sharp interface between two gases of differing density and pressure, brought into contact. In general, shock tests violate the inviscid assumption of the SPH fluid equations, wherein artificial viscosity is required to suitably simulate this type of phenomena. Accordingly, the Sod problem provides an appropriate validation test for the viscosity model, especially as is it accompanied by an analytic solution. 

Conventionally, the Sod test is measured with one interface at the center of a domain of unit length with reflective boundaries. In our setup, we initialize two Riemann interfaces located at $x=0.5$ and $x=1.5$ in a periodic domain $x\in [0,2)$. This configuration ensures that the resulting shocks will not interfere up to the final time $t=0.15$, where the measurements are made over the isolated domain $x\in [0,1)$, effectively replicating the traditional setup.

\begin{figure*}[tp]
    \includegraphics[width=\textwidth]{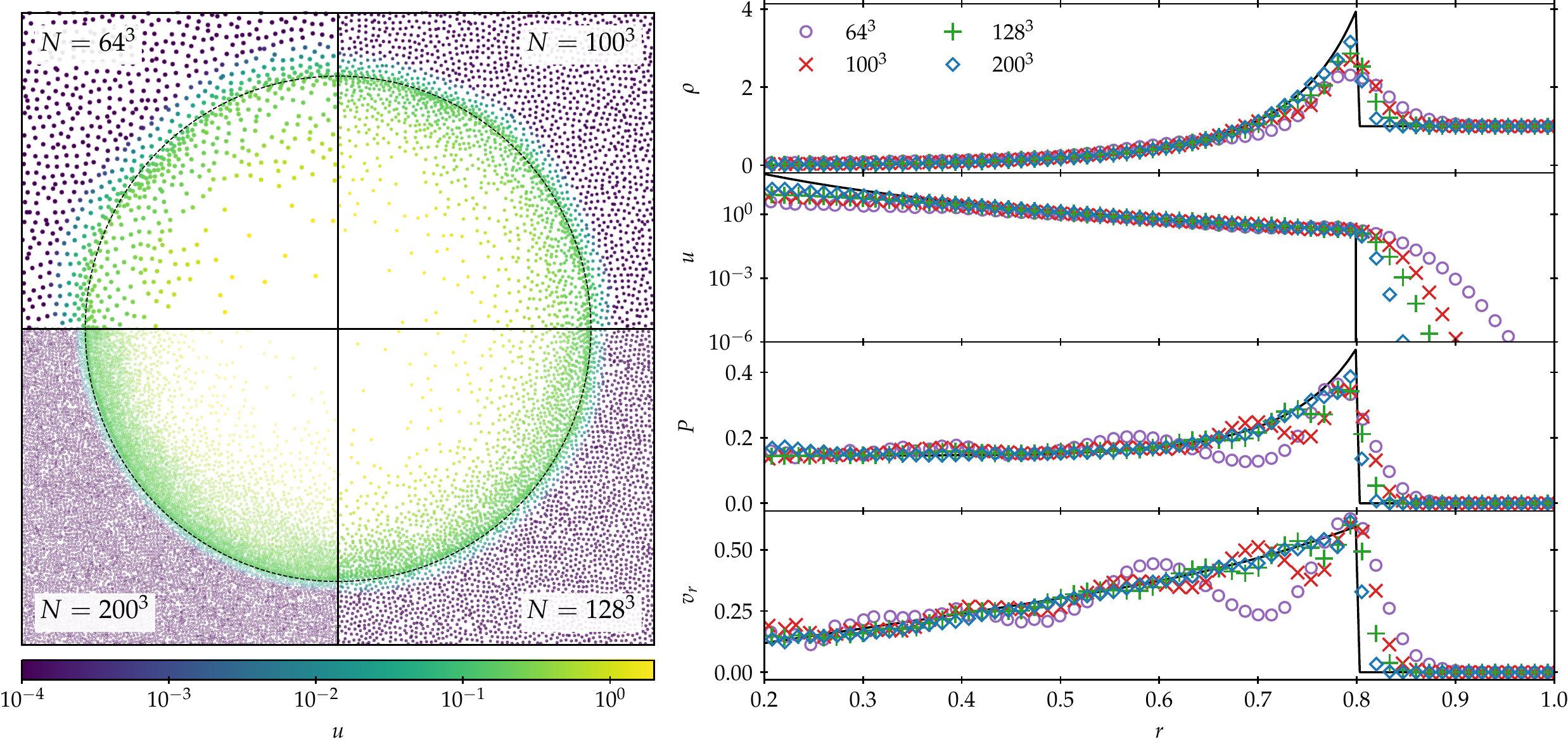}
    \caption{Results of the Sedov-Taylor blast wave problem. \textbf{Left:} particle distribution at time $t_\mathrm{target} \simeq 0.402$ in a slice of thickness $2/N$, colored by the specific energy $u$ for four different resolution runs. The point size is scaled with the particle mass. \textbf{Right:} radial profiles of density, specific energy, pressure and radial velocity, obtained using the median value for each of the 60 bins across a domain $r\in[0.2, 1.0]$. The simulations converge to the analytic solution (black line) at the target time.}
    \label{fig:sedov}
\end{figure*}

\begin{figure}[tp]
    \includegraphics[width=\columnwidth,height=\columnwidth]{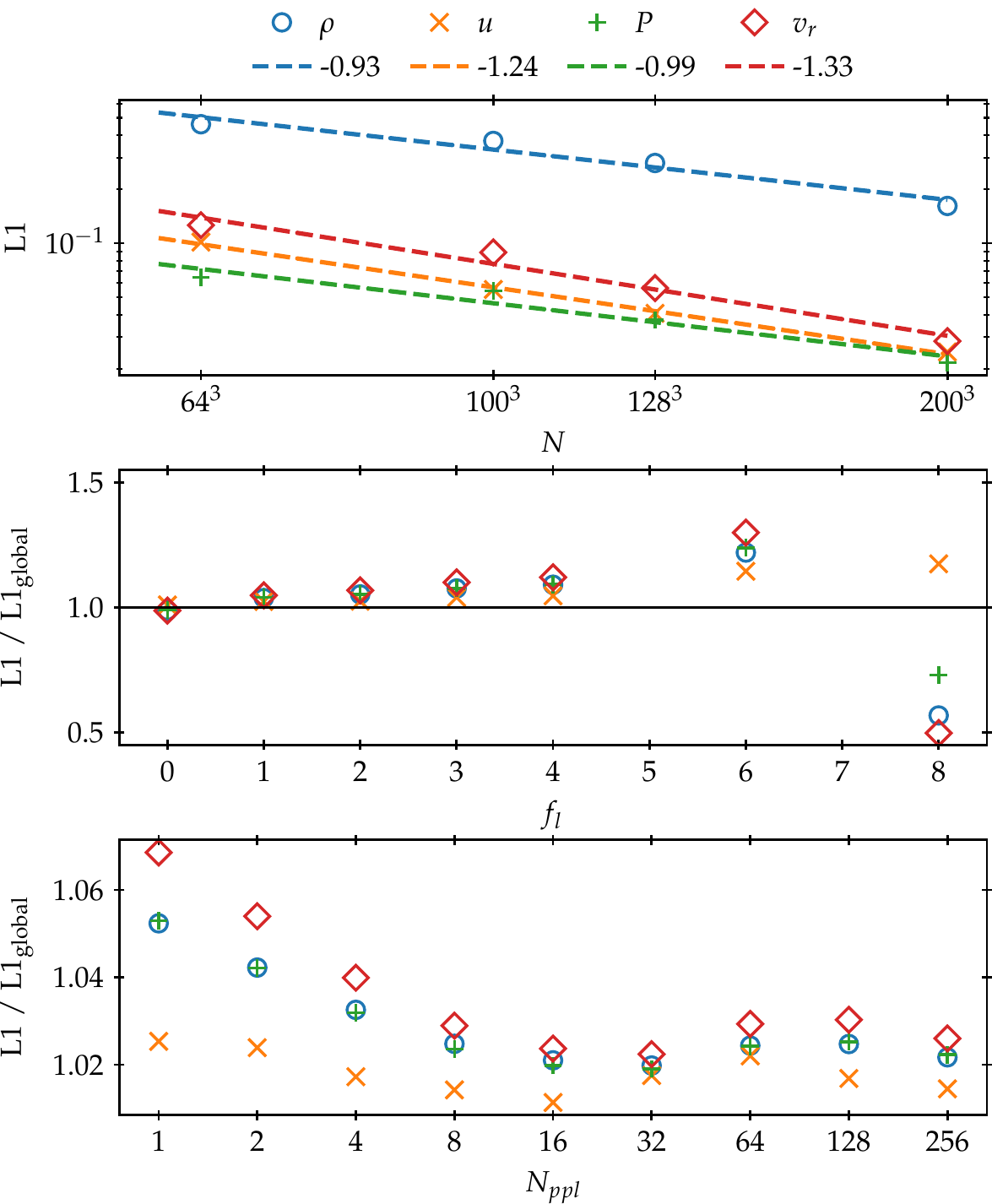}
    \caption{Convergence of the Sedov-Taylor blast wave with respect to the analytic solution. \textbf{Default parameters:} ${N=100^3}$ resolution, ${N_{ppl} = 256}$ particles per leaf, $f_l = 2$ limiter level. \textbf{Top:} varying resolution \textbf{Middle:} varying limiter  \textbf{Bottom:} varying leaf size.
    The dashed lines in the top panel show the best-fit power law $N^\alpha$ of the L1-norm for each particle property, with exponents given in the legend; the results exhibit the expected linear ($N^{-1}$) convergence. The middle and bottom panels measure the L1-norm ratio compared to a simulation with global time stepping at resolution $N=100^3$; the measurements indicate our default parameter values for $f_l$ and $N_{ppl}$ are conservatively selected. }
    \label{fig:sedov_convergence}
\end{figure}

The transition between the high and low pressure monatomic gases is made continuous by initializing the density and pressure according to
\begin{equation}\label{eq:sod_init}
    f(x) =  f_\mathrm{h} + (f_\mathrm{l} - f_\mathrm{h})
        \begin{cases}
            \left[ 1 + e^{\frac{0.5-x}{\Delta}} \right]^{-1} & x \in [0, 1);\\
            \left[ 1 + e^{\frac{x-1.5}{\Delta}} \right]^{-1} & x \in [1, 2),
        \end{cases}
\end{equation}
where $f$ stands for $\rho$ and $P$, ${\rho_\mathrm{h} = 1}$, ${\rho_\mathrm{l} = 1/8}$, ${P_\mathrm{h} = 1}$, and ${P_\mathrm{l} = 1/8}$ (same temperature on both sides). The interface smoothing scale $\Delta$ is set to $\Delta_\mathrm{low}/2$, where $\Delta_\mathrm{low}$ is the unperturbed particle spacing in the low density region. 

We initialize the 1D problem first, using ${\rho_\mathrm{l} = 1/2}$. To keep constant masses across the interface, we integrate \cref{eq:sod_init} to determine the target cumulative mass function, which is then inverted to specify the SPH particle positions. At each sample location in the 1D distribution, we then construct a 2D grid of particles in the other dimensions, such that the 3D density matches \cref{eq:sod_init} with ${\rho_\mathrm{l} = 1/8}$. We remark that \CRKSPH\ does not require a continuous interface to correctly resolve the Sod test, where we find indistinguishable results when using a step-function initial condition.

Fig.~\ref{fig:sodproblem} shows the properties of the system at time ${t=0.15}$ in the isolated domain, as well as the analytic solution. The Sod result includes three primary features, a propagating rarefaction wave (between ${x\sim0.3}$ and ${x\sim0.5}$), contact discontinuity (density jump at $x\sim 0.6$), and shock-front ($x\sim 0.8)$. We evolve three resolutions, ${N_x=50+25}$, ${100+50}$, and ${200+100}$, labeled by the number of particles along the $x$-axis in the ${[0, 1)}$ interval, split by the initial Riemann interface. The simulations converge to the analytic prediction with increasing particle count, where the shock-front and contact discontinuity are well resolved. The small scatter (``Gibbs phenomenon'') in velocity, is commonly seen with high order methods near discontinuities, and was measured similarly in \citetalias{frontiere2017}. We emphasize that the figure includes all points in the 3D distribution. No scatter is visible at a given $x$-coordinate, illustrating that the 1D evolution is preserved in our expansion to three dimensions. 

In Fig.~\ref{fig:sod_convergence}, we show the L1-norm convergence towards the analytic solution with increasing resolution, fit to a power-law $N^\alpha$. We anticipate first-order convergence ($N^{-1}$) for problems involving discontinuities, given the inclusion of artificial viscosity. Encouragingly, we find that the deviations of all hydrodynamic quantities scales approximately linearly with resolution, albeit with entropy converging slightly slower than the others.

\subsubsection{Sedov-Taylor Blast Wave}
\label{sec:sedovblast}

The Sedov-Taylor explosion \citep{Sedov1946, Taylor1950} simulates a spherical shock generated by the injection of thermal energy in a static uniform background. 
The difficult setup requires the ability to simulate shock-fronts while simultaneously maintaining spherical symmetry and conservation laws. The blast wave problem has also been used to test the accuracy of hierarchical time step integrators, and served as motivation for the incorporation of time step limiters (\citealt{saitoh2009}), as discussed in Section~\ref{sec:timestep}.

For our experiment, we initialize equal-mass SPH particles using a glass distribution in a cube of side-length ${L=2}$. The density of the monatomic gas is set to unity with null pressure and velocity.
At the center of the box, we inject a total energy of $E_0 = 1$ smoothed across half the kernel extent; consistent with \citetalias{frontiere2017}, we have found point-like injections provide qualitatively identical results as the smoothed initial conditions presented here. In total, we set up four different simulations with ${N=64^3,100^3, 128^3 \;\text{and}\; 200^3}$ particles.

The system is then evolved until the shock-front reaches a radius of ${R_\mathrm{target}=0.8}$, corresponding to a time ${t_\mathrm{target} \simeq 0.402}$. In the left panel of Fig.~\ref{fig:sedov}, we show the particle distribution in a thin slice of depth ${\Delta_z = 2/N}$ through $z=0$, colored by the specific energy to differentiate the hot (inside) and cold (outside) gas in regard to the bubble. The shock wave matches the analytic radius well. In the right panel, we plot the median value of density, specific energy, pressure, and radial velocity measured in evenly spaced radial bins. 
With increasing resolution, the shock-front is resolved more sharply, and the measured peak density and pressure converge towards the analytic solution.
The oscillations of pressure and radial velocity in the post-shock region, a common artifact in SPH codes (see, e.g., \citealt{Hu2014, wadsley2017}), decay as the shock is better resolved. The deviation close to the origin, particularly in the internal energy, is attributed to the difficulty of resolving a near-vacuous solution with low particle sampling.   

As was done in the \hyperref[sec:sodproblem]{Sod shock tube test}, we measure the L1-norm convergence in the top panel of Fig.~\ref{fig:sedov_convergence}; once again, we reassuringly find an approximate $~N^{-1}$ power law between the particle data and analytic solution for all properties.

We briefly turn to an investigation of the time step limiter. The middle panel of Fig.~\ref{fig:sedov_convergence} shows the ratio of measured L1-norms as a function of limiter level $f_l$, when compared to simulations where global time stepping was employed. In this experiment the resolution was fixed to $N=100^3$ particles, with leaf size $N_{ppl}=256$. Our results indicate that limiter values $f_l\in[0,4]$ achieve accurate solutions, where higher levels begin to degrade. This is consistent with the recommendation of \cite{saitoh2009} to use $f_l = 2$. As noted in Section \ref{sec:timestep}, our limiters are enforced at the leaf level (rather than particle), and are thus more conservative than the procedure described in \cite{saitoh2009}. This effect is explored in the bottom panel of the figure, where the limiter is fixed at $f_l = 2$ for the same resolution ($N=100^3$), and we vary the number of particles per leaf $N_{ppl}$. We find that using $N_{ppl} > 16$ particles reaches the noise floor, however the extent that increasing leaf size can assist in the accuracy is limited. We conclude that our default settings of $f_l = 2$ and $N_{ppl} \in [128,256]$ used throughout the paper are certainly suitable, if not overly conservative.

\begin{figure*}[tp]
    \includegraphics[width=\textwidth]{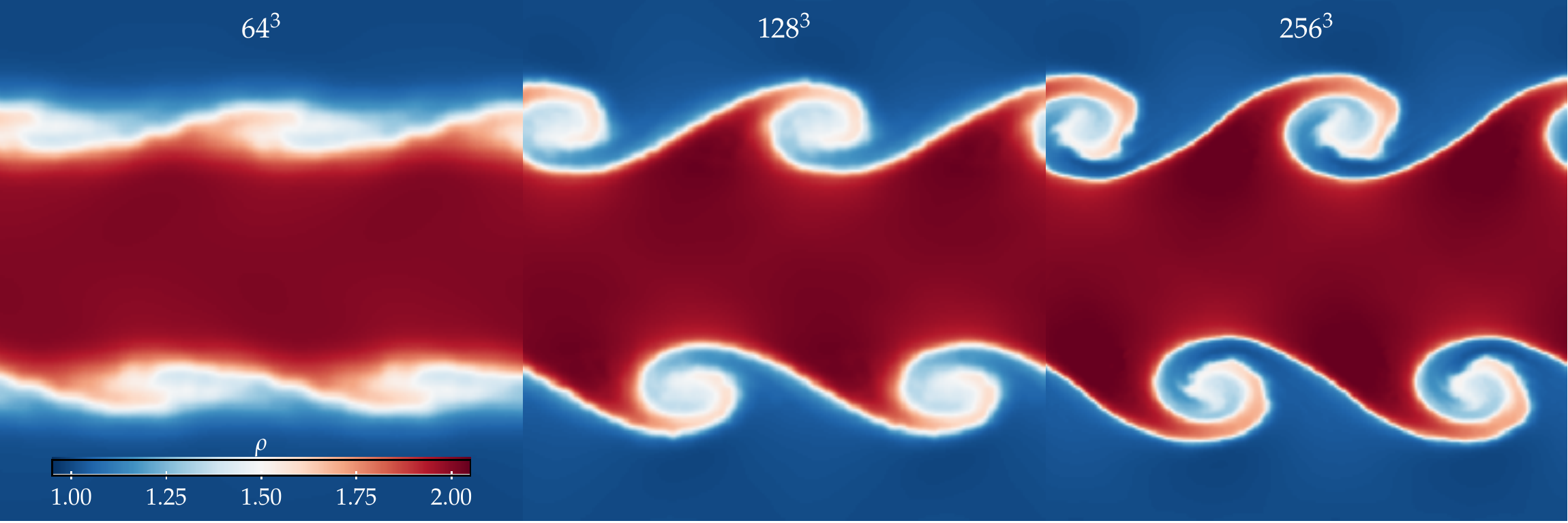}
    \caption{Density field at time $t=2.0$ for the Kelvin-Helmholtz instability test measured on the $xy$-plane for $N=64^3$, $128^3$, and $256^3$ particles. At the lowest resolution, the vortices are barely resolved, leading to over-diffusion of the primary instability. With increasing resolution, fluid-mixing is more prevalent, capturing the roll-up of the primary vortices, in addition to secondary instabilities becoming visible. 
    }
    \label{fig:khdensity}
\end{figure*}

\begin{figure}[htp]
    \includegraphics[width=\columnwidth]{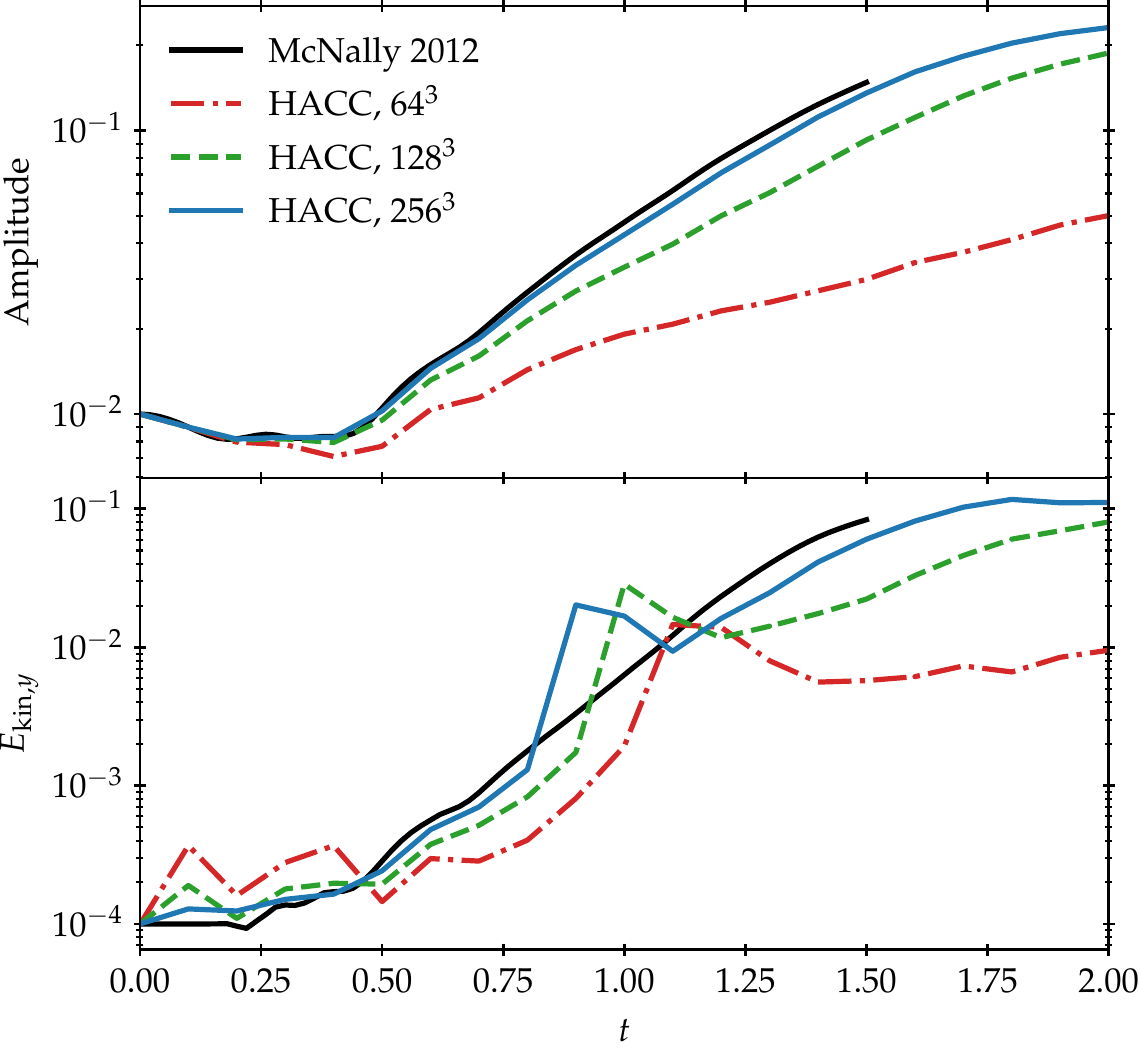}
    \caption{The time-evolution of the $y$-mode amplitude and the maximum kinetic energy along the $y$-axis of the Kelvin-Helmholtz instability. In addition to the three \HACC-simulations of $N=64^3,$ $128^3,$ and $256^3$ particles, we include the high-resolution result by \citet{Mcnally2012}. The simulations approach this reference with improving resolution, as the vortices and turbulence are increasingly refined. 
    }
    \label{fig:khenergy}
\end{figure}

\subsubsection{Kelvin-Helmholtz Instability}\label{sec:kelvinhelmholtz}
Fluid instabilities play a crucial role in mixing processes and production of turbulence relevant to both astrophysics and general hydrodynamics. The Kelvin-Helmholtz (KH) instability considers a velocity-induced shearing flow across a perturbed fluid interface, and has historically exposed the tendency of traditional SPH methods to suppress fluid mixing (see, e.g., \citealt{Agertz2007}). The presence of surface tension (``E0'') errors, in addition to the over-activity of traditional artificial viscosity prescriptions, are primary contributors to the erroneous solutions measured with unaltered SPH. As discussed in \citetalias{frontiere2017}, \CRKSPH\ directly addresses these concerns by utilizing accurate reproducing kernels in combination with a novel artificial viscosity treatment, summarized in Section \ref{sec:HydroEqns}.

The KH setup examines the shearing flow across two fluids with differing densities at pressure equilibrium. A single instability mode is excited by adding a velocity perturbation perpendicular to the fluid interface. The shear is converted into vorticity that cascades to turbulence due to secondary instabilities. For our test, we use the smooth KH initial conditions described in \citet{Mcnally2012}. In particular, we place a monatomic gas with uniform pressure $P=2.5$ in a periodic box with side-length $L=1$. The interface along the $y$-axis is characterized by a smooth transition in density ($\rho$) and $x$-velocity ($v_x$), according to
\begin{equation}
    f(y) = 
        \begin{cases} 
            f_1 - f_m e^{(y-1/4)/\Delta}, & y \in [0, 1/4];\\
            f_2 + f_m e^{(1/4-y)/\Delta}, & y \in [1/4, 1/2);\\
            f_2 + f_m e^{(y-3/4)/\Delta}, & y \in [1/2, 3/4);\\
            f_1 - f_m e^{(3/4-y)/\Delta}, & y \in [3/4, 1),\\
        \end{cases}
\end{equation}
where $f$ stands for $\rho$ and $v_x$, ${f_m = (f_1 - f_2)/2}$, ${\rho_1 = 1}$, ${\rho_2 = 2}$, ${v_1 = 0.5}$, ${v_2 = -0.5}$, and ${\Delta = 0.025}$.

A perturbation in the velocity along the $y$-axis is added as 
\begin{equation}
    v_y(x) = \delta_y \sin(2 \pi x / \lambda),
\end{equation}
with amplitude $\delta_y = 0.01$ and wavelength $\lambda = 1/2$. The SPH particles are placed on a regular 3D grid with resolutions $N=64^3$, $128^3$, and $256^3$, using masses initialized by the $\rho(y)$ density constraint, and null velocities in the $z$-dimension. 

Fig.~\ref{fig:khdensity} shows the evolved density at $t=2$ on a slice perpendicular to the $z$-axis. 
The vortical roll-ups are induced with increasing detail as the resolution improves.
To quantify the convergence, we measure the growth of the $y$-velocity mixing mode according to the definition in \cite{Mcnally2012} (Eqs. 10-13 in that reference) and the evolution of the maximum kinetic energy along the $y$-axis, $E_{\mathrm{kin}, y} = \max_i [\frac{1}{2} m_i v_{y,i}^2]$. The quantities are shown in Fig.~\ref{fig:khenergy}, where we also include the high-resolution results provided by \citet{Mcnally2012}. At low resolution, the amplitude of the $y$-mode is highly underestimated, especially at late times, which is also visible from the missing vorticity and turbulence in the density snapshot. As the resolution is increased, the $y$-mode amplitude converges to the \citet{Mcnally2012} curve. We observe a similar trend in $E_{\mathrm{kin}, y}$, however, as the kinetic energy measurement is based on a single particle, it is inherently more noisy than the averaged $y$-mode amplitude. The scatter feature around $t=1$ occurs at roughly the classical instability growth rate $\tau_\text{KH}$ (\citealt{Chandra1961}), where non-linear effects begin to occur, and has been previously observed in \citetalias{frontiere2017}. 


\begin{figure}[tbp]
    \includegraphics[width=\columnwidth]{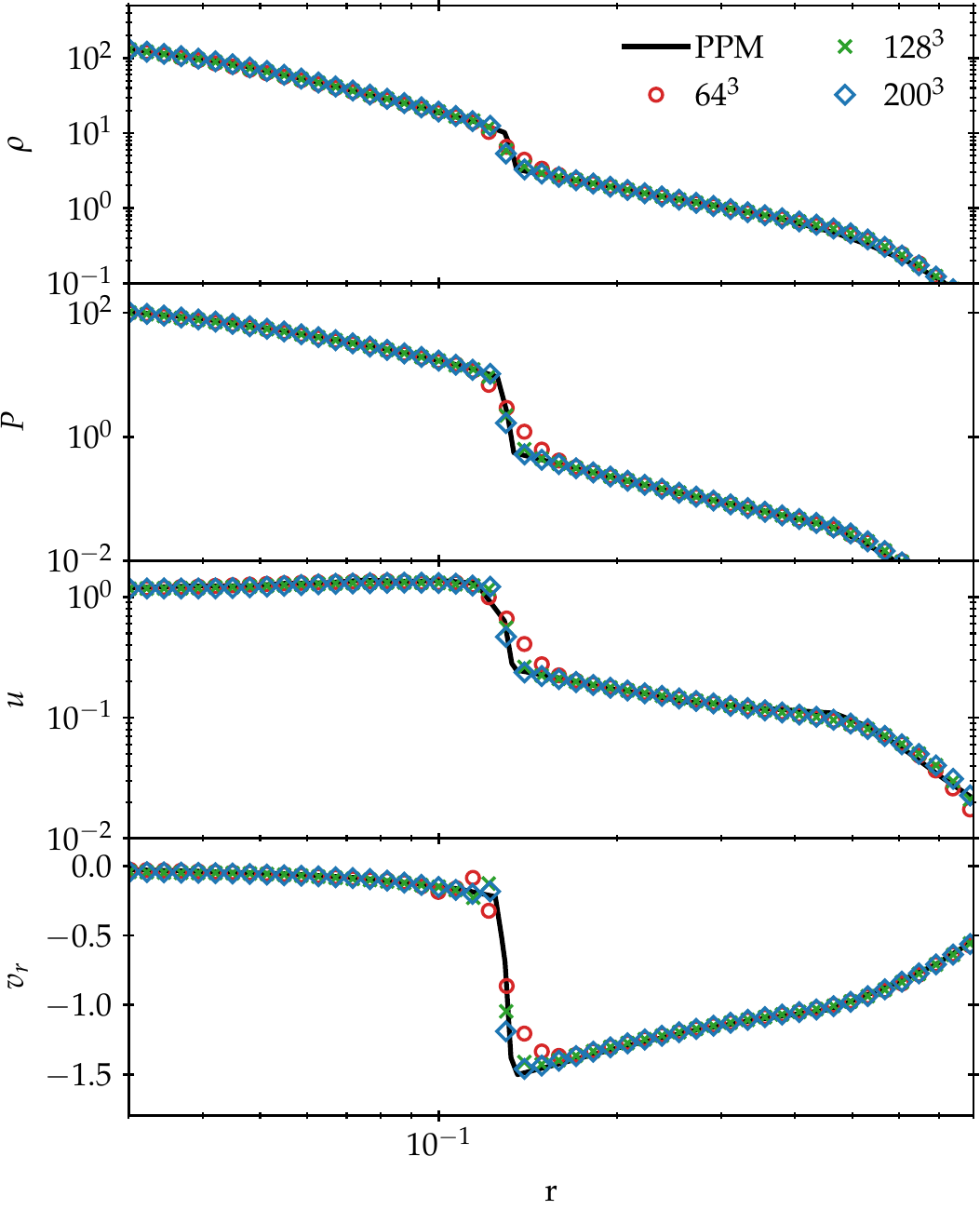}
    \caption{Radial profiles for density, pressure, specific energy, and radial velocity at time $t = 0.77$ for the Evrard collapse problem with resolutions $N=64^3$, $100^3$, and $200^3$ particles. The solid line shows the results from the 1D PPM calculation by \citet{Steinmetz1993}. The simulations converge well with increasing resolution.
    }
    \label{fig:evrard}
\end{figure}

\subsection{Gravitation Coupling Experiments}\label{sec:evaluation_gravhydro} 
We now consider examples of self-gravitating systems when combined with hydrodynamics. 
In particular, we present the \CRKHACC\ solutions for the gravitational (Evrard) collapse of a gas sphere in Section \ref{sec:evrard}, in addition to a plane-wave Zel'dovich pancake test in a baryon-only universe (Section \ref{sec:plane-wave}). We further analyze cosmological cluster simulations in Section \ref{sec:cluster}, including the well-studied Santa Barbara (\citealt{Frenk1999}) and nIFTy (\citealt{sembolini2016nifty}) cluster comparisons. For a detailed investigation of \CRKSPH\ performance in modeling Keplerian and pressure-supported rotating disks, see \cite{Raskin2016}; accretion disk simulations are insightful tests of angular momentum conservation, where \CRKSPH\ is shown to significantly reduce momentum transport errors compared to a number of SPH prescriptions. 

\subsubsection{Evrard Collapse}\label{sec:evrard}

\begin{figure}[tbp]
    \includegraphics[width=\columnwidth]{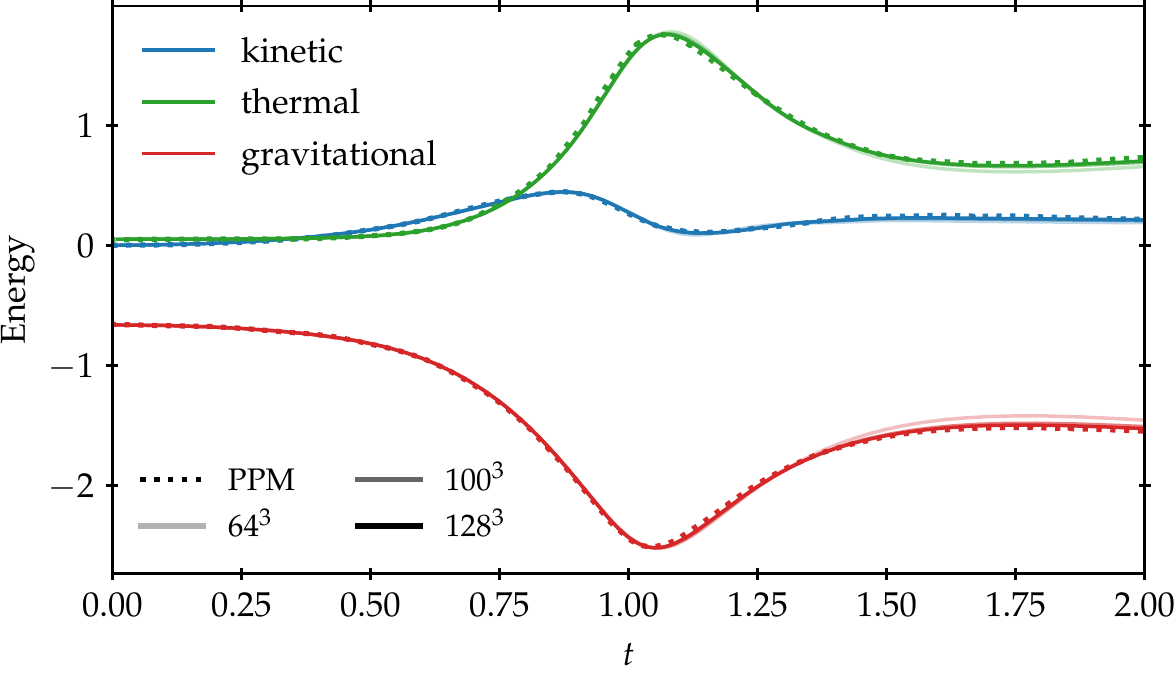}
    \caption{Evolution of the kinetic, thermal, and gravitational energy for the Evrard collapse problem. Initially, gravitational potential energy is converted to kinetic in-fall energy, which in turn is shock-heated into thermal energy. As in Fig.~\ref{fig:evrard}, \CRKHACC\ convergences nicely to the 1-D PPM results of \citet{Steinmetz1993}. 
    }
    \label{fig:evrard_energy}
\end{figure}

The Evrard spherical collapse test is both a simple and instructive self-gravitating experiment extensively used to evaluate astrophysics codes (e.g., \citealt{Springel2010,Hopkins2015,wadsley2017,cabezon2017}). The problem, stated in natural units with $G=1$, considers a spherical gas distribution with an initial $\rho \propto r^{-1}$ density profile that collapses under self-gravity. 
The resulting heat generated by the compression forms an outwards moving shock.
Over the entire evolution, gravitational energy is converted into kinetic and thermal energy, allowing for an insightful test of the interplay between the gravity and hydrodynamic solver, as well as the preservation of conservation laws.

\begin{figure*}[tbp]
    \centering
    \includegraphics[width=0.8\textwidth]{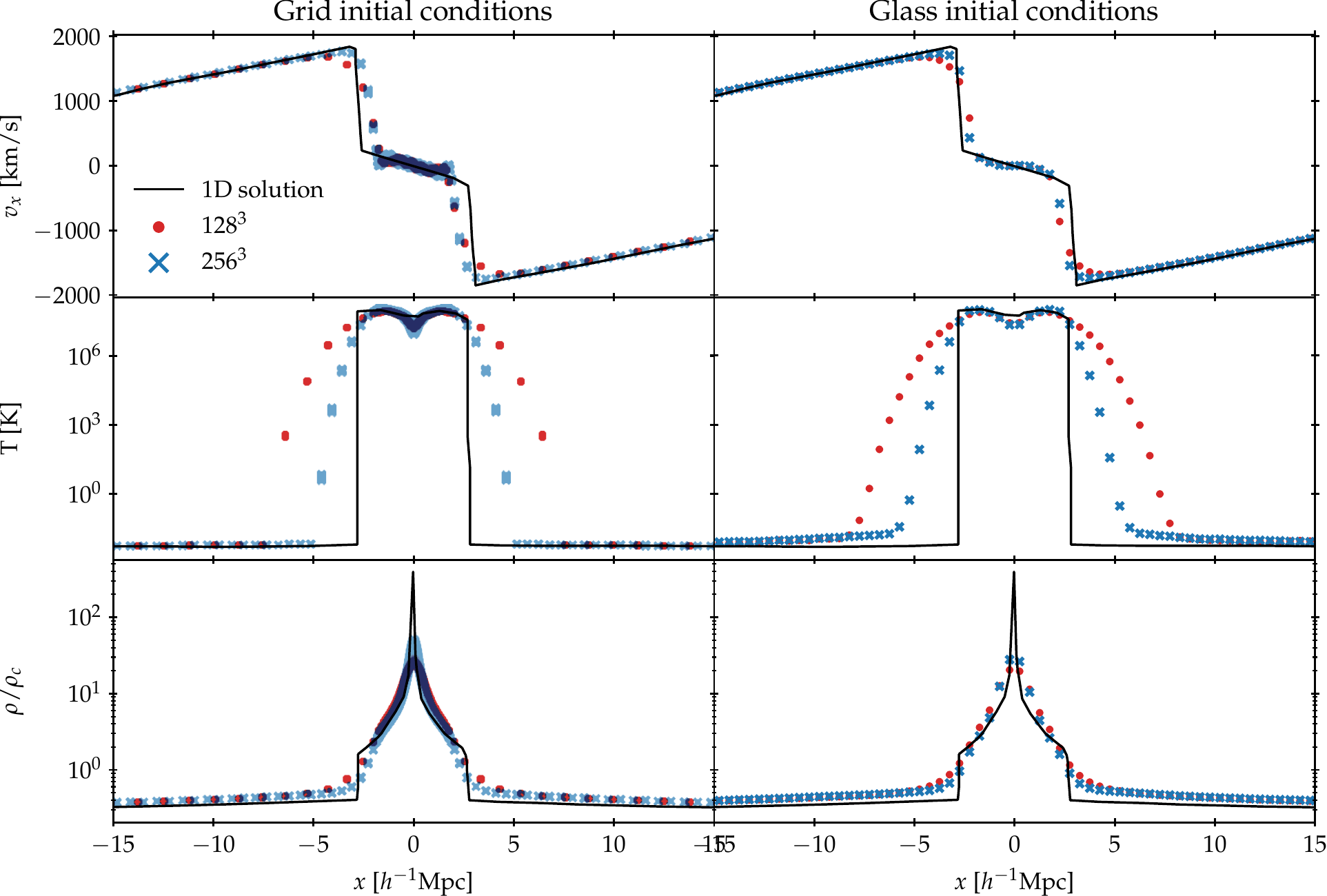}
    \caption{Results of a plane-wave collapse (Zel'dovich pancake) test problem with grid (left) and glass (right) initial conditions. We display the velocity along the plane wave, the gas temperature, and the normalized density at $z=0$. The grid panel includes all simulation particles, with the lack of scatter indicating that the planar symmetry is preserved throughout the volume. For the glass setup, we show the median values in narrow distance bins perpendicular to the plane of collapse. In black, we include outputs from a high-resolution 1D simulation taken from \citet{Hopkins2015}, which the \CRKHACC\ runs converge towards. }
    \label{fig:planewave}
\end{figure*}

To set up the initial SPH positions, we distort a particle glass distribution such that the density is $\rho(r) \propto r^{-1}$, achieved with a radial transformation $r \mapsto r^{3/2}$.\footnote[2]{The differential mass in a uniform distribution $dM \propto r^2 dr$ is mapped to a desired mass $dM' \propto r'dr'$, implying a radial transformation $r^2 dr \propto r'dr' \Rightarrow r' \propto r^{3/2}$.}
Particles outside a unit sphere are removed, and the equal masses of the remaining points are rescaled such that the total mass is $M=1$. The monatomic gas is initialized with zero velocity and a uniform specific internal energy of $u = 0.05$. We run the Evrard collapse problem in three resolutions, starting from a $N=64^3$, $100^3$, and $200^3$ glass.
The systems are evolved to $t=2.0$ in a large periodic box of size $L=90$ to minimize periodicity effects.

Fig.~\ref{fig:evrard} shows the radial properties of the evolved system at time $t=0.77$. This snapshot was previously studied by \citet{Steinmetz1993} who carried out a high-resolution 1D integration in spherical coordinates with a piecewise parabolic method (PPM). 
We include their results for density, pressure, and radial velocity in the figure. At the sharp shock-front, \CRKSPH\ converges well to the PPM curve with increasing resolution. 
Outside the shock wave, the measurements are consistent for all resolutions. 

In Fig.~\ref{fig:evrard_energy}, we show the evolution of the kinetic, thermal, and gravitational energy. 
We find close agreement with the \citet{Steinmetz1993} measurement for all resolutions, with the caveat that the $64^3$ simulation slightly underestimates the thermal energy at late times. The results showcase accurate conversion and conservation of energy for the collapse test.    

\subsubsection{Plane-wave Collapse}\label{sec:plane-wave}

Structure in the universe forms through gravitational collapse of primordial density perturbations. 
The 3-dimensional nature of these random fluctuations results in a process that is anisotropic, where there is a major axis along which these systems collapse first \citep{Zeldovich1970}.
Examining the evolution of a single Fourier mode, i.e., Zel'dovich pancake, allows us to study this phenomenon in an idealized setting \citep[see also][]{Shandarin1989}.

We consider a diagonal plane-wave collapse in a purely baryonic Einstein--de Sitter universe with $\Omega_m = \Omega_b = 1$,  $\Omega_\Lambda = 0$, and $H_0 = 100h$~km~s$^{-1}$~Mpc$^{-1}$. 
Starting from a uniform Lagrangian distribution, either a grid or glass, in a cube with side-length $L$, we displace configurations of $N=128^3$ and $256^3$ particles by applying the Zel'dovich approximation \citep{Zeldovich1970}; we use an initial potential of the form $\phi(\mathbf{q}) = - A_\phi \cos\left(\mathbf{k}\mathbf{q}\right)$, where $\mathbf{q}$ are the Lagrangian coordinates and $\mathbf{k} = 2\pi / L \, (1,1,1)$ is a diagonal wave vector with wavelength $\lambda = L / \sqrt{3}$.
The resulting perturbed comoving position $\mathbf{x}$, velocity $\dot{\mathbf{x}}$, density $\rho$, and temperature $T$ can be written as
\begin{align}
    \mathbf{x}(\mathbf{q}, a)       &= \mathbf{q} - A_x a \mathbf{k} \sin(\mathbf{k}\mathbf{q}), \\
    \dot{\mathbf{x}}(\mathbf{q}, a) &= - H_0 a^{-1/2} A_x \mathbf{k} \sin(\mathbf{k}\mathbf{q}), \\
    \rho(\mathbf{q}, a)             &= \rho_c (\det \; \mathrm{d}\mathbf{x}/\mathrm{d}\mathbf{q})^{-1} \nonumber \\
                                    &= \rho_c (1 - A_x a \mathbf{k}^2 \cos(\mathbf{k}\mathbf{q}))^{-1}, \\
    T(\mathbf{q}, a)                &= T_0 (\rho / \rho_c)^{\gamma - 1},
\end{align}
where ${A_x = 2/(3H_0^2) A_\phi}$ is the amplitude of the perturbation. Requiring shell-crossing to occur at time $a_\times$ (i.e., $\rho(\mathbf{0}, a_\times) \to \infty$) implies ${A_x = (L / 2 \pi)^2 \, (3 a_\times)^{-1}}$.
We initialize the canonical test problem with ${\lambda = \unit{64}{\Mpch}}$, $h=0.5,$ and $a_\times = 0.5$ (consistent with \citealt{bryan1995, trac2004, Springel2010,Hopkins2015}) at a starting redshift $z=200$ and an initial gas temperature $T_0 = 396$~K (corresponding to 100~K at $z=100$).

In Fig.~\ref{fig:planewave}, we show the velocity along the axis of collapse, the gas temperature, and the overdensity at $z = 0$ (well after the crossing time) for both grid and glass initial distributions. In the case of the glass initial conditions, we bin the particles according to their radial distance to the plane of collapse and present the median values of the particle state measurements. For the grid results, we show all particles in the figure, where the lack of visible scatter highlights the preservation of symmetry in the \CRKSPH formalism. 

For both initial distributions, we find similar convergence with the high-resolution solution of a 1D PPM simulation taken from \citet{Hopkins2015}; moreover, our results are consistent with the modern SPH glass measurements performed in that work. During shell-crossing, a massive shock is induced that produces a temperature peak many orders of magnitude larger than the cold initial medium. \CRKSPH\ establishes well-behaved velocity profiles of the inner post-shock region, in addition to capturing the shock-fronts favorably with increasing resolution. Outside the active region, the simulations correctly follow the linear solution. 

Grid initial conditions for the Zel'dovich pancake are notoriously difficult for Lagrangian codes (given the collapse of a single dimension), and yet, the \CRKHACC\ solver performs well in the diagonal setting imposed here.
In general, \CRKSPH\ demonstrates similar robustness in a number of hydrodynamic problems (as investigated in \citetalias{frontiere2017}), where discontinuous and gridded initial configurations nevertheless yielded accurate results.

\subsubsection{Cluster Comparisons}
\label{sec:cluster}

\begin{figure*}[htp]
    \centering
    \includegraphics[width=0.32\textwidth]{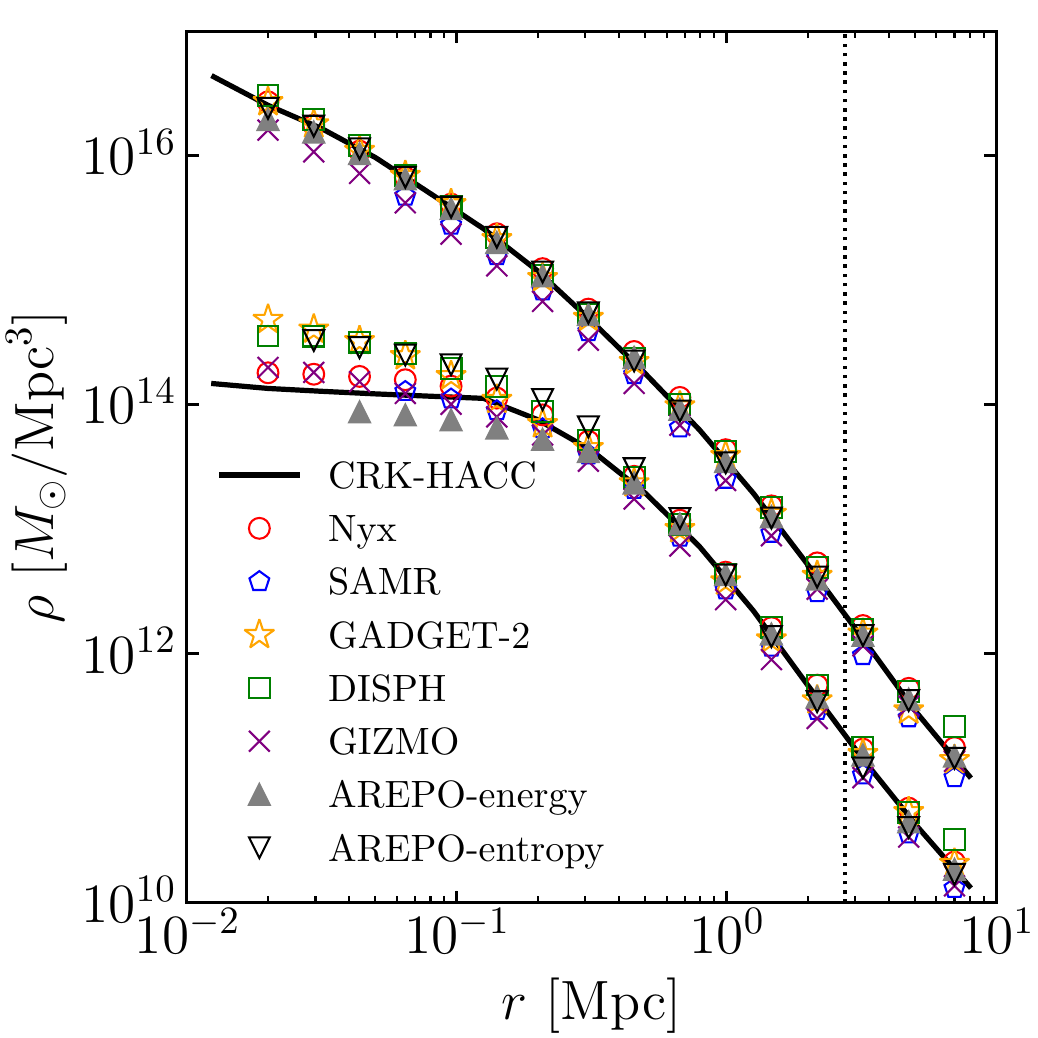}
    \includegraphics[width=0.32\textwidth]{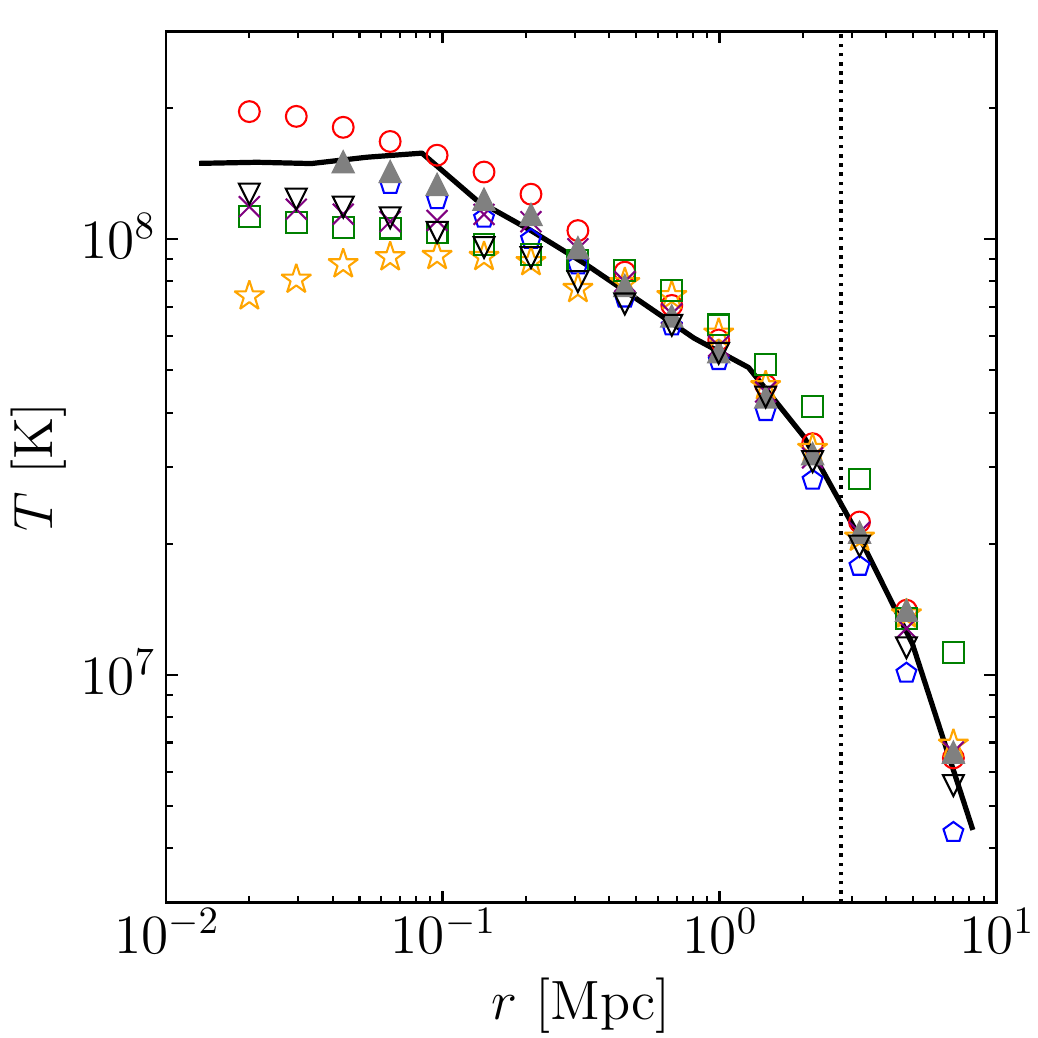}
    \includegraphics[width=0.32\textwidth]{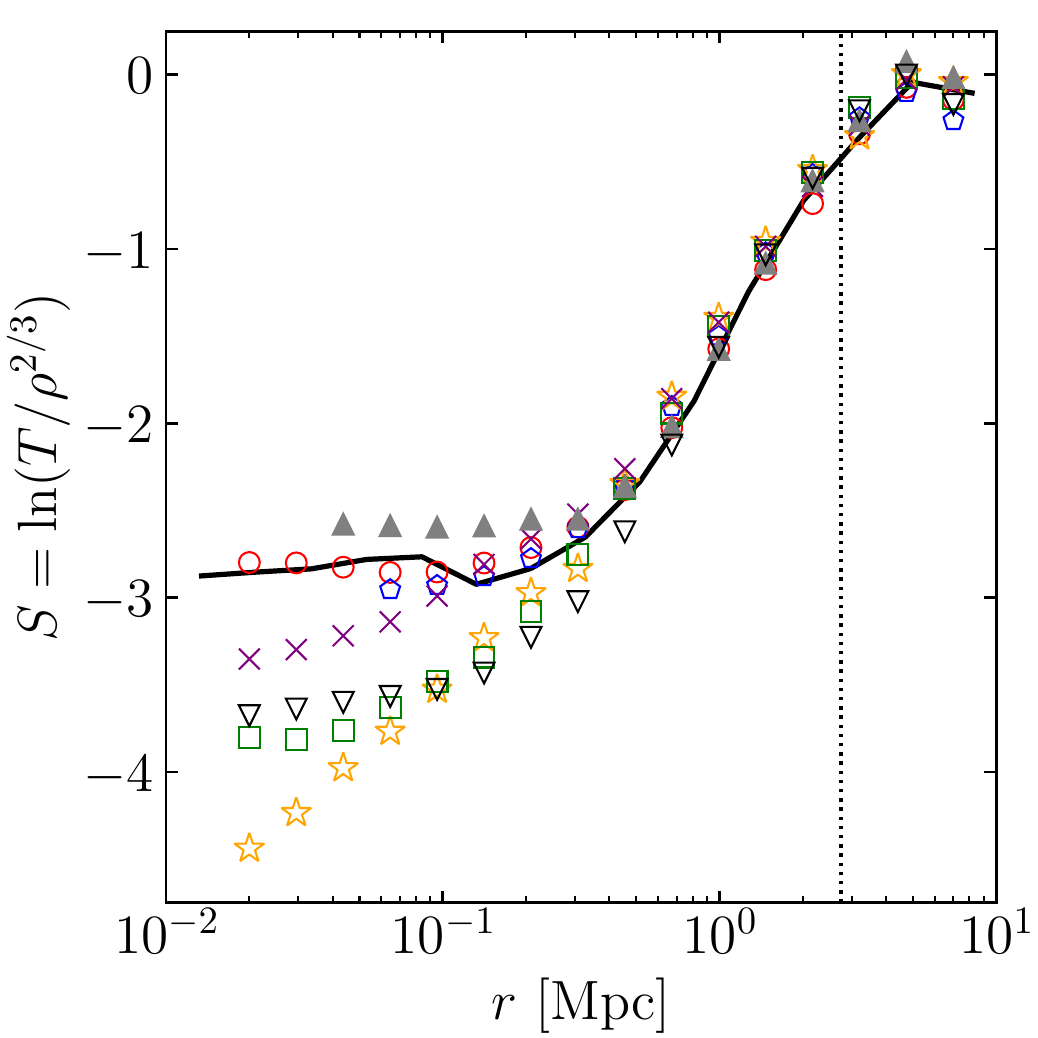}
    \caption{Radial profiles showing density (left), temperature (center), and entropy (right)
    for the Santa Barbara cluster (\citealt{Frenk1999}) at redshift $z = 0$ for a number of Lagrangian (\CRKHACC, \smaller{GADGET-2}, \smaller{DISPH}), Eulerian (\smaller{Nyx}, \smaller{SAMR}), and hybrid (\smaller{GIZMO}, \smaller{AREPO}) solvers. In the case of \smaller{AREPO} we show results from both the strict energy conservation and dual entropy formulations. The lower (upper) set of data points in the left panel correspond to the baryon (dark matter) density profile, while the vertical dotted line in each panel denotes $R_{200} = 2.75\ {\rm Mpc}$. Notably, the \CRKHACC measurements are consistent with the grid-based approaches. 
    }
    \label{fig:sbprofiles}
\end{figure*}

\begin{figure*}[htp]
    \centering
    \includegraphics[width=0.32\textwidth]{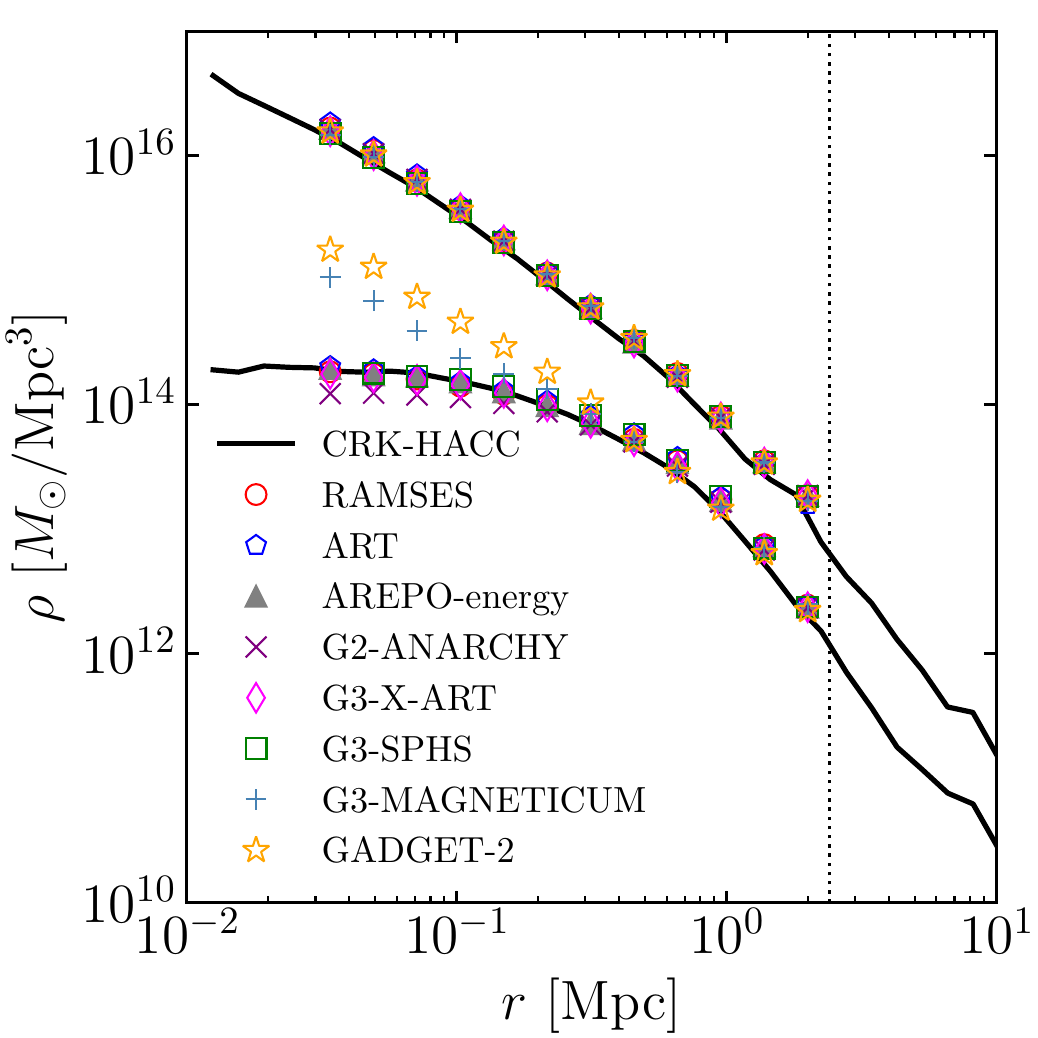}
    \includegraphics[width=0.32\textwidth]{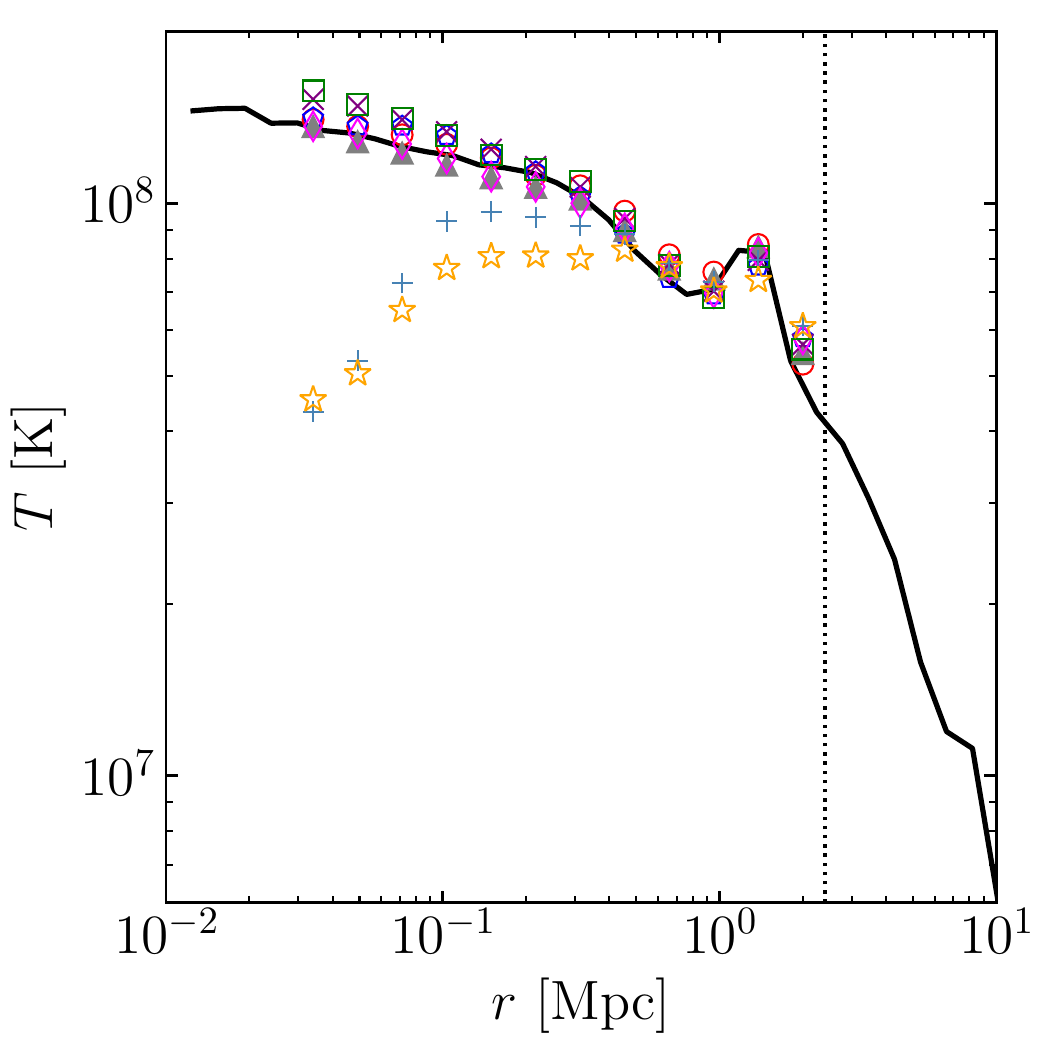}
    \includegraphics[width=0.32\textwidth]{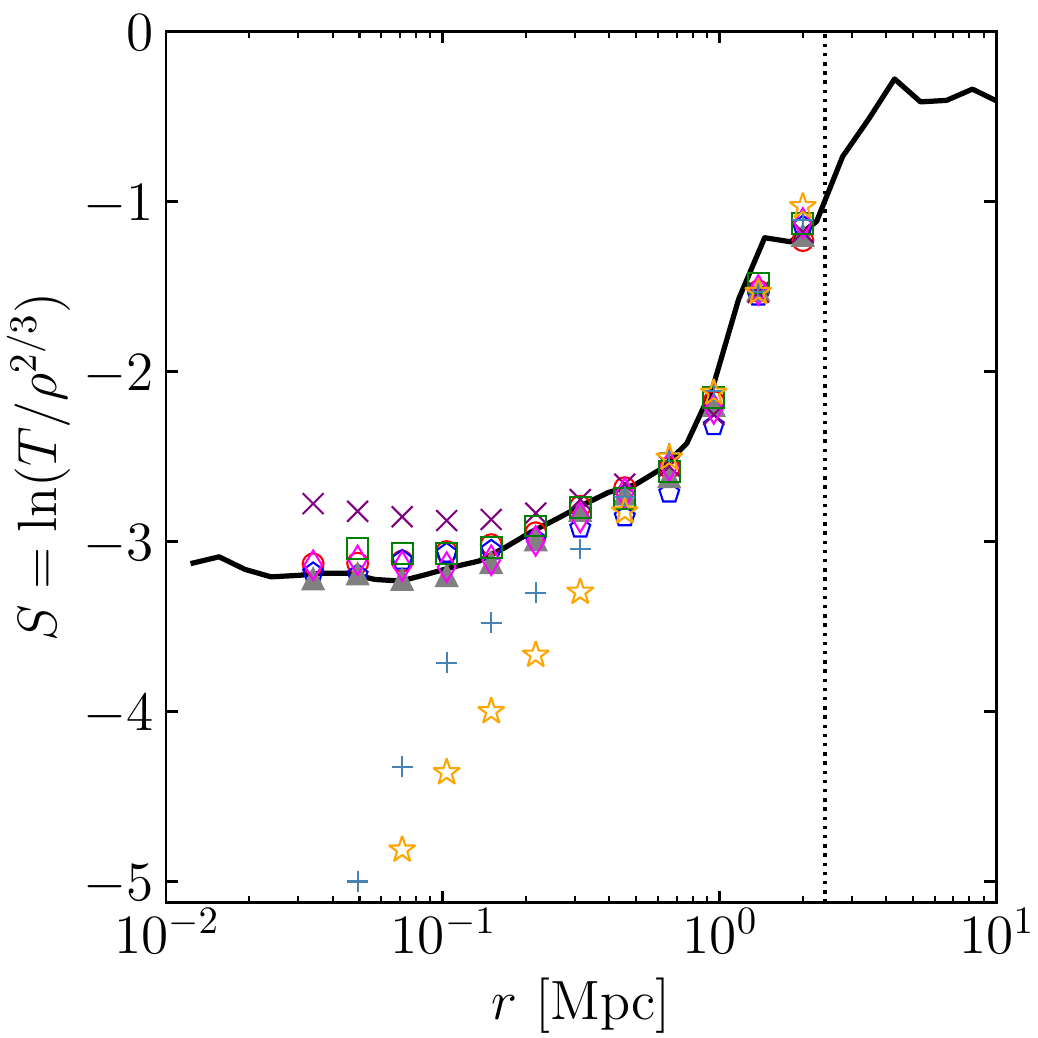}
    \caption{Radial profiles showing density (left), temperature (center), and entropy (right)
    for the nIFTy cluster (\citealt{sembolini2016nifty}) at redshift $z = 0$ for a variety of Lagrangian (\CRKHACC, \smaller{GADGET-2}, \smaller{G2-ANARCHY}, \smaller{G3-X-ART}, \smaller{G3-SPHS}, \smaller{G3-MAGNETICUM}), Eulerian (\smaller{RAMSES}, \smaller{ART}), and hybrid (\smaller{AREPO}) solvers. The lower set of data points in the left panel correspond to the baryon density profile from the non-radiative run while the upper set of points correspond to the total matter density profile in the gravity-only run. In each panel the vertical dotted line denotes the halo radius, $R_{200} = 2.41\ {\rm Mpc}$, identified with \CRKHACC. As opposed to the discrepancies seen in Fig.~\ref{fig:sbprofiles}, most of the methods are in closer relative agreement for the nIFTy comparison. The \CRKHACC results are again consistent with the mesh-codes, and are amongst the grouping of modern SPH codes.   
    }
    \label{fig:nfprofiles}
\end{figure*}

\begin{figure*}[htp]
    \centering
    \includegraphics[width=0.32\textwidth]{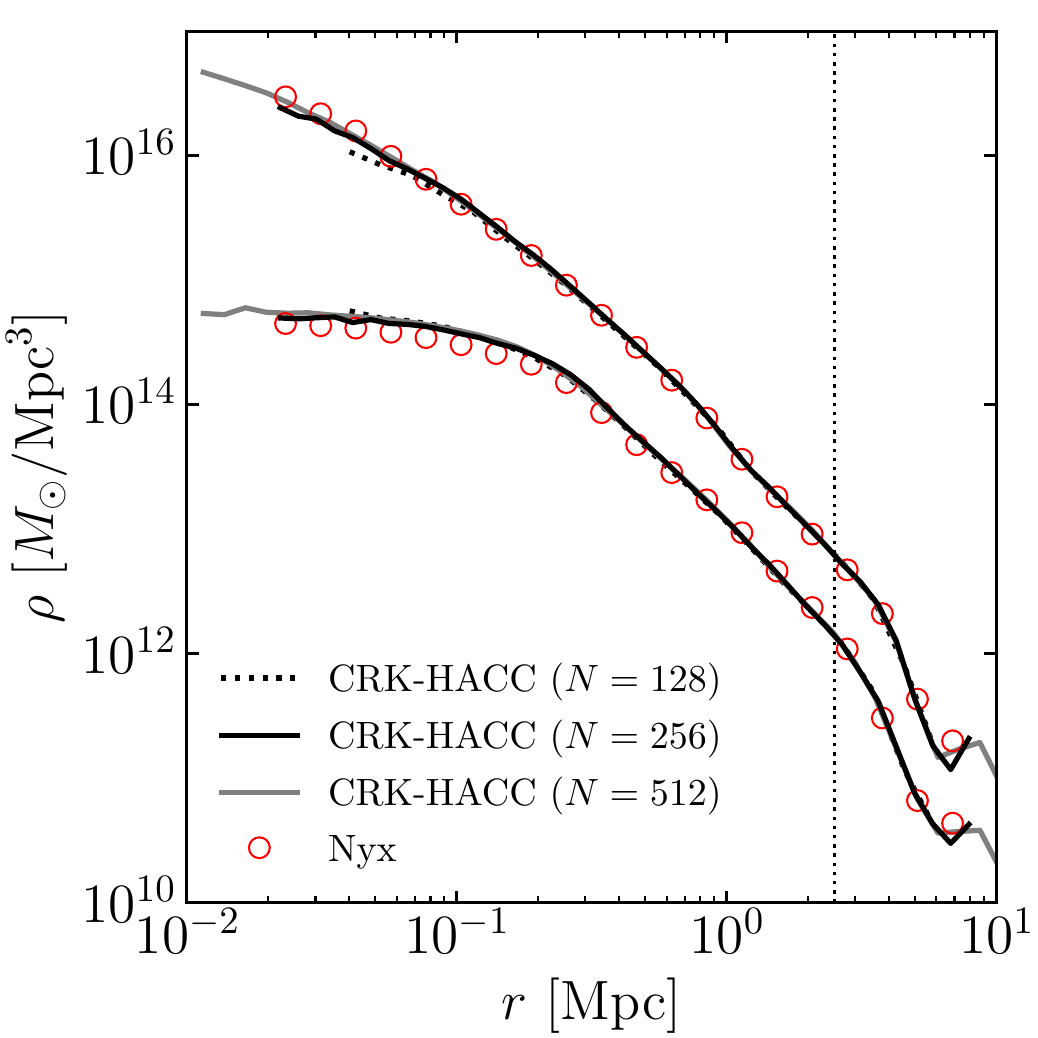}
    \includegraphics[width=0.32\textwidth]{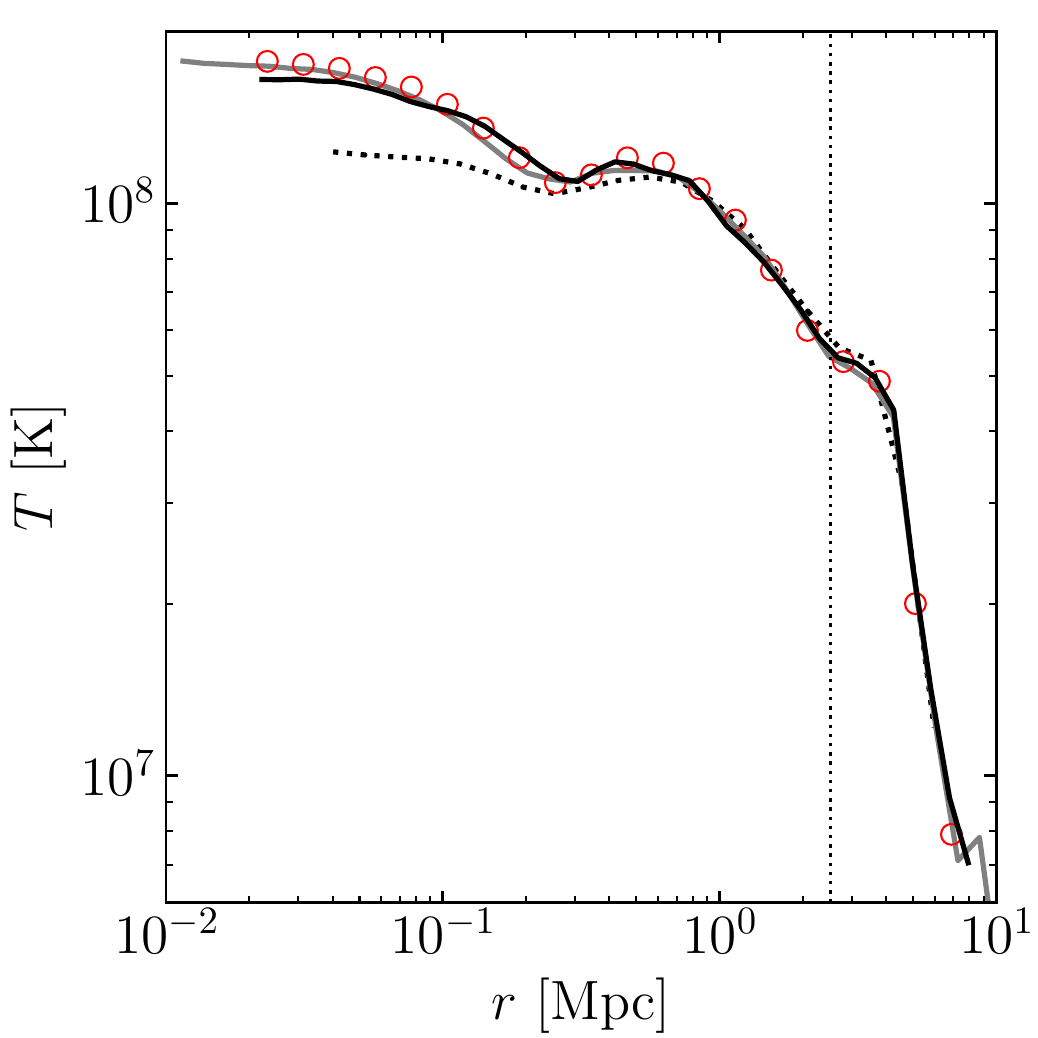}
    \includegraphics[width=0.32\textwidth]{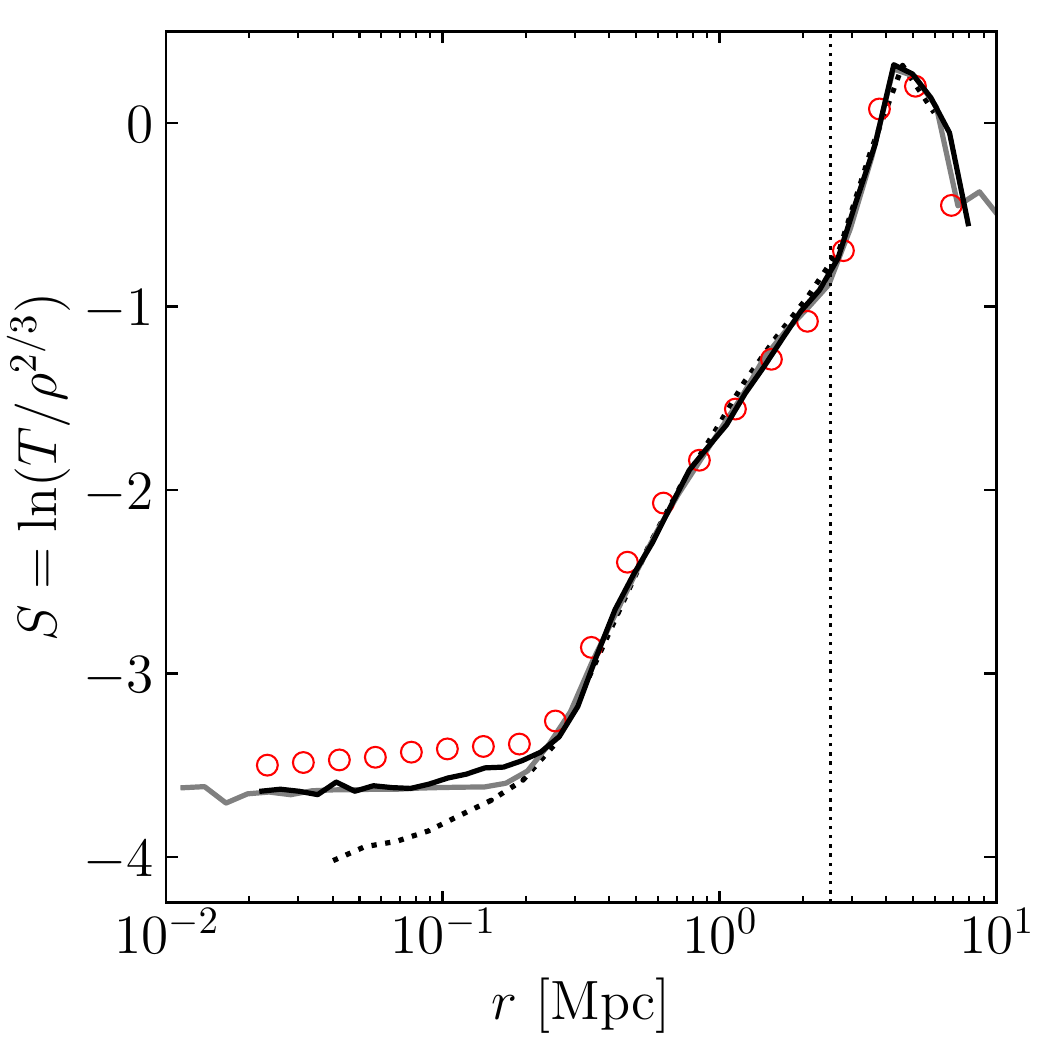}
    \caption{Radial profiles showing density (left), temperature (center), and entropy (right)
    for the Argonne cluster run at redshift $z = 0$. The lower (upper) set of
    data points in the left panel correspond to the baryon (dark matter) density profile, while the vertical dotted line in each panel denotes $R_{200} = 2.52\ {\rm Mpc}$ found by \CRKHACC. As in Fig.~\ref{fig:sbprofiles}, the \CRKHACC\ solutions are closely consistent with the Eulerian approach (\smaller{Nyx}), where the results appear to be numerically converged.  
    }
    \label{fig:acprofiles}
\end{figure*}

We present results from three sets of cosmological simulations constructed to study profiles of massive clusters. 
The first example is the well-known Santa Barbara (SB) cluster \citep{Frenk1999}, with initial conditions specified in \citet{heitmann2005}. The setup evolves ${N=2\times256^3}$ dark matter plus baryon particles in a box of side-length $L=\unit{32}{\Mpch}$,
at starting redshift $z = 63$. The assumed cosmology is SCDM with 
($\Omega_c$, $\Omega_b$, $n_s$, $\sigma_8$, $h$) = (0.9, 0.1, 1.0, 0.9, 0.5), where the dark matter and baryon mass resolutions are $m_\text{DM} = \unit{4.88 \times 10^8}{\massh}$ and $m_\text{gas} = \unit{5.42 \times 10^7}{\massh}$, respectively. 

A more recent code comparison was performed by the nIFTy collaboration in \cite{sembolini2016nifty}. As opposed to the SB test, with initial conditions chosen to deliberately produce a statistically rare cluster at the center of a small box, the nIFTy setup utilizes the zooming technique of \cite{klypin2001} to extract a cluster from a larger domain of edge-length $L=\unit{1}{\Gpch}$. Specifically, the ICs are taken from the MUSIC-2 dataset\footnote[2]{\url{https://music.ft.uam.es/}} built from the Multidark simulation (\citealt{prada2012}). The cosmology is WMAP-7+BAO+SNI \citep{Komatsu_2011} with ($\Omega_m$, $\Omega_b$, $\Omega_\Lambda$, $n_s$, $\sigma_8$, $h$) = (0.27, 0.0469, 0.73, 0.95, 0.82, 0.7). Gravity-only and hydrodynamic runs were both performed, which correspond to a mass resolution of ${m_\text{DM} = \unit{1.09 \times 10^9}{\massh}}$ for the former, and ${m_\text{DM} = \unit{9.01 \times 10^8}{\massh}}$ and ${m_\text{gas} = \unit{1.9 \times 10^8}{\massh}}$ for the latter. 

The final simulation is called the Argonne cluster, which represents a 
modern version of the SB setup to be used for convergence studies. The initial conditions follow the procedure of \cite{hoffman1991}, similar to the SB arrangement. We employ the best-fit \mbox{WMAP-7} $\Lambda$CDM
cosmology with ($\Omega_c$, $\Omega_b$, $\Omega_\Lambda$, $\Omega_\nu$,
$w$, $n_s$, $\sigma_8$, $h$) = (0.22, 0.0448, 0.7352, 0, -1, 0.963, 0.8, 0.71) in a volume of width $L=\unit{64}{\Mpch}$. The ICs seed
a random field at $z=200$ on a $2048^3$ grid to accommodate particle counts up to
$2\times1024^3$ without incurring interpolation error from the finite mesh. Initial configurations for $N=2\times128^3,$ $2\times256^3,$ and $2\times512^3$ particles have been made publicly available for comparison.\footnote[7]{\url{https://cosmology.alcf.anl.gov/static/ArgonneCluster/ICs}} The mass resolution of the middle ($2\times256^3$) run corresponds to a dark matter mass $m_\text{DM} = \unit{9.54 \times 10^8}{\massh}$, and baryon mass $m_\text{gas} = \unit{1.94 \times 10^8}{\massh}$.

We begin with an investigation of the SB cluster. The simulation ran with a gravitational softening length set to $r_{\rm soft}=\unit{1.25}{\kpch}$. 
The main cluster was extracted using an FOF algorithm applied only to the dark matter particles. Spherical shells were built outwards from the gravitational potential center in order to construct radial profiles for both particle species.
The radius at which the interior density falls below $200\rho_c$ was found to be $R_{200} = \unit{2.75}{Mpc}$ with an enclosed mass of 
$M_{200} = \unit{1.21\times10^{15}}{M_\odot}$. These numbers are in
agreement with the values reported in \citet{Frenk1999} measured for a variety of solvers, albeit on the upper range of the data. As described in that work, cluster property measurements, in general, have all been found to be reasonably consistent among the different codes studied. 

Radial profiles for density, temperature, and entropy of the SB cluster at $z=0$
are shown in Fig.~\ref{fig:sbprofiles}. We have included measurements pulled from the literature
for a number of cosmological codes that employ a diverse set of hydrodynamic solvers. 
In particular, we compare results for
Eulerian AMR methods (\smaller{Nyx}; \citealt{almgren2013}, \smaller{SAMR}; \citealt{Frenk1999}), Lagrangian SPH 
schemes (\smaller{GADGET-2}; \citealt{Springel2005}, \smaller{DISPH}; \citealt{Saitoh2016}), and 
hybrid approaches (\smaller{GIZMO}; \citealt{Hopkins2015}, \smaller{AREPO}; \citealt{Springel2010}). In the case of \smaller{GIZMO}, we show the profiles produced by the meshless finite-mass (MFM) solver, and for \smaller{AREPO} we consider both the strict energy conservation and dual entropy formulations (marked appropriately in Fig.~\ref{fig:sbprofiles}). 

Consistent with the findings of \cite{Frenk1999} and subsequent investigations, the range of methods agree on both the complete dark matter profile, and gas measurements outside the cluster central region ($r \gtrsim \unit{0.5}{Mpch}$). However, for the inner domain ($r \lesssim \unit{0.5}{Mpch}$), 
deviations between solvers are apparent, the most prominent being the entropy profile; these conflicts have not been resolved with increasing resolution (e.g., \citealt{mitchell2009}). As traditionally observed, the Eulerian solvers (\smaller{Nyx} and \smaller{SAMR}) demonstrate a systematically higher entropy ``core,'' as opposed to the unmodified SPH codes (\smaller{GADGET-2}). These tensions have been somewhat alleviated by modern SPH methods (\smaller{DISPH}) and hybrid approaches (\smaller{AREPO-}entropy and \smaller{GIZMO}), however, there is still a significant offset in magnitude. Interestingly, \CRKHACC\ appears to produce consistent measurements with the mesh-based approaches, in contrast to the other schemes. 

One partial explanation for the discrepancy of the solvers (e.g., \citealt{Springel2010,Hopkins2015}), is the adherence of SPH codes to fluid mixing, thus suppressing the entropy profile. Additionally, it is further argued that grid codes are more susceptible to artificial heating, seeded by noise from the gravity solver, as well as over-mixing and diffusing entropy. In accordance, observe the \smaller{AREPO-}energy solution in Fig.~\ref{fig:sbprofiles}, the subject of which is described in detail in \cite{Springel2010}. Briefly, by disabling a dual entropy/energy switch used to suppress artificial heating incurred by cold fluids moving at a high Mach number, \smaller{AREPO} will generate similar entropy profiles as grid codes, supporting the claim that these cores originate from numerical processes. These results are reaffirmed in the findings of \cite{Hopkins2015}, where \smaller{GIZMO} can artificially recover the entropy cores by introducing enhanced diffusion via shutting off similar entropy/energy switches. The conservative conclusion of these examinations would be that the ``true'' solution lies between the unmodified SPH and Eulerian profiles, however, the degree to which is unclear. 

Despite these conjectures, \CRKHACC\ appears to produce consistent measurements with the mesh-based approaches without any modifications. Indeed, one of the benefits of the \CRKSPH\ solver is the improvement of fluid mixing in SPH simulations, although the extent to which it agrees with the Eulerian codes is notable. It is insightful to compare the \CRKHACC results with \smaller{DISPH}. Similar to the \CRKSPH methodology, \smaller{DISPH} utilizes a modern version of artificial viscosity, and promotes improved fluid mixing by utilizing energy-based SPH weighting functions (see \citealt{Hopkins2012} for a generalized description of this approach). Unsurprisingly, \smaller{DISPH} demonstrates improved solutions for the SB cluster, with profiles similar to \smaller{AREPO-}entropy, yet does not fully realize entropy cores as \CRKHACC does. 

A critical observation was made in \cite{Saitoh2016}, where \smaller{DISPH} could be altered to agree with the Eulerian codes if it included artificial conductivity treatments -- a common approach that introduces dissipation to smooth discontinuities in thermal energy (e.g., \citealt{Price2008}), akin to viscosity models for kinetic energy. In fact, multiple earlier studies have concluded that artificial conductivity can resolve the entropy profile tensions for SPH methods (e.g., \citealt{wadsley2008,Read2012,Power2014,biffi2015,Hopkins2015,beck2016}). While some of these investigations required unphysical conductivity models, which in turn degraded the SPH scheme, others found sophisticated switches to improve these effects. The culmination of this effort is conveyed well by the nIFTy comparison cluster, which we illustrate in Fig.~\ref{fig:nfprofiles}.

Similar to the SB project, the nIFTy collaboration includes results from a number of modern cosmology solvers. For comparison, we extracted profiles for both grid-based approaches (\smaller{RAMSES}; \citealt{teyssier2002}, and \smaller{ART}; \citealt{rudd2008}) and ``modern" SPH variants of \smaller{GADGET} (\smaller{G2-ANARCHY}; see Appendix A in \citealt{schaye2015eagle}, \smaller{G3-X-ART}; \citealt{beck2016}, \smaller{G3-SPHS}; \citealt{Read2012}, and \smaller{G3-MAGNETICUM}; \citealt{hirschmann2014}). We further include \smaller{AREPO-}energy and \smaller{GADGET-2} (referred to as \smaller{G2-X} in \citealt{sembolini2016nifty}) as was done in our previous SB investigation. The \smaller{AREPO-}entropy formulation was not utilized in this study.

Fig.~\ref{fig:nfprofiles} shows the radial profiles, where we measured the halo mass to be $M_{200} = \unit{1.59\times10^{15}}{M_\odot}$
with $R_{200} = \unit{2.41}{Mpc}$. While the traditional SPH approach once again displays similar tensions with the Eulerian codes for the entropy profile, a majority of the modern methods are in relative agreement, including the \CRKHACC results. However, there were a couple of SPH methods in the nIFTy investigation that produced suppressed profiles (e.g., the \smaller{G3-MAGNETICUM} solution in Fig.~\ref{fig:nfprofiles}). The primary listed contributor to the discrepancy is differing sophistication of artificial conductivity models. In fact, all of the modern SPH approaches required some form of conductivity to achieve their results, as described above. 

In light of this fact, the \CRKHACC findings are intriguing. The \CRKSPH framework displays consistent profiles with the other methods, yet it does not include conductivity in the hydrodynamic solver. While conductivity would be simple to add to the \CRKSPH methodology, we chose to avoid the dangers of potentially resolving unphysical entropy transport -- particularly as our fluid and self-gravitating experiments do not indicate a necessity for it (similarly stated in \citetalias{frontiere2017}).

To conclude, we briefly inspect the convergence of the \CRKHACC results, to ensure that the SB and nIFTy measurements are not seeded by unresolved numerical artifacts. We utilize the Argonne cluster for this purpose, simulated with both \CRKHACC\ and the AMR code \smaller{Nyx}.
In the case of \CRKHACC, we performed convergence runs of $N=2\times128^3$, $2\times256^3$,  $2\times512^3$ dark matter plus baryon particles and set the gravitational softening length to 1/16 the mean inter-particle separation ($r_{\rm soft}=\unit{15.6}{\kpch}$). Regarding \smaller{Nyx}, we ran using $N_{\text{DM}}=256^3$ dark matter
particles in conjunction with a hydro mesh of extent $N_\text{cell}=256^3$ that allowed for 4 levels of grid refinement. 

Fig.~\ref{fig:acprofiles} shows radial profiles of the Argonne cluster at $z = 0$ for
both codes. The halo mass is calculated to be $M_{200} = \unit{1.88\times10^{15}}{M_\odot}$
with $R_{200} = \unit{2.52}{Mpc}$. Reassuringly, we find that the internal profiles measured between \CRKHACC and \smaller{Nyx} are in close agreement. While the smallest simulation slightly undershoots the \smaller{Nyx} entropy core, the higher resolution \CRKHACC runs match well. 
Thus, \CRKHACC appears to be numerically converged to a consistent solution with Eulerian solvers, and achieves this result without the use of artificial conductivity. We intend to investigate these findings further in the future, where we suspect the explanation is likely connected to the compatible energy formalism in \CRKSPH. 

\section{Scale-free Cosmology Simulations}
\label{sec:selfsim}

Validation of cosmological simulations is complicated by the inherently non-linear evolution of gravitational collapse, precluding an analytic prediction. However, imposing scale-free initial conditions results in self-similar structure formation, where conformance to theoretical scaling relations can be measured to uniquely probe solver accuracy. This methodology pairs well with standard code comparisons, such as the investigations of the previous section, which provide guidance in gauging the relative systematic variances between different numerical methods. 

The theoretical motivation for self-similar solutions traces back to the early results of \cite{efstathiou1979} as well as derivations in \cite{peebles1980}. Further, \cite{kaiser1986} studied the application of similarity to predict the evolution of galaxy clusters. In general, scale-free simulations have a long history in N-body measurements (e.g., \citealt{efstathiou1988,colombi1995,jain1998,smith2003, widrow2009,ludlow2016,benhaiem2017,joyce2021} to name a few). 

The concept is to construct a cosmological simulation where the only relevant physical scale
is the amplitude of density fluctuations. Following the approach outlined in \citet{owen1998adiabtic}, this can be achieved by instantiating a particle distribution in an Einstein--de Sitter (EdS) universe that is displaced by a Gaussian random field of initial power spectra 
\begin{equation}\label{eq:power}
P(k) = A_0 k^{n_s}, 
\end{equation}
with normalization $A_0$ and spectral index $n_s$.
The initial conditions and background growth are both free of any physical scale, and, therefore, produce self-similar growth in time. 
Moreover, we expect numerical artifacts to not obey the theoretical scaling relations, where measurements of departure from self-similarity can highlight the spatial resolution limitations of a solver.

We focus here on simulations with gravity and non-radiative hydrodynamics, as both phenomena do not introduce any additional fundamental scales. 
One can further expand this approach to model radiative gas, which is explored for a particular family of power-law cooling functions in \cite{owen1998cooling}. 
The subsequent analysis is restricted to similarity measurements of integrated halo properties and profiles, where we aim to quantify the adherence to predicted scaling relations of hydrodynamic quantities. 

To determine self-similarity, we first need to identify the underlying (non-linear) gravitational collapse scale, where we denote measurements with an asterisk; namely, mass ($M_*$), density ($\rho_*$), radius ($R_*$), temperature ($T_*$), and entropy ($S_*$). Due to the hierarchical nature of structure formation, the non-linear extent will increase with time. Self-similarity dictates that objects normalized by the collapsed scales will match corresponding structures at later times. For instance, the scaled temperature, $T/T_*(t)$, should agree for halos of fixed scaled mass $M/M_*(t)$ (but not fixed $M$) at all times $t$. While technically we still lack analytic predictions to compare the structure of specific collapsed objects, we now are capable of confirming temporal self-similarity over the entire evolution.

In the following analysis, we employ a scale-free cosmological simulation to investigate self-similarity using the combined gravity and non-radiative hydrodynamic solvers in \CRKHACC. 
The simulation contains $N=2\times512^3$
dark matter plus baryon particles in a box of width $L=\unit{40}{\Mpch}$, with corresponding mass resolutions
$m_\text{DM} = \unit{1.06 \times 10^8}{\massh}$ and $m_\text{gas} = \unit{2.65 \times 10^7}{\massh}$.
The initial conditions are
set at $z = 200$ using a power-law power spectrum (\cref{eq:power}) with spectral index $n_s = -2$ and normalization
determined by $\sigma_8 = 0.5$. The cosmology is specified by an EdS universe with $\Omega_b = 0.2$,
$\Omega_c = 0.8$, and $h = 0.5$. The gravitational softening length is fixed to 
$r_{\rm soft} = \unit{4.88}{\kpch}$, and we use \textsc{CosmoTools} (see Section \ref{sec:analysistools}) to identify SO halos with $\Delta_c = 200$ relative to the critical density of the universe at a range of simulation snapshots.

We begin the investigation in Section~\ref{sec:ss_relations}, by providing a derivation of the scaling relations imposed by scale-free initial conditions.
We then use a gravity-only simulation in
Section~\ref{sec:ss_go} to determine the resolution limits where self-similarity breaks down due
to numerical artifacts. Finally, we present self-similar measurements when hydrodynamics is additionally modeled in Section~\ref{sec:ss_results}.

\subsection{Scaling Relations}
\label{sec:ss_relations}

The following derivation of the non-linear scale quantities (e.g., $M_*$) considers specifically a spectral index $n_s = -2$; however, the procedure follows identically for any arbitrary choice of $n_s$. 
By definition, $M_*$ is inferred from the scale at which gravitational collapse occurs. 
The spherical collapse model in an EdS universe predicts this scale is reached for density perturbations above a critical value of $\delta_c = 1.686$ \citep{Lacey1993}. We can calculate $M_*(a)$ by smoothing the linear density field with filters of varying mass to measure where the density first exceeds $\delta_c$.

We follow the approach outlined in \citet{Lacey1993}, whereby the variance in the density field
for mass scale $M$ is computed by integrating over the power spectrum $P(k,a)$ filtered with a spherical
kernel of radius $R(M)$:
\begin{equation}
\sigma^2(M,a) = \frac{1}{2\pi^2} \int_0^\infty k^2 P(k,a) W[R(M)k]^2 \, \mathrm{d}k.
\end{equation}
Here, $W(x) \equiv (3/x^3)(\sin{x}-x\cos{x})$ is the filter, and $R(M) = (3M/4\pi\bar{\rho})^{1/3}$, with $\bar{\rho}$ denoting the mean matter density of the universe. 
We can simplify this expression by substituting ${P(k,a) = a^2A_0k^{-2}}$, where we make use of the 
fact that the growth factor is simply $a$ in an EdS universe. 
In the case where $n_s = -2$, we are left with an integral over the filter function ${\int_0^\infty W(Rk)^2 dk}$ = $3\pi/(5R)$, allowing us to write 
\begin{equation}
\sigma^2(M,a) = \frac{3a^2A_0}{10\pi R(M)}.
\end{equation}
The normalization $A_0$ is determined by imposing the constraint that $\sigma^2(M,1)$ evaluates to $\sigma_8^2$ when smoothed on the length scale $R = \unit{8}{\Mpch}$.
With this requirement, we find that $A_0 = 80\pi\sigma_8^2/3$ and, thus,
arrive at the final result:
\begin{equation}
\sigma^2(M,a) = a^2\frac{8\sigma_8^2}{R(M)}.
\label{eq:sigma2ma}
\end{equation}

The non-linear mass scale is evaluated by setting $\sigma^2(M_*,a) = \delta_c^2$ and solving 
\cref{eq:sigma2ma} for $M_*$.  We then associate $M_*$ with a density scale,
$\rho_*$, by making the assumption that the non-linear mass collapses to form spherical objects with an overdensity $\Delta_c$. We choose a value of $\Delta_c = 200$ to facilitate comparisons with the convention used in our halo-finder, though we
note that the exact choice of $\Delta_c$ is unimportant in terms of identifying self-similar relations.
The length scale $R_*$ is set as the radius of a sphere of mass $M_*$ and density $\rho_*$, while the non-linear temperature is taken to be the virial quantity $T_* = (\mu m_p)/(2k_B)(GM_*/R_*)$. For the molecular weight $\mu$, we assume a constant value of $\mu=0.59$ for fully ionized gas. 
Finally, entropy is calculated following the definition 
$S_* = {\rm ln}(T_*/\rho_*^{2/3})$. Evaluating all of these relations for the use case of 
$\sigma_8 = 0.5$ yields the following results expressed in proper units:
\begin{align}
M_* &= 4.05\times 10^{11} a^6\ h^{-1}M_\odot, \nonumber\\
\rho_* &= 5.55\times 10^{13} a^{-3}\ h^2 M_\odot {\rm Mpc}^{-3},\nonumber\\
R_* &= 0.12 a^3\ h^{-1}{\rm Mpc},\nonumber\\
T_* &= 5.19\times 10^5 a^3\ {\rm K},\nonumber\\
S_* &= -7.94 + 5{\rm ln}(a).
\label{eq:ss-scalingrelations}
\end{align}
With the scaling quantities defined, we now proceed with an analysis of the convergence limits in the scale-free simulations. 

\subsection{Convergence Limits}
\label{sec:ss_go}

The objective of this investigation is to provide measurements of temporal self-similarity for halos spanning a broad range in both mass and time. 
As mentioned previously, the finite resolution of numerical simulations will introduce non-physical artifacts that violate the assumptions required for self-similar growth. Therefore, we must first determine a minimum length $R_{\rm conv}$ and mass $M_{\rm conv}$ at which we would expect to find converged measurements of self-similarity.

An evident non-physical measure introduced in the simulation is the gravitational softening length, $r_{\rm soft}$. 
Another related characteristic scale is the extent at which mass segregation effects occur due to the artificially strong gravitational forces between the light baryon and heavy dark matter particles \citep[e.g.,][]{Efstathiou1981,Binney2002,Power2016,emberson2019}.
The hydrodynamic force resolution 
set by the SPH neighbor count also introduces a nonphysical scale, albeit one that is spatially 
adaptive to the local baryon density. 
Based on the analysis in \citet{emberson2019}, we expect that the coarsest of these limitations is set by mass segregation, which was reported to occur at lengths a few times larger than $r_{\rm soft}$, as well as in halos containing fewer than roughly three thousand particles. 

A straightforward way to measure the scales impacted by mass segregation is to perform a gravity-only simulation. In this case, we expect the baryons to trace the dark matter, as they both follow the same gravitational evolution. Hence, the baryon fraction measured within halos should equal the universal mean, $f_b = \Omega_b/\Omega_m$, and any departures from this value will signify that those scales have failed to converge due to numerical artifacts. Therefore, to obtain a measure of the converged quantities $R_{\rm conv}$ and $M_{\rm conv}$, we run a gravity-only realization with the same initial conditions as the non-radiative target simulation. 
We then measure the baryon fraction within
halos and use departures from the mean value to set the convergence limits in our analysis.

\begin{figure}[htp]
    \centering
    \includegraphics[width=\linewidth]{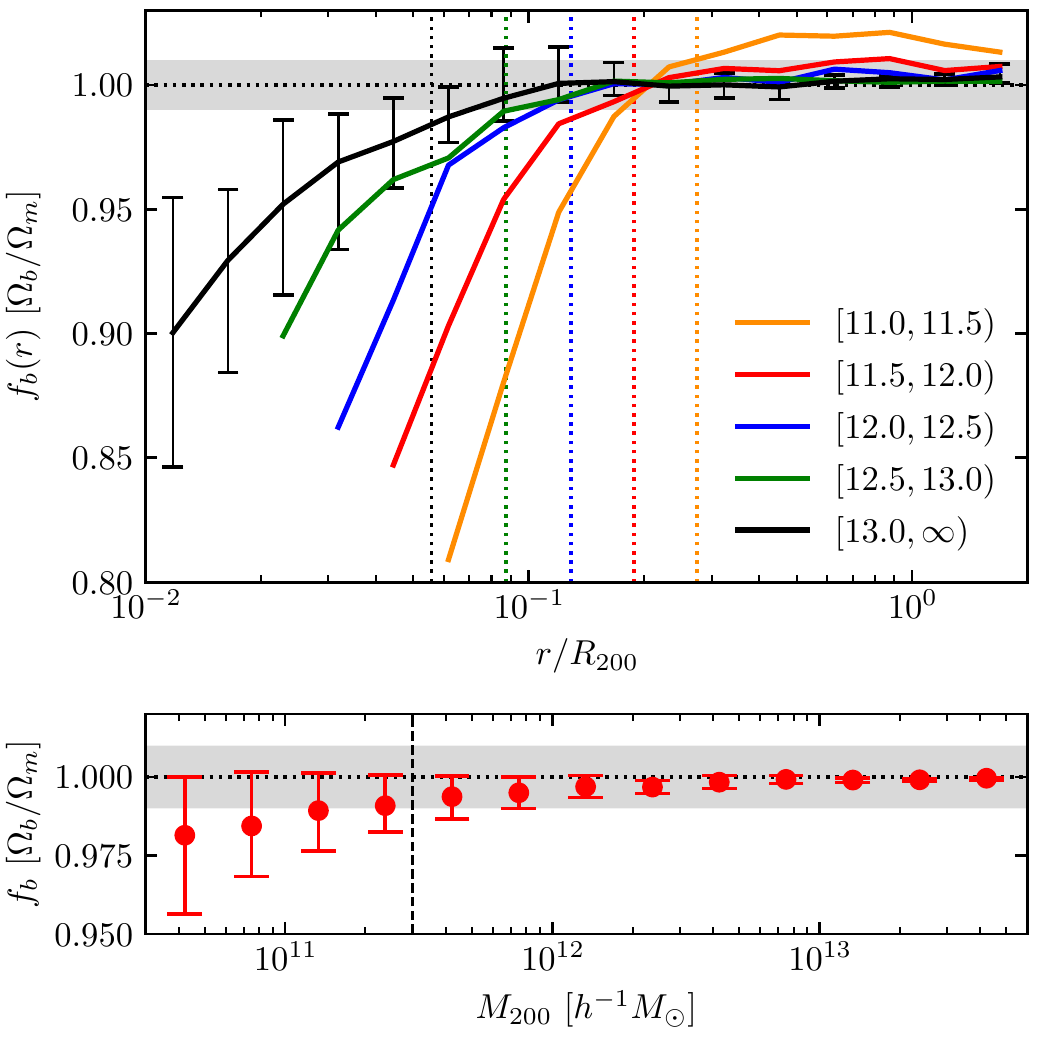}
    \caption{(Top) Radial profiles of the baryon fraction normalized to the universal mean at $z = 0$
    for the gravity-only scale-free simulation. 
    The solid colored lines trace the median profile for halos stacked in five mass bins. The legend lists the bin widths in units of ${\rm log}(M_{200})$.
    For the most massive bin, we also show the 25th
    and 75th percentiles from the stack as error bars. The vertical dotted lines indicate the location
    of the convergence length $R_{\rm conv} = \unit{24.4}{\kpch}$ for each mass bin of the corresponding
    color. (Bottom) Global baryon mass fraction in units of the universal mean as a function of $M_{200}$ at $z = 0$. The red circles denote the median in each bin, while the error bars show the 25th
    and 75th percentiles of the distribution. The vertical dashed black line indicates the converged mass scale $M_{\rm conv} = \unit{3\times10^{11}}{\massh}$. In both panels, the gray shaded band
    highlights $\pm1\%$ deviations from the universal mean.}
  \label{fig:self-similar-gasfraction}
\end{figure}

In the top panel of Fig.~\ref{fig:self-similar-gasfraction}, we show radial profiles of the baryon fraction stacked over halos in different mass bins. 
To avoid potential scatter from major merger events, we consider only relaxed halos for which the halo potential minimum and center of mass deviate by no more than $7\%$ of the SO radius $R_{200}$. 
The result is that the four largest mass bins agree well with the universal baryon fraction at $r \sim R_{200}$ but show a steady decline in $f_b$ with decreasing radius.
The vertical dotted lines indicate a fixed length of $5r_{\rm soft} = \unit{24.4}{\kpch}$ for each mass bin, highlighting a roughly constant physical deviation scale. 
For the four largest mass bins, this scale approximately separates where the baryon fraction begins to diverge at the percent level from the universal mean. 
Therefore, we set $R_{\rm conv} = \unit{24.4}{\kpch}$ as the
minimum converged length scale in our simulations. 

It is important to recognize that the lowest mass bin in the top panel of Fig.~\ref{fig:self-similar-gasfraction} deviates
from the universal baryon fraction on scales larger than $R_{200} > R_{\rm conv}$. In this case, the lack of 
convergence is related to insufficient particle sampling to constrain the mass segregation to the halo core.
We examine this explicitly in the bottom panel of Fig.~\ref{fig:self-similar-gasfraction},
where the global baryon fraction is measured as a function of halo mass. 
It is apparent that large mass halos agree well with the universal mean while the smallest objects are
systematically low. From this result, we set the minimum converged mass to 
$M_{\rm conv} = \unit{3\times10^{11}}{\massh}$, which corresponds to $\sim2300$ times the combined
dark matter plus baryon particle mass.
With $R_{\rm conv}$ and $M_{\rm conv}$ determined, we now investigate self-similar scaling when coupling the hydrodynamic force solver. 

\subsection{Self-similar Measurements}
\label{sec:ss_results}
For our final examination, we explore temporal self-similarity of integrated halo measurements, as well as spherically averaged radial profiles for the non-radiative scale-free simulation.
In accordance with the previous section, we include only
relaxed halos, discard masses below $M_{\rm conv} = \unit{3\times10^{11}}{\massh}$,
and truncate radial profiles below $R_{\rm conv} = \unit{24.4}{\kpch}$. 
If this selection process successfully mitigates the contributions from numerical artifacts, 
we expect to find self-similar evolution over all remaining mass and length scales.

\begin{figure*}[htp]
\centering
\includegraphics[width=0.32\textwidth]{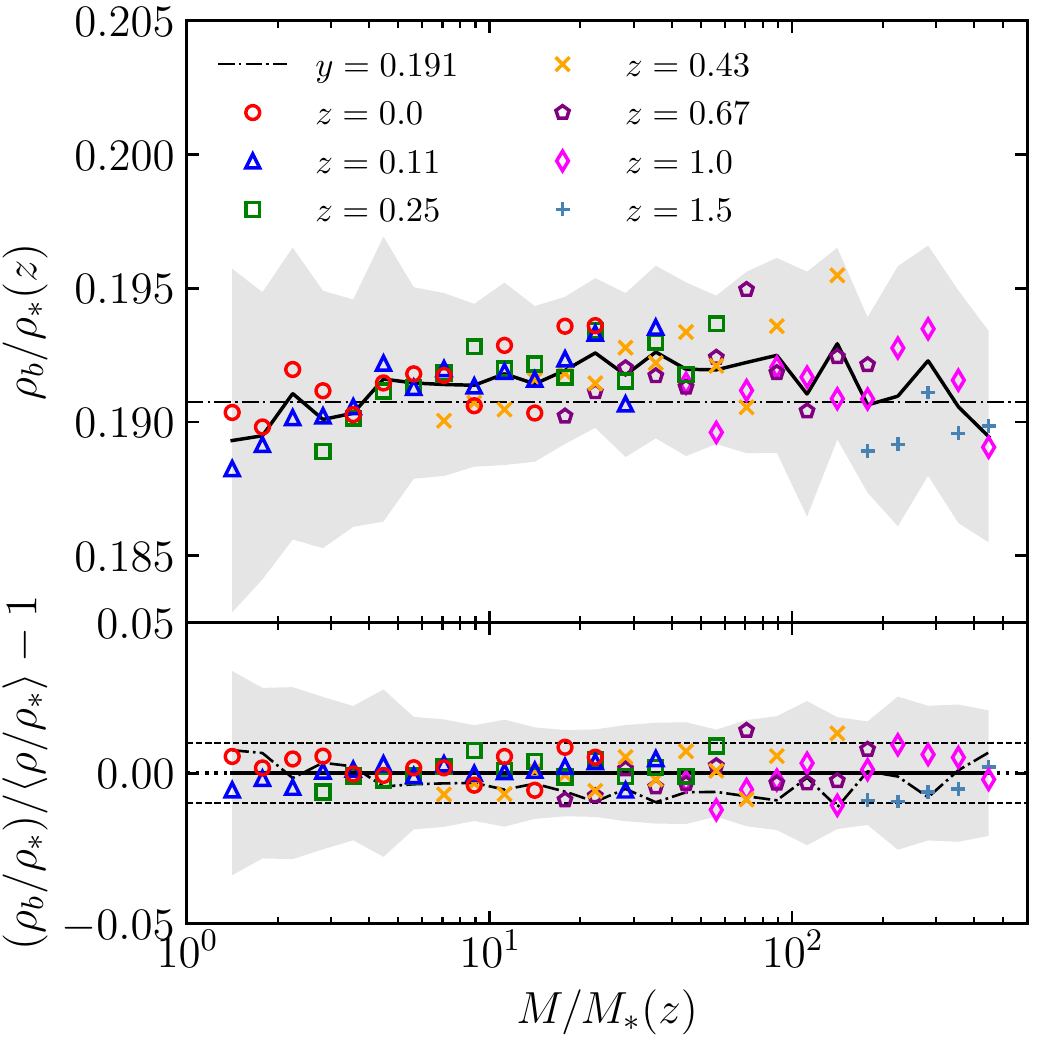}
\includegraphics[width=0.32\textwidth]{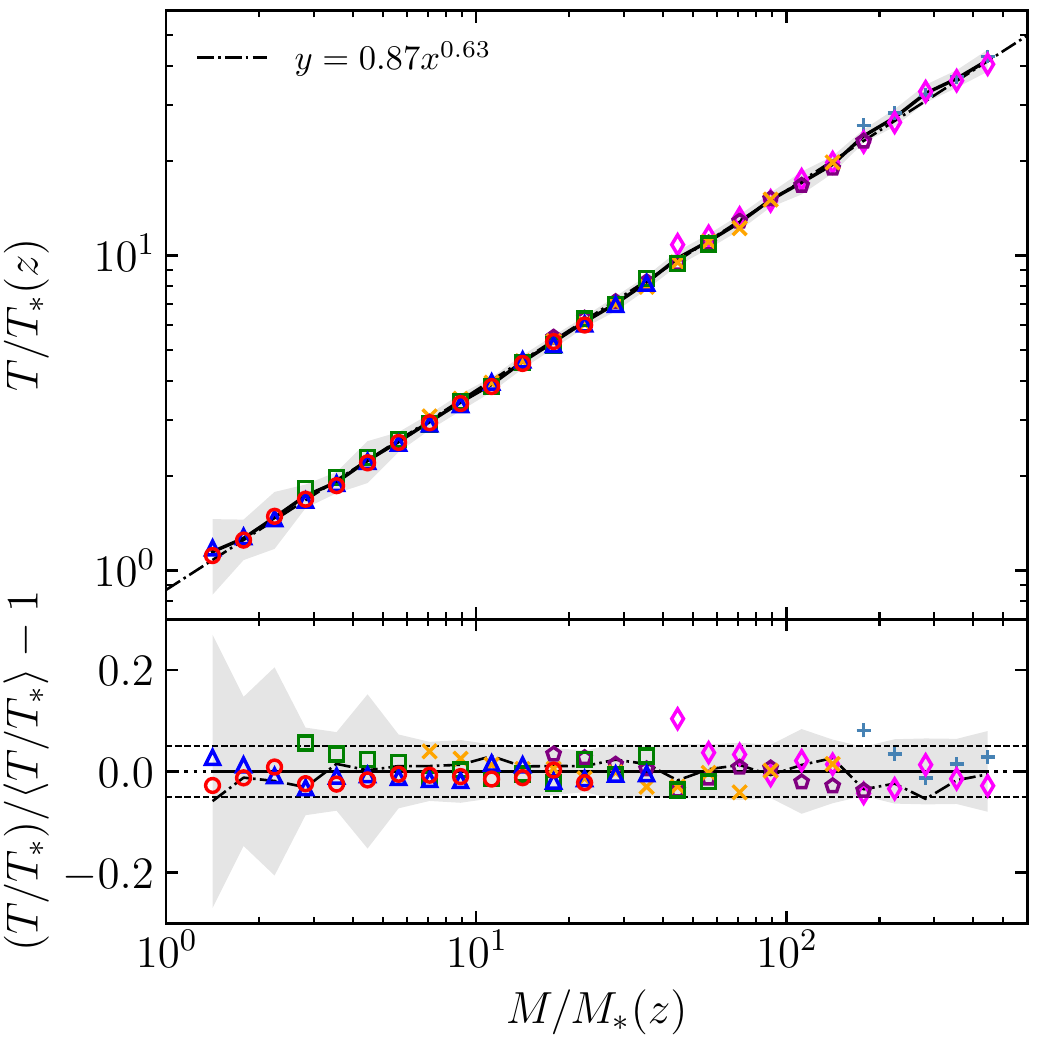}
\includegraphics[width=0.32\textwidth]{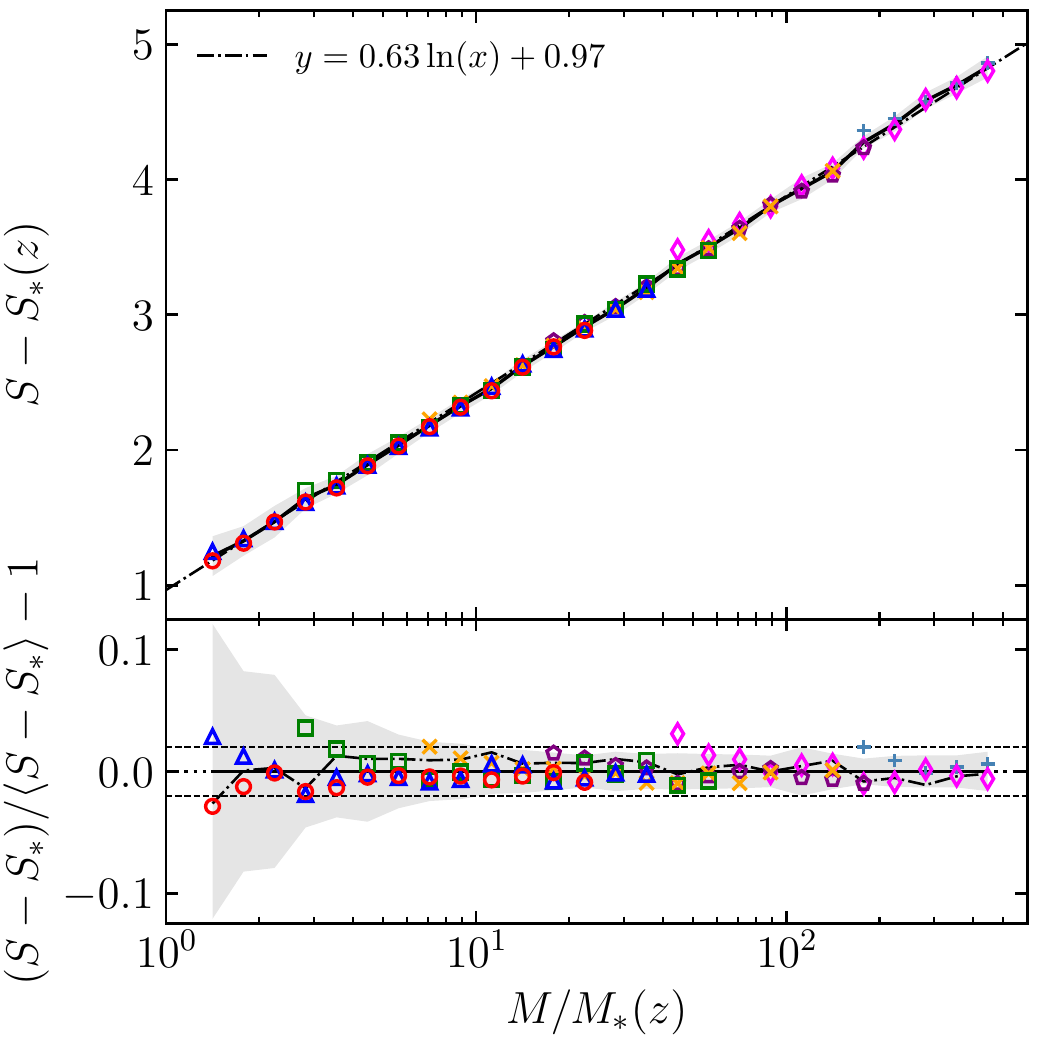}
    \caption{Integrated halo baryon density (left), temperature (center), and entropy (right)
    normalized to the corresponding non-linear scale defined in \cref{eq:ss-scalingrelations}
    for redshifts listed in the legend. 
    Each marker denotes the mean value for all halos in a given
    redshift and mass bin consisting of at least 10 halos. 
    We only include results from bins containing more than one redshift sample.
    The solid black line traces the mean value of each bin for all redshifts, while the shaded gray band shows the corresponding standard deviation. The dash-dotted black line represents a least-squares regression fit of the data for each quantity. 
    The bottom panel shows the ratio of all measurements to the mean curve, where the horizontal dashed lines indicate a $\pm1\%$, $\pm5\%$, and $\pm2\%$ spread for the left, center, and right panels, respectively.
    }
    \label{fig:ssglobals}
\end{figure*}

We begin with an investigation of the global density, temperature, and entropy integrated over
the full halo volume. In the case of density, we specifically measure the baryon component, $\rho_b = 3M_{200,b}/(4\pi R_{200}^3)$, where
$M_{200,b}$ is the baryon mass contained within $R_{200}$. 
Regarding temperature, we calculate the mass-weighted average of all baryon particles within the halo, which in turn determines the entropy as $S = {\rm ln}(T/\rho_b^{2/3})$.
To gauge self-similarity of the integrated measurements, we use the following procedure.

At each simulation snapshot, we separate the halos into scaled mass bins that are evenly distributed in log space by $\Delta {\rm log}(M/M_*) = 0.1$ (henceforth $M$ denotes $M_{200}$).
For bins containing at least 10 halos, we record the mean and standard deviation $\sigma$ of the integrated density, temperature, and entropy normalized by the corresponding non-linear values listed in \cref{eq:ss-scalingrelations}.
Finally, we collect results over all snapshots and discard any mass bins that do not contain at least two redshift samples.

The outcome of this procedure is shown in Fig.~\ref{fig:ssglobals}. For each integrated quantity, we plot the mean value of a given mass bin at multiple redshifts listed in the legend. 
We further average each bin over all corresponding redshift values and trace the result with a solid black curve. 
Similarly, the gray shading centered on the curve denotes the standard deviation obtained in quadrature from the individual $\sigma$-measurements at each redshift. 

The test for self-similarity is determined by considering each fixed $M/M_*$ mass bin and confirming that the individual redshift samples are in reasonable agreement. 
To facilitate this comparison, the bottom panels of Fig.~\ref{fig:ssglobals} show the ratio of individual redshift measurements and the averaged value in each bin.
For almost all normalized masses, we find that the redshift samples are well 
contained within their combined $1\sigma$ scatter. The variation in $\rho_b/\rho_*$, $T/T_*$, and $S-S*$ is roughly $1\%$, $5\%$, and $2\%$, respectively.
The mass bins with the highest constraints on self-similarity are those with the most overlapping redshift samples; in particular, mass bins near $M/M_* \simeq 20$ contain the five lowest redshift values, ranging from $z = 0$ up to $z = 0.67$. From \cref{eq:ss-scalingrelations}, $M_* \propto a^6$, indicating that halos within these bins differ by a factor of $\sim20$ in mass.

We reiterate that the analysis is performed on halos at fixed $M/M_*$, and that the scale-free setup does not generally predict self-similar growth of objects at different $M/M_*$ values.
Nevertheless, in cosmological simulations, halos tend to evolve into characteristic density profiles \citep{Navarro1997} where it is reasonable to assume that some level of self-similarity will exist across mass bins. 
In fact, this assumption has been used in the past to model the evolution of galaxy clusters \citep[e.g.,][]{kaiser1986}. 
To determine the extent of similarity across different mass bins, we can measure the scaling behavior across each $M/M_*$ sample. 

The result of this investigation is characterized by the black dash-dotted lines in Fig.~\ref{fig:ssglobals},
which represent fitted scaling relations made across $M/M_*$ bins. Regarding density, the data is well approximated by a constant ratio
of $\rho_b/\rho_*(z) = 0.191$. 
Given that the universal mean baryon fraction is $\Omega_b/\Omega_m = 0.2$, halos in the scale-free simulation are evolving to a constant proportion $f_b = 0.191/0.2 = 0.955$.
Interestingly, this finding is consistent with the $f_b$ measurements produced by the cosmological non-radiative {\em Borg Cube} simulation, which were recorded to be $\sim95\%$ that of the universal mean \citep{emberson2019}.
For temperature, the power-law 
$T/T_* = 0.87 (M/M_*)^{0.63}$ tightly characterizes the data, which is in close agreement with the relation
${T/T_* \propto (M/M_*)^{2/3}}$
that one would expect
if self-similarity was assumed to hold across $M/M_*$\footnote[2]{Following from $T \propto M/R,$ with $R \propto M^{1/3}$.}.
A comparable result was observed in 
the $n_s = -1$ scale-free simulations of \citet{owen1998adiabtic}, who found a mass-temperature
power-law exponent of $0.6\pm0.1$. 
Finally, the entropy fit was determined to be ${(S-S_*) = 0.63\,{\rm ln}(M/M_*) + 0.97}$, which is consistent with inserting the density and temperature power-laws into the definition $S= {\rm ln}(T/\rho^{2/3})$. 
Hence, we indeed observe a high degree of self-similarity across $M/M_*$, despite not being a generally analytic prediction of the scale-free initial conditions.

\begin{figure*}[htp]
\centering
\includegraphics[width=0.32\textwidth]{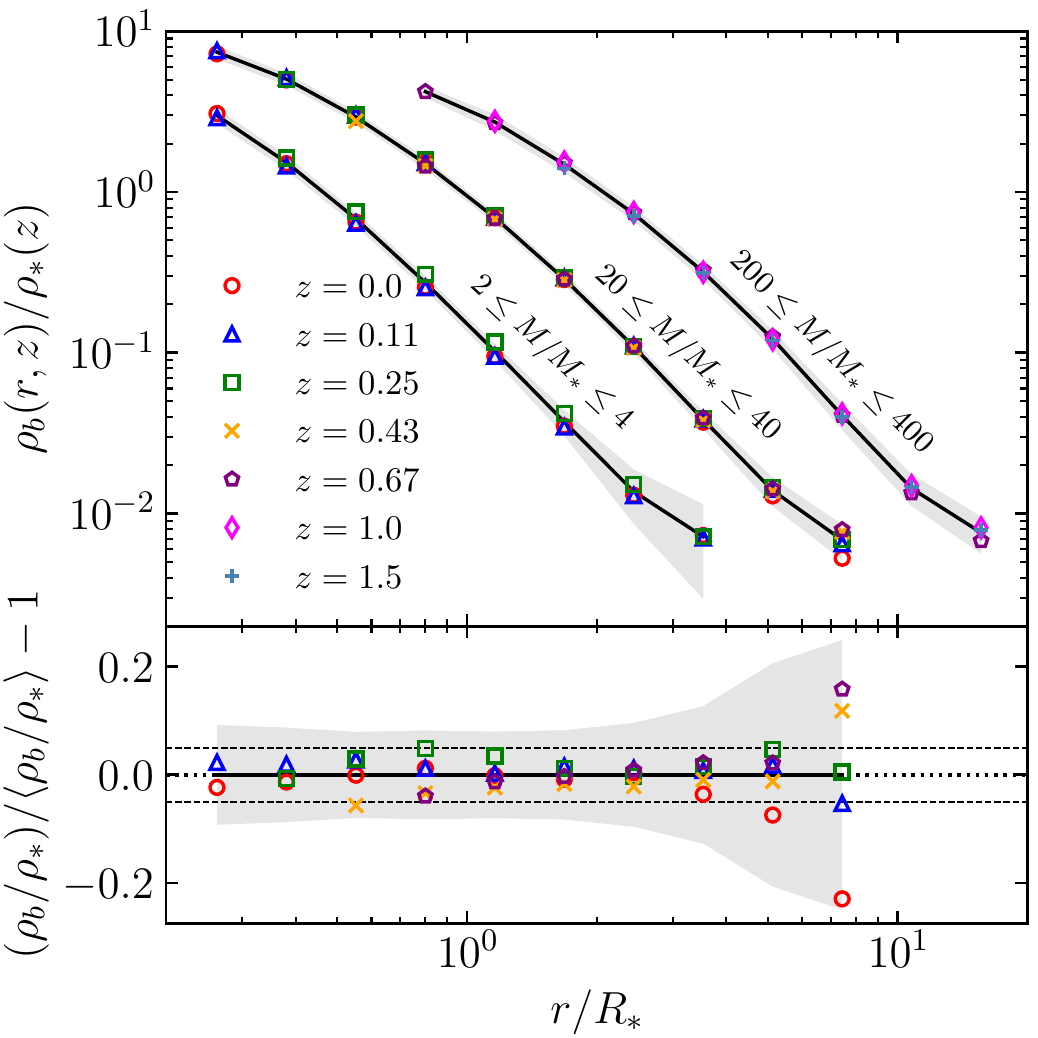}
\includegraphics[width=0.32\textwidth]{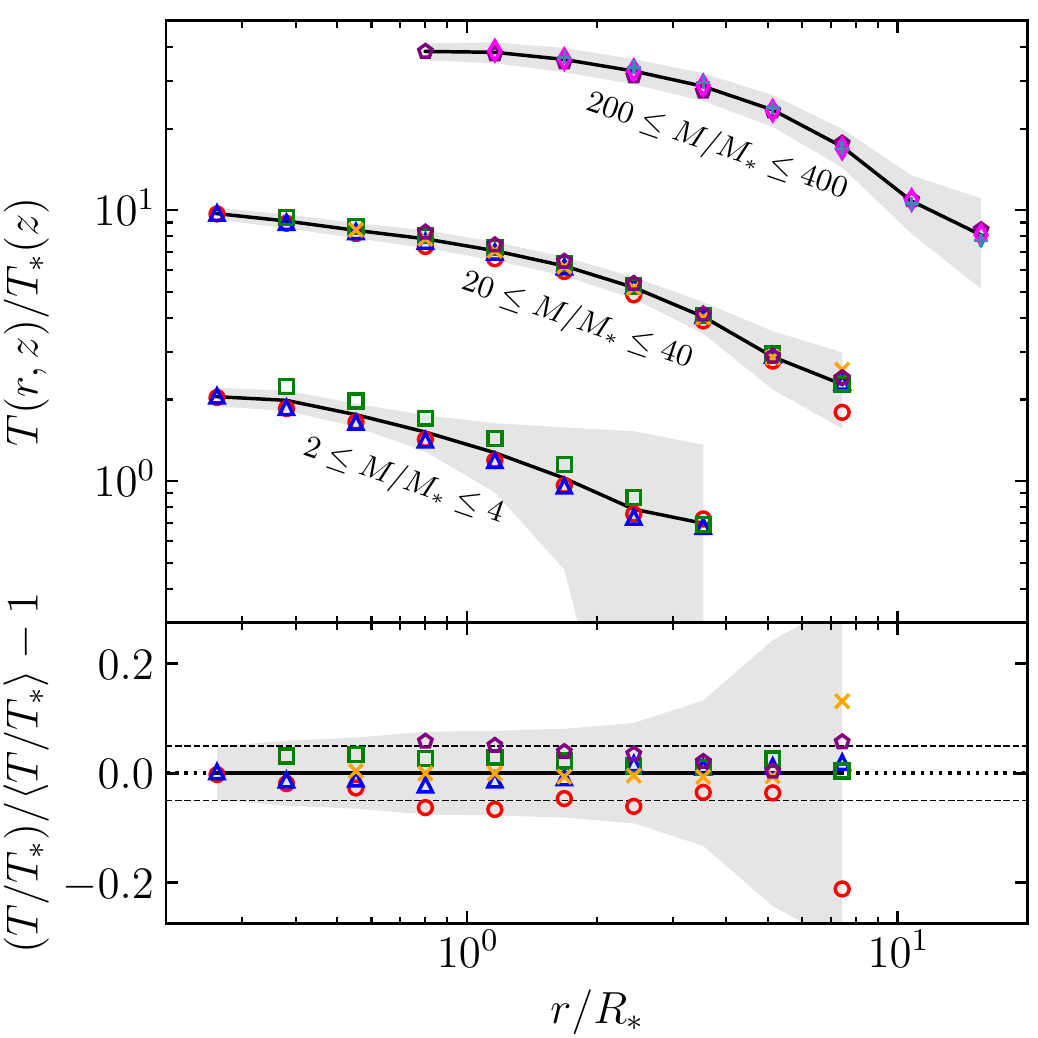}
\includegraphics[width=0.32\textwidth]{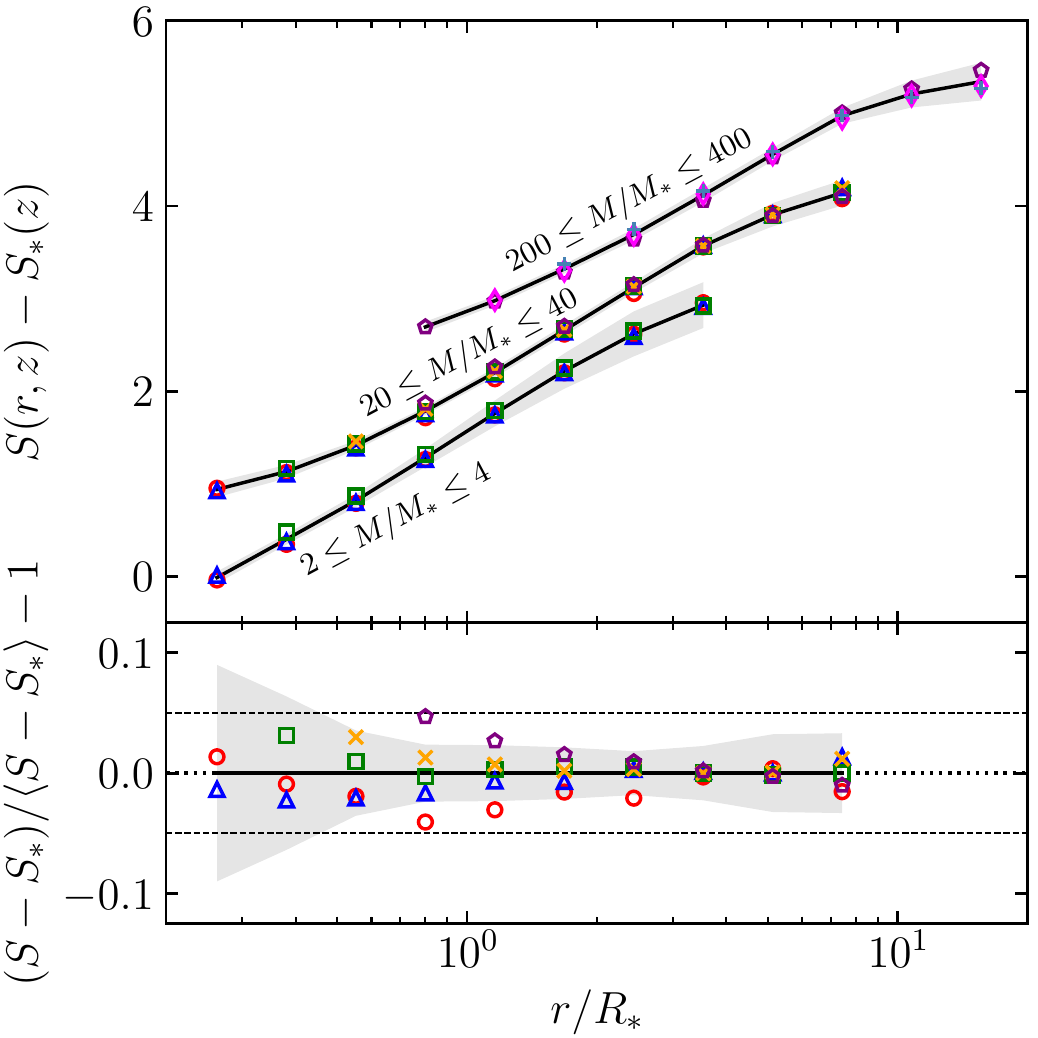}
\caption{Stacked radial profiles of baryon density (left), temperature (center), 
and entropy (right) normalized to the non-linear scale (\cref{eq:ss-scalingrelations}) for redshifts 
detailed in the legend. We show results for three separate $M/M_*$ mass
bins, which are separately labeled in each plot. The markers denote the mean stack for halos
at a fixed redshift within a given mass bin, while the solid black line and gray shading trace the combined stack and its standard deviation. 
The bottom panels show the ratio with the combined stack profile of the middle mass bin, where the horizontal dashed lines indicate a $\pm5\%$ spread.
}
\label{fig:ssprofiles}
\end{figure*}

We finish the scale-free evaluation by considering spherically averaged radial profiles of halos with fixed $M/M_*$ at different redshifts.
For this analysis, we focus on halos grouped into
three mass bins: i) $2 \leq M/M_* \leq 4$, ii) $20 \leq M/M_* \leq 40$, and  
iii) $200 \leq M/M_* \leq 400$. 
From \cref{eq:ss-scalingrelations}, these bins roughly correspond to halo masses of $M=10^{12}$, $10^{13}$, and $\unit{10^{14}}{\massh}$, respectively, at redshift $z = 0$. For each halo, we calculate radial profiles of baryon density, temperature, and entropy along a common $r/R_*$ axis.
We then produce stacked measurements by computing mean profiles at each corresponding redshift, in addition to averaging across all redshifts, where we determine the standard deviation in quadrature. 

The profile measurements are shown in Fig.~\ref{fig:ssprofiles}. 
Similar to the integrated halo quantities, we find that the individual redshift results agree within their combined $1\sigma$
scatter for each mass bin. 
In the bottom panels, we highlight the scatter in the redshift measurements with respect to the combined stacked profile (over all redshifts) for the middle mass bin. 
We have chosen this bin as it displays the strongest test of self-similarity, containing five redshift samples spanning an order of magnitude in physical halo mass.
In this case, we measure agreement at the $5\%$ level for each fixed $r/R_*$ bin, except 
in the outermost regions of the density and temperature profiles where the variance in the
combined stack is also enhanced. We have calculated comparable results in the other two mass bins, indicating that radial self-similarity is achieved over three decades in $M/M_*$ and
nearly two decades in $r/R_*$.

\begin{figure}[htp]
    \centering
    \includegraphics[width=0.90\linewidth]{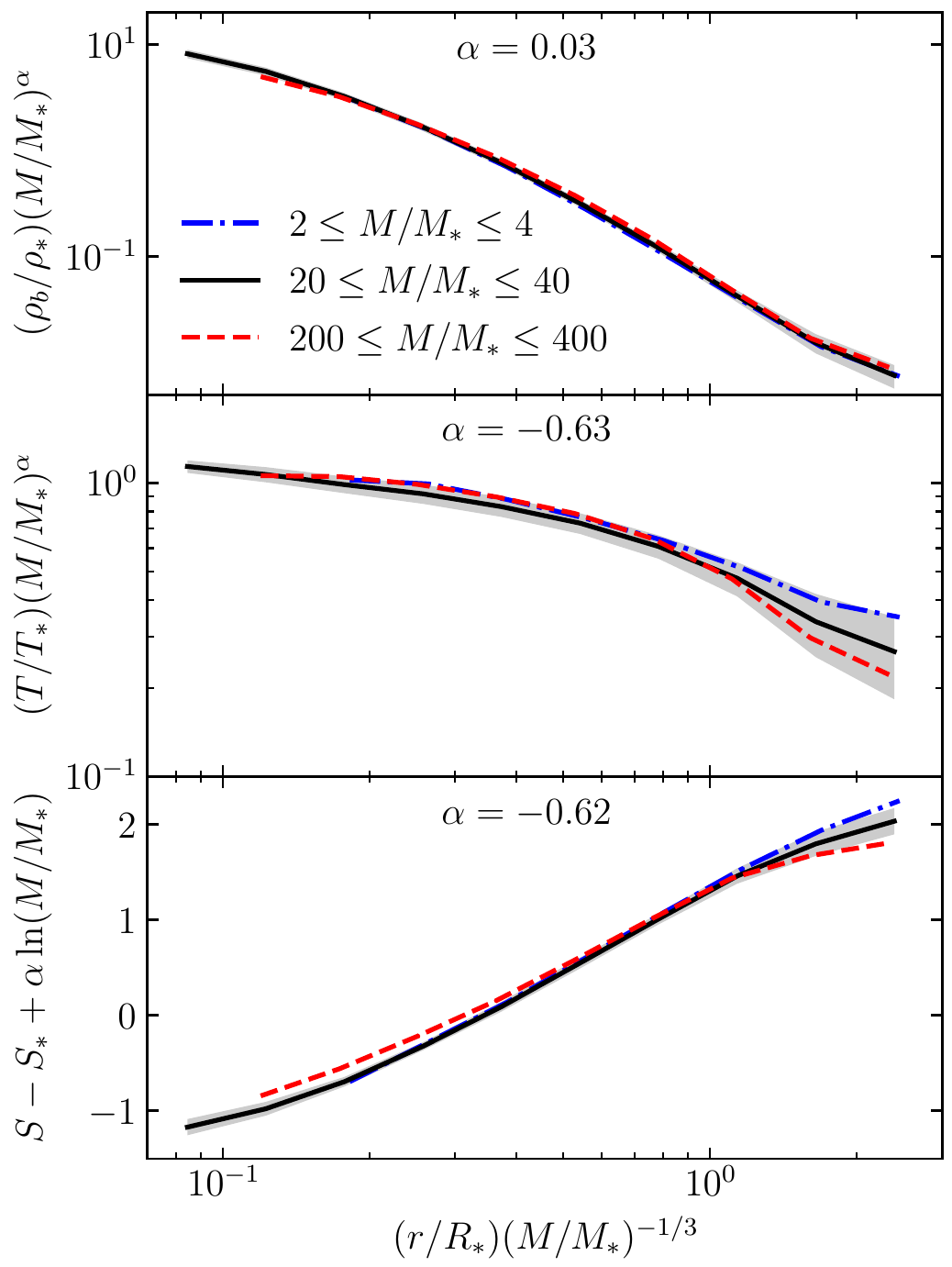}
    \caption{Stacked radial profiles showing baryon density (top), temperature (center), and
    entropy (bottom) for three $M/M_*$ bins indicated in the legend. The radial profiles
    have been transformed in the $x$-axis by $(M/M_*)^{-1/3}$ and scaled in the $y$-axis using
    a procedure that minimizes the L1-norm between the three curves.}
  \label{fig:self-similar-radialss}
\end{figure}

As was done for the integrated quantities, we investigate potential self-similarity across mass bins, in this case by taking the three solid black curves
in Fig.~\ref{fig:ssprofiles} and performing two operations.
First, we use the fact that $R \propto M^{1/3}$ to shift the $x$-axis
of each profile via $(r/R_*)(M/M_*)^{-1/3}$, where $M$ is taken as the midpoint of each bin.
Second, for the density and temperature profiles, we consider
arbitrary scaling in the $y$-axis by $(Y/Y_*)(M/M_*)^\alpha$, where $Y$ denotes either
$\rho_b$ or $T$ and $\alpha$ is a free parameter. In the case of entropy, this is modified
as a vertical translation, $(S-S_*) + \alpha\,{\rm ln}(M/M_*)$. We optimize for $\alpha$ by performing
a uniform search to determine values that minimize the L1-norm between the
three sets of curves. We find $\alpha = 0.03$, $-0.63$, and $-0.62$ for $\rho_b$,
$T$, and $S,$ respectively. 
Fig.~\ref{fig:self-similar-radialss} shows the stacked radial profiles for the three $M/M_*$ bins when scaled in this manner,
where we observe a high degree of self-similarity in a mass range spanning two orders of magnitude at any fixed redshift. 
The optimized $\alpha$ values also match well with the 
fitting functions obtained in Fig.~\ref{fig:ssglobals}, which correspond to 
$\alpha = 0$, $-0.63$, and $-0.63$ for $\rho_b$, $T$, and $S$, respectively.

In summary, we have found a high degree of temporal self-similarity for both integrated
halo measurements and stacked radial profiles. 
Predicted scaling relations are obeyed at the $1\%$, $5\%$, and $2\%$ level, for the global baryon density, temperature, and entropy, respectively, as well as at the $\sim5\%$ level for the profiles of each quantity. 
Additionally, we have observed further similarity between halos  of varying $M/M_*$, although the presence of this feature is not a general outcome of the scale-free derivation. 
\section{Conclusions and Future Work}
\label{sec:conclude}
In this compendium we introduced the \CRKHACC framework, an enhancement of the highly performant cosmology code \HACC to include the capability of modeling hydrodynamics. We summarized the primary software components that were developed for this extension (Section~\ref{sec:addons}) as well as established solver correctness on a variety of standard validation tests in both the idealized fluid and cosmological domains (Section~\ref{sec:evaluaton}). Of particular note is the cluster comparison investigation (Section~\ref{sec:cluster}), where \CRKHACC\ displayed close agreement with other modern SPH and Eulerian-based solvers despite not including artificial conductivity in our methodology. 
We further analyzed scale-free simulations (Section~\ref{sec:selfsim}), a useful exercise that can be repeated for cross-code validation, and has extensions that can incorporate radiative cooling models. As predicted by theory, the \CRKHACC\ solver demonstrates self-similarity of converged structure-forming quantities, such as integrated halo measurements and profiles. 

The accuracy afforded by the \CRKSPH\ approach combined with the performance of the \HACC\ design will be important for properly including baryonic effects in upcoming large-volume cosmological simulations targeting survey predictions. \CRKHACC is optimized to exploit current and future supercomputing hardware, particularly GPU-accelerated systems. These capabilities will facilitate next-generation hydrodynamic simulations that encompass Gpc-scale volumes, providing excellent statistics of massive halos, and enabling numerous synthetic sky measurements. 

While we have outlined the primary elements of the \CRKHACC\ framework in this paper, there are multiple components that warrant further explanation and examination. For example, our validation was limited to the adiabatic solver, neglecting the inclusion of subgrid sources and radiative effects. However, \CRKHACC\ has been furnished with a number of galaxy formation models (summarized in Section \ref{sec:subgrid}) that will be necessary to evolve baryonic processes important for structure formation. The complexities of formulating and calibrating subgrid models, particularly for cosmological probes, come with their own set of challenges and will be the focus of upcoming studies. Additionally, the incorporation and GPU-acceleration of the baryon-specific analysis pipelines (briefly discussed in Section \ref{sec:analysistools}) is interesting and useful in its own right, and will be the subject of study in a full performance analysis report. 

There are further augmentations in \CRKHACC currently under development as well. These include neutrino extensions, as well as the capability to investigate primordial non-Gaussianity effects. As discussed in Section \ref{sec:ICs}, there are a number of intricacies involved with cosmological initial conditions (such as perturbation order and starting redshift), as well as additional considerations for multi-species and individual transfer functions. Circumstances are further complicated by the incorporation of massive neutrino cosmologies (e.g., \citealt{Zennaro2017}), where we intend to conduct comprehensive convergence investigations of these effects as well.

\acknowledgments

We thank Mike Owen and Cody Raskin for their continued collaboration and guidance, as well as their contributions in both the origination and extensions of \CRKSPH. SH acknowledges inspiring past discussions with Bryan (Bucky) Kashiwa in Los Alamos National Laboratory's Theoretical Division. We recognize the efforts of \HACC team members Hal Finkel, Patricia Larsen, Vitali Morozov, Adrian Pope, Esteban Rangel, and Tom Uram as well as helpful discussions and continuing collaboration with members of the \smaller{Nyx} team: Ann Almgren, Sol\`ene Chabanier, Zarija Luki\'c, Hannah Ross, and Jean Sexton. 

Argonne National Laboratory's work was supported under the U.S. Department of Energy contract DE-AC02-06CH11357. This research was supported by the Exascale Computing Project (17-SC-20-SC), a collaborative effort of the U.S. Department of Energy Office of Science and the National Nuclear Security Administration. This work used resources of the Oak Ridge Leadership Computing Facility, which is a DOE Office of Science User Facility supported under Contract DE-AC05-00OR22725. Additionally, this study utilized resources of the Argonne Leadership Computing Facility, which is a DOE Office of Science User Facility supported under Contract DE-AC02-06CH11357. This research also used resources of the National Energy Research Scientific Computing Center (NERSC), a U.S. Department of Energy Office of Science User Facility located at Lawrence Berkeley National Laboratory, operated under Contract No. DE-AC02-05CH11231 using NERSC award m3921 in 2021/22. CAFG was supported by NSF through grants AST-1715216, AST-2108230, and CAREER award AST-1652522; by NASA through grant 17-ATP17-0067; by STScI through grant HST-AR-16124.001-A; and by the Research Corporation for Science Advancement through a Cottrell Scholar Award. Lastly, NF would like to thank his mother for her contributions to the literacy of the paper; NF and co-authors hasten to add that no blame attaches to her for any remaining infelicities of language.


\bibliography{references,bib}
\bibliographystyle{apj}

\end{document}